\documentclass[12pt]{article} 
\usepackage[sectionbib]{natbib}
\usepackage{array,epsfig,fancyheadings,rotating}
\usepackage{xcolor}

\usepackage[CJKbookmarks=true,
bookmarksnumbered=true,
bookmarksopen=true,
colorlinks=true,
citecolor=blue,
linkcolor=blue,
anchorcolor=blue,
urlcolor=blue]{hyperref}

\usepackage{sectsty, secdot}
\sectionfont{\fontsize{12}{14pt plus.8pt minus .6pt}\selectfont}

\subsectionfont{\fontsize{12}{14pt plus.8pt minus .6pt}\selectfont}

\usepackage{amsmath}
\usepackage{amssymb}
\usepackage{amsfonts}
\usepackage{multirow}
\usepackage{amsthm}

\usepackage{xr}
\makeatletter
\newcommand*{\addFileDependency}[1]{
  \typeout{(#1)}
  \@addtofilelist{#1}
  \IfFileExists{#1}{}{\typeout{No file #1.}}
}
\makeatother

\newcommand*{\myexternaldocument}[1]{%
    \externaldocument{#1}%
    \addFileDependency{#1.tex}%
    \addFileDependency{#1.aux}%
}

\myexternaldocument{Supp_aux}

\usepackage[utf8]{inputenc}
\usepackage{amsmath}
\usepackage{bm}
\usepackage{amsthm}
\usepackage{comment}
\usepackage{amssymb}
\usepackage{enumerate}
\usepackage{epsfig}
\usepackage{graphics}
\usepackage[letterpaper]{geometry}
\usepackage{algpseudocode}
\usepackage{algorithm}
\usepackage{color,xcolor}
\usepackage{mathrsfs}
\usepackage{enumitem}
\usepackage{multirow}
\usepackage{booktabs} 
\usepackage {setspace}
\newcommand{\RNum}[1]{\uppercase\expandafter{\romannumeral #1\relax}}
\newcommand{\rNum}[1]{\lowercase\expandafter{\romannumeral #1\relax}}

\usepackage{subfigure}
\usepackage{caption}

\usepackage{titlesec}
\titlespacing*{\section}
{0pt}{1ex}{0.5ex}
\titlespacing*{\subsection}
{0pt}{0.5ex}{0.25ex}

\usepackage{soul}
\usepackage{tikz}
\usetikzlibrary{shapes,arrows}
\tikzstyle{startstop} = [rectangle,rounded corners, minimum width=3cm, text width = 3cm,minimum height=1cm,text centered, draw=black]
\tikzstyle{io} = [trapezium, trapezium left angle = 70,trapezium right angle=110,minimum width=4cm,minimum height=1cm, text width = 3.5cm,text centered,draw=black]
\tikzstyle{process} = [rectangle,minimum width=6cm, minimum height=1cm,text centered,text width =5cm,draw=black]
\tikzstyle{decision} = [diamond,aspect = 3,text centered,draw=black]
\tikzstyle{arrow} = [thick,->,>=stealth]
\tikzstyle{arrowdash} = [dashed,->,>=stealth]
\tikzstyle{arrowsum} = [dotted, line width=1.5pt,->,>=stealth]
\tikzstyle{arrowbig} = [thick, line width=2pt,->,>=stealth]

\tikzstyle{item} = [rectangle,rounded corners, minimum width=1.25cm, text width = 1.25cm, minimum height=1cm,text centered, draw=black]
\tikzstyle{itemlong} = [rectangle,rounded corners, minimum width=5.75cm, text width = 5.75cm, minimum height=1cm,text centered, draw=black]

\makeatletter
\def\thm@space@setup{\thm@preskip=0pt
\thm@postskip=0pt}
\makeatother
\newtheoremstyle{newstyle}      
{} 
{} 
{\mdseries} 
{} 
{\bfseries} 
{.} 
{ } 
{} 

\theoremstyle{newstyle}

\newtheorem{assumption}{Assumption}

\newtheorem{remark}{Remark}

\newtheorem{thm}{Theorem}
\newtheorem{defn}{Definition}

\newcounter{subassumption}[assumption]
\renewcommand{\thesubassumption}{(\textit{\alph{subassumption}})}
\makeatletter
\renewcommand{\p@subassumption}{\theassumption}
\makeatother
\newcommand{\subassumption}{
  \refstepcounter{subassumption}%
  \thesubassumption~\ignorespaces}
  
\newcounter{sublemma}[lemma]
\renewcommand{\thesublemma}{(\textit{\alph{sublemma}})}
\makeatletter
\renewcommand{\p@sublemma}{\thelemma}
\makeatother

\newcounter{subsublemma}[sublemma]

\makeatletter
\renewcommand{\p@subsublemma}{\thelemma\thesublemma}
\makeatother

\def\PTD{\operatorname{PTD}}
\def\NOtwo{$\rm NO_2$}
\def\NOx{$\rm NO_x$}
\def\Var{\mathrm{var}}
\def\Var{\mathrm{Var}}

\def\cov{\mathrm{cov}}

\def\dist{\mathrm{dist}}

\def\FDP{\operatorname{FDP}}

\def\FDR{\operatorname{FDR}}

\def\tm{\tilde{m}}

\def\wtt{\widetilde{t}}

\def\whbeta{\widehat{\beta}}
\def\whtau{\widehat{\tau}}
\def\whrho{\widehat{\rho}}
\def\whsig{\widehat{\sigma}}

\def\whR{\widehat{R}}
\def\whG{\widehat{G}}

\def\whT{\widehat{T}}
\def\whr{\widehat{r}}

\def\whV{\widehat{V}}
\def\whS{\widehat{S}}

\def\whpi{\widehat{\pi}}
\def\whmu{\widehat{\mu}}

\def\whsig{\widehat{\sigma}}

\def\1f{\mathbf{1}}

\def\tf{\mathbf{t}}

\def\Xf{\mathbf{X}}

\def\wf{\mathbf{w}}

\def\Rb{\mathbb{R}}
\def\Zb{\mathbb{Z}}
\def\Nb{\mathbb{N}}
\def\Pb{P}
\def\Eb{E} 

\def\bepsilon{\bm{\epsilon}}

\def\Scal{\mathcal{S}}

\def\Fcal{\mathcal{F}}
\def\Hcal{\mathcal{H}}
\def\Vcal{\mathcal{V}}

\def\Tcal{\mathcal{T}}

\def\Ncal{\mathcal{N}}
\def\argmax{\mathop{\mbox{arg\,max}}}

\def\tScal{\widetilde{\mathcal{S}}}

\def\sumstSm{\sum_{s\in\tScal_m}}
\def\sumsSm{\sum_{s\in\Scal_m}}

\def\op{o_{\Pb}}

\def\opn1{o_{\Pb_n}(1)}

\usepackage{enumitem}

\newcommand{\linsuin}[1]{{\bf\textcolor{red}{[Linsui(note): #1]}}}

\def\spacingset#1{\renewcommand{\baselinestretch}%
{#1}\small\normalsize} \spacingset{1}

\begin{document}

  \title{Powerful Spatial Multiple Testing via Borrowing Neighboring Information}
  \author{Linsui Deng$^{1}$,
  Kejun He$^{1}$\thanks{Correspondence: \href{mailto:kejunhe@ruc.edu.cn}{kejunhe@ruc.edu.cn} and \href{mailto:zhangxiany@stat.tamu.edu}{zhangxiany@stat.tamu.edu}.}~,
  Xianyang Zhang$^{2}$\footnotemark[1] 
  \\
$^{1}$The Center for Applied Statistics, Institute of Statistics and Big Data, \\ Renmin University of China, Beijing, China \\
$^{2}$Department of Statistics, Texas A\&M University, College Station, USA\\
}
  \date{}
  \maketitle



\begin{abstract}
Clustered effects are often encountered in multiple hypothesis testing of spatial signals. In this paper, we propose a new method, termed \textit{two-dimensional spatial multiple testing} (2d-SMT) procedure, to control the false discovery rate (FDR) and improve the detection power by exploiting the spatial information encoded in neighboring observations. The proposed method provides a novel perspective of utilizing spatial information by gathering signal patterns and spatial dependence into an auxiliary statistic. 2d-SMT rejects the null when a primary statistic at the location of interest and the auxiliary statistic constructed based on nearby observations are greater than their corresponding cutoffs. 2d-SMT can also be combined with different variants of the weighted BH procedures to improve the detection power further. A fast algorithm is developed to accelerate the search for optimal cutoffs in 2d-SMT. In theory, we establish the asymptotic FDR control of 2d-SMT under weak spatial dependence. Extensive numerical experiments demonstrate that the 2d-SMT method combined with various weighted BH procedures achieves the most competitive performance in FDR and power trade-off.

\end{abstract}

\noindent%

\spacingset{1.71} 

\section{Introduction}\label{sec:intro}

Large-scale multiple testing with spatial structure has become increasingly important in various areas, e.g., Functional Magnetic Resonance Imaging research, genome-wide association studies, environmental studies, and astronomical surveys. 
The essential task is identifying locations that exhibit significant deviations from the background to build scientific interpretations. 
Since thousands or even millions of spatially correlated hypotheses tests are often conducted simultaneously, incorporating informative spatial patterns to provide a powerful multiplicity adjustment for dependent multiple testing is becoming a significant challenge.

There has been a growing literature on spatial signal detection with false discovery rate control \citep[FDR,][]{Benjamini1995}. \cite{Heller2006} and \cite{Sun2015} proposed to perform multiple testing on cluster-wise hypotheses by aggregating location-wise hypotheses to increase the signal-to-noise ratio. \cite{Benjamini2007}, \cite{Sun2015} and \cite{Basu2018} defined new error rates to reflect the relative importance of hypotheses associated with different clusters, e.g, a hypothesis related to a larger cluster is more important than the one associated with a smaller cluster. \cite{Scott2015} considered a two-group mixture model with the prior null probability dependent on the auxiliary spatial information. \cite{Yun2020} proposed a spatial-adaptive FDR-controlling procedure by exploiting the mirror conservatism of the null p-values and the spatial smoothness under the alternative. 
\cite{Tansey2018} enforced spatial smoothness by imposing a penalty on the pairwise differences of log odds of hypotheses being signals  between adjacent locations.
Along a related line, \cite{Genovese2006} suggested to weight p-values or equivalently assign location-specific cutoffs by leveraging the tendency of hypotheses being null. This idea has been further developed in some recent papers to include different types of structural and covariate information. See, e.g., \cite{Ignatiadis2016}, \cite{Ang2018}, \cite{Cai2021}, \cite{zhang2022covariate}, and \cite{cao2022optimal}.

In many applications, signals tend to exhibit in clusters. As a result, hypotheses around a non-null location are more likely to be under the alternative than under the null. 
One way to account for spatially clustered signals is to screen out the locations where the average signal strength of the neighbors captured by an auxiliary statistic is weak \citep{Shen2002}. The locations passing the screening step are subjected to further analysis. This procedure suffers from the so-called selection bias as the downstream statistical inference needs to account for the selection effect from the screening step. A simple remedy is sample splitting \citep{Wasserman2009, Liu2022:PCScreen_Knockoff}, where a subsample is used to perform screening, and the remaining samples are utilized for the downstream inference. Sample splitting is intuitive and easy to implement, but it inevitably sacrifices the detection power because it does not excavate complete information. 
Furthermore, there are often no completely independent observations to conduct sample splitting in spatial settings.


We propose a new method named the two-dimensional spatial multiple testing (2d-SMT) procedure that fundamentally differs from the existing spatial multiple testing procedures. 2d-SMT consists of a two-dimensional rejection region built on two statistics, an auxiliary statistic $T_1(s)$ and a primary statistic $T_2(s)$, for each hypothesis. 
The first dimension utilizes the auxiliary information constructed from the neighbors of a location of interest to perform feature screening, which helps to increase the signal density and lessen the multiple testing burdens in the second dimension. The second dimension then uses a statistic computed from the data at the location of interest to pick out signals. 2d-SMT declares the hypothesis at location $s$ to be non-null if $T_1(s)\geq t_1$ and $T_2(s)\geq t_2$. The optimal cutoffs $t_1^\star$ and $t_2^\star$ are chosen to achieve the maximum number of discoveries while controlling the FDR at the desired level. 2d-SMT involves three main ingredients, designed to improve its robustness and efficiency: (1) accounting for the dependence between the auxiliary statistic and the primary statistic, which alleviates the selection bias; (2) borrowing spatial signal information through an empirical Bayes approach; and (3) accelerating the search for the bivariate cutoff through an efficient algorithm.
In a related study, \cite{Zhang2021} proposed the 2dFDR approach to detect the association between omics features and covariates of interest in the presence of confounding factors, borrowing information from confounder-unadjusted test statistics to boost the power in testing with confounder-adjustment. In contrast, 2d-SMT is designed for the spatial multiple testing by borrowing information from neighboring observations. 

The contribution of this work lies in its innovative methodology, theoretical analysis, and a new searching algorithm. First, 2d-SMT explores spatial information from a completely different perspective compared to the existing weighted procedures. It thus can be combined with these methods to improve power further. 
Examples include the group BH procedure \citep[GBH,][]{hu2010}, independent hypothesis weighting \citep[IHW,][]{Ignatiadis2016}, structure adaptive BH algorithm \citep[SABHA,][]{Ang2018}, and locally adaptive weighting and screening approach \citep[LAWS,][]{Cai2021}. 
The readers are referred to Section~\ref{sec:vary_null} for more details. 
Second, our asymptotic analysis allows weak spatial dependence, which goes beyond the independence assumption required by the existing empirical Bayes theory, thereby broadening its application scope. To the best of our knowledge, this is the first analytical framework where dependent observations are allowed within the context of empirical Bayes theory. 
Third, we develop an algorithm to overcome the computational bottleneck in finding the 2d cutoff values without sacrificing accuracy, which can be applied to procedures using two-dimensional rejection regions, including 2dFDR and 2d-SMT.

The rest of the paper is organized as follows. 
Section~\ref{sec:method} develops the 2d-SMT procedure, including the oracle procedure, the feasible procedure with estimated covariance structure, and the extension by combining it with various weighted BH procedures. Section~\ref{sec:imp_detail} discusses some implementation details. Section~\ref{sec:theory} establishes the asymptotic FDR control of the 2d-SMT procedure. In Sections~\ref{sec:simu} and \ref{sec:real_data}, extensive simulation studies and an analysis of ozone data demonstrate the effectiveness of the 2d-SMT procedure. 
Section~\ref{sec:discussion} concludes and points out a few future research directions.  Some additional details of the numerical experiments and technical proofs are presented in an online supplementary file.




\setcounter{equation}{0} 
\section{Method}\label{sec:method}
\allowdisplaybreaks
Consider a random field $\{X(s):s\in \Scal\}$ defined on a spatial domain $\Scal\subseteq \mathbb{R}^K$ with $K\geq 1$ that takes the form of $X(s)=\mu(s)+\epsilon(s)$,
where $\mu(s)$ is an unobserved process of interest and $\epsilon(s)$ is a mean-zero Gaussian process. The model is prevalent in spatial multiple testing across various domains, such as fMRI \citep{Heller2006}, environment study \citep{Sun2015}, temperature data analysis \citep{Huang2021}.
We are interested in examining whether $\mu(s)$ belongs to an indifference region $\mathcal{A}$. For example, $\mathcal{A}=\{\mu\in\mathbb{R}: \mu\leq\mu_0\}$ for a one-sided test and $\mathcal{A}=\{\mu\in\mathbb{R}: |\mu|\leq\mu_0\}$ for a two-sided test, where $\mu_0$ is some pre-specified value. 
The unobserved process $\mu(s)$ and the indifference region $\mathcal{A}$ induce a background statement $\theta(s)=\boldsymbol{1}\{\mu(s)\not\in \mathcal{A}\}$ on the spatial domain $\mathcal{S}$.
We define $\Scal_0=\{s\in\mathcal{S}:\theta(s)=0\}$ and $\Scal_1=\{s\in\mathcal{S}:\theta(s)=1\}$ as the sets of null and non-null locations respectively. 

We focus on the point-wise analysis, 
testing the hypothesis $\mathcal{H}_{0,s}:\mu(s)\in\mathcal{A}$ versus $\mathcal{H}_{a,s}:\mu(s)\notin \mathcal{A}$. 
At each location $s\in\mathcal{S}$, 
we make a decision $\delta(s)$, where $\delta(s)=1$ if $\mathcal{H}_{0,s}$
is rejected and $\delta(s)=0$ otherwise. Let $\Delta_1=\{s\in\mathcal{S}:\delta(s)=1\}$ be the set of rejections associated with the decision rule $\delta$. The false discovery rate (FDR) is defined as 
$$
\text{FDR}=\Eb\left(\frac{|\Delta_1\cap \mathcal{S}_0|}{1\vee |\Delta_1|}\right),   
$$
\noindent
where $|\cdot|$ denotes the cardinality of a 
set and $\Delta_1\cap \mathcal{S}_0$ is the set of false discoveries. 
We next present a spatial multiple testing procedure that borrows neighboring information to improve the signal
detection power 
without sacrificing the FDR control.



\subsection{Motivation}\label{sec:mov}
For clarity, we focus our discussions on the one-sided test for the rest of the paper, i.e., 
$$\mathcal{H}_{0,s}:\mu(s)\leq 0\quad\text{versus}\quad \mathcal{H}_{a,s}:\mu(s)>0.$$
\noindent
For each location $s\in\Scal$, we define a set of its neighbors as $\mathcal{N}(s)\subseteq \Scal\backslash \{s\}$. Because of the spatial dependence and smoothness encountered in many real applications, the set of neighboring observations $\{X(v):v\in\mathcal{N}(s)\}$ is expected to provide useful side information on determining the state of $\theta(s)$. To formalize this idea, we  consider two statistics, namely the auxiliary statistic 
\begin{equation}\label{eq:defT1}
	T_{1}(s)=\frac{1}{\tau(s)}\sum_{v\in \mathcal{N}(s)}X(v)
\end{equation}
\noindent based on the averaged observed values in the neighborhood of $s$ and the primary statistic $T_{2}(s)=\sigma^{-1}(s)X(s)$ based on the observation from the location of interest, where $\sigma^2(s)=\Var\{\epsilon(s)\}$ and $\tau^{2}(s)=\sum_{v,v'\in \mathcal{N}(s)} \cov\left\{\epsilon(v),\epsilon(v')\right\}$. 
For $s\in\Scal$, 
we have $T_{1}(s)=\xi(s)+V_1(s)$ and $ T_{2}(s)=\sigma^{-1}(s)\mu(s)+V_{2}(s)$ 
where $\xi(s)=\tau^{-1}(s)\sum_{v\in\mathcal{N}(s)}\mu(v)$ and\vskip-1em$$\left(\begin{array}{c}V_{1}(s) \\V_{2}(s)\end{array}\right)\sim\mathcal{N}\left(\left(\begin{array}{c}0 \\0\end{array}\right),\left(\begin{array}{cc}1 & \rho(s) \\\rho(s)  & 1\end{array}\right)\right)$$\vskip-0em\noindent with $\rho(s)= \{\sigma(s)\tau(s)\}^{-1} \sum_{v\in \mathcal{N}(s)} \cov\{\epsilon(s),\epsilon(v)\}$. 
Our method is motivated by the following two-stage procedure.
At stage~1, we use the auxiliary statistic $T_1(s)$ to screen out the nulls based on the belief that observations around a non-null location tend to take larger values on average. At stage~2, we use $T_2(s)$ to pick out signals among those who survive from stage~1. Given the cutoffs $(t_1,t_2)$ of these two stages, the procedure can be described as follows:
\begin{itemize}[leftmargin=60pt,itemsep=0pt,parsep=0pt,topsep=0pt]
    \item[Stage 1.] Use the auxiliary statistic $T_1(s)$ to determine a preliminary set of signals $\mathcal{D}_1=\left\{s\in\mathcal{S}: T_1(s)\geq  t_1\right\}$.
    \item[Stage 2.] Reject $\mathcal{H}_{0,s}$ for $T_2(s)\geq  t_2$ and $s\in \mathcal{D}_1$. As a result, the final set of discoveries is given by $\mathcal{D}_2=\left\{s\in\mathcal{S}: T_1(s)\geq  t_1,T_2(s)\geq  t_2\right\}$.
\end{itemize}
Since the screening step reduces the multiple testing burden for the second stage, we expect the above method to be more powerful than the traditional method based only on the primary statistic. Indeed, the 1d rejection region is a special case of the 2d rejection region $\left\{s\in\mathcal{S}: T_1(s)\geq  t_1,T_2(s)\geq t_2\right\}$ by setting $t_1\leq \min_{s\in\mathcal{S}} T_1(s)$, i.e., the first stage preserves all locations. If we select the two cutoffs one by one, then the choice of $t_2$ should consider the selection effect from the first stage. 
Here we propose a new method to address this issue by simultaneously selecting the two cutoffs, which we name the \textit{2-dimensional} (2d) procedure.

\subsection{Approximation of the false discovery proportion}\label{sec:appFDP}
Note that $\mathcal{H}_{0,s}$ is rejected when $T_{1}(s)\geq t_1$ and $T_{2}(s)\geq t_2$. 
Recalling that $\mathcal{S}_0$ is the set of true nulls, the false discovery proportion ($\FDP$) is then given by
$$
\FDP(t_1,t_2)=\frac{\sum_{s\in\mathcal{S}_0}\1f\left\{T_{1}(s)\geq t_{1}, T_{2}(s)\geq t_{2}\right\}}{1\vee R(t_1,t_2)}\leq 
\frac{\sum_{s\in\mathcal{S}_0}\1f\left\{V_{1}(s)+\xi(s)\geq t_{1}, V_{2}(s)\geq t_{2}\right\}}{1\vee R(t_1,t_2)},
$$
\noindent where $R(t_1,t_2)=\sum_{s\in \mathcal{S}} \1f\left\{T_{1}(s)\geq t_{1}, T_{2}(s)\geq t_{2}\right\}$ corresponds to the total number of rejections, and the inequality holds because $\mu(s)\leq 0$ for $s\in\mathcal{S}_0$. 
Motivated by the law of large numbers, we follow the approaches in multiple testing \citep{Benjamini1995,Storey2002}, further developed for detecting spatial signals 
\citep{Benjamini2007,Sun2015,Cai2021}, to substitute the numerator of the right-hand side (RHS) of the above inequality by its expected value. This substitution leads to an asymptotic upper bound of FDP as  
\begin{equation}\label{eq:up1}
\begin{aligned}
\FDP(t_1,t_2)  &\lesssim \frac{\sum_{s \in \mathcal{S}_0}  \Pb\left\{V_{1}(s)+\xi(s)\geq t_{1},V_{2}(s)\geq t_{2}\right\} }{1\vee R(t_1,t_2)} \\ 
&:= \frac{\sum_{s \in \mathcal{S}_0}L\left\{t_{1}, t_{2},\xi(s), \rho(s)\right\}}{1\vee R(t_1,t_2)}, 
\end{aligned}  
\end{equation}
\noindent where $L\left\{t_{1}, t_{2},\xi(s), \rho(s)\right\}=\Pb\left\{V_{1}(s)+\xi(s)\geq t_{1},V_{2}(s)\geq t_{2}\right\}.$
The major challenge here is the estimation of the expected number of false rejections given by $\sum_{s \in \mathcal{S}_0}L\left\{t_{1}, t_{2},\xi(s), \rho(s)\right\}$, which involves a large number of nuisance parameters $\xi(s)$.
To overcome this difficulty, we adopt an empirical Bayes viewpoint 
to borrow spatial information across different locations and directly estimate the expected number of false rejections without estimating individual $\xi(s)$ at each location explicitly.

\subsection{Nonparametric empirical Bayes}\label{sec:non_eb}
{ Let $G_{\Scal_0}$ be the empirical distribution of $\{\xi(s): s\in \Scal_0\}$. The expected number of false rejections in \eqref{eq:up1} can be approximated by $\sum_{s \in \mathcal{S}_0}\int L\left\{t_{1}, t_{2},x, \rho(s)\right\}d G_{\Scal_0}(x)$. However, directly estimating $G_{\Scal_0}$ is challenging as the auxiliary statistics $\{T_1(s):s\in\mathcal{S}\}$ blend information from both the null and alternative hypotheses. To overcome this difficulty, we observe that $\xi(s)$ typically takes greater values under the alternative than under the null and $L\left\{t_{1}, t_{2},x, \rho(s)\right\}$ is a monotonically increasing function of $x$. Thus, we have
\begin{equation}\label{equ:FDGS0}
 \begin{split}
     \sum_{s \in \mathcal{S}_0}\int L\left\{t_{1}, t_{2},x, \rho(s)\right\}d G_{\Scal_0}(x) & \lesssim  
 \sum_{s \in \mathcal{S}_0}\int L\left\{t_{1}, t_{2},x, \rho(s)\right\}d G_{\Scal}(x) 
 \\ 
 &\leq  \sum_{s \in \mathcal{S}}\int L\left\{t_{1}, t_{2},x, \rho(s)\right\}d G_{\Scal}(x), 
 \end{split}
 \end{equation}
 \noindent where $G_{\Scal}$ is the empirical distribution $\{\xi(s): s \in \Scal\}$.  Consequently, we aim to estimate $G_{\Scal}$ based on the whole set of auxiliary statistics $\{T_1(s):s\in\mathcal{S}\}$ through the nonparametric empirical Bayes (NPEB) approach.}


The estimation in NPEB can be achieved by maximizing the marginal distribution of $T_1(s)=\xi(s)+V_1(s)$, which is given by
$f_{G_{\Scal}}(x)=\int \phi(x-u) d G_{\Scal}(u)$ with $\phi$ denoting the density function of the standard normal distribution. 
Classical empirical Bayes methods often assume independence among the observations, which is violated in our case due to spatial dependence. 
To reduce the dependence, we select a subset $\widetilde{\mathcal{S}}$ of $\mathcal{S}$ such that any two
points in $\widetilde{\mathcal{S}}$ have a distance larger than some cutoff $c_0$ (so that the dependence between $T_1(v)$ and $T_1(v')$ for any $v,v'\in \widetilde{\mathcal{S}}$ is sufficiently weak).
Following \cite{kiefer1956consistency} and \cite{Zhang2009}
, we consider the general maximum likelihood estimator (GMLE) defined as
\begin{equation}\label{equ:GMLE_hat}
	\widetilde{G}_{\widetilde{\mathcal{S}}}=\argmax_{G\in\mathcal{G}} \sum_{s\in\widetilde{\mathcal{S}}} \log f_{G}\{T_{1}(s)\},
\end{equation}
\noindent where $\mathcal{G}$ represents the set of all probability distributions on $\mathbb{R}$ and $f_{G}(x) = \int \phi(x-u) dG(u)$ is the convolution between $G$ and $\phi$. {Our theoretical analysis in Lemma~\ref{lemma:cnvrg_emp_bayes_cp} of the supplement shows that the estimated GMLE is close to the limit of the empirical distribution of $\{\xi(s):s\in\Scal \}$, denoted by $G_0$; see Assumption~\ref{ass:emp_bayes} for a formal definition of $G_0$.} 
The optimization in \eqref{equ:GMLE_hat} can be cast as a convex optimization problem that can be efficiently solved by modern interior point methods \citep{koenker2014convex}. 

\subsection{2d spatial multiple testing procedure}\label{sec:2dSMT}
We now describe a procedure to select the two cutoffs simultaneously. {In view of \eqref{eq:up1}--
\eqref{equ:GMLE_hat}, 
we consider an approximated upper bound for $\mathrm{FDP}(t_1,t_2)$ given by 
\begin{equation}\label{eq:up2}
\widetilde{\text{FDP}}(t_1,t_2):=\frac{\sum_{s\in\mathcal{S}} \int L\left\{t_{1}, t_{2},x,\rho(s)\right\} d \widetilde{G}_{\widetilde{\mathcal{S}}}(x) }{1\vee R(t_1,t_2)}.
\end{equation}
As shown in Lemma~\ref{lemma:cnvrg_discovery} of the supplement, $\sum_{s \in \mathcal{S}}\int L\left\{t_{1}, t_{2},x, \rho(s)\right\}d G_0(x)$ can be consistently estimated by the numerator of \eqref{eq:up2}.}  
For a desired FDR level $q\in(0,1)$, the 2d-SMT procedure chooses the optimal cutoff such that 
$(\widehat{t}^\star_1,\widehat{t}^\star_2)=\argmax_{(t_1,t_2)\in\mathcal{F}_q} R(t_1,t_2)$,
where $\mathcal{F}_q=\{(t_1,t_2)\in\mathbb{R}^2:\widetilde{\text{FDP}}(t_1,t_2)\leq q \}$. 

{We argue that the 2d-SMT procedure is generally more powerful than the classical BH procedure based on the primary statistics alone. When setting $t_1=-\infty$, 2d-SMT is equivalent to BH, as it fails to exclude any hypotheses at the first stage so that signal detection relies solely on the second stage, and our FDP estimator in \eqref{eq:up2} is equivalent to the FDP estimator in the BH procedure, i.e.,
$$\frac{\sum_{s \in \mathcal{S}}\int  \Pb\left\{V_{1}(s)+x\geq -\infty,V_{2}(s)\geq t_{2}\right\}d \widetilde{G}_{\widetilde{\mathcal{S}}}(x) }{1\vee R(-\infty,t_2)} = \frac{\sum_{s \in \mathcal{S}} \Pb\left\{V_{2}(s)\geq t_{2}\right\} }{1\vee \sum_{s\in\Scal}\boldsymbol{1}\left\{T_{2}(s)\geq t_{2}\right\}} \, .
$$
\noindent The 2d-SMT procedure has the flexibility to choose an additional cutoff $t_1$ to maximize the number of rejections and guarantees to make more rejections. Section \ref{sec:theory} will show 2d-SMT has asymptotic FDR control; thus more rejections typically translate into a higher power.

{ Similar to the FDP estimator of the BH procedure, the conservatism of \eqref{eq:up2} arises partly from expanding the index set  $\Scal_0$ of the summation in \eqref{eq:up1} to $\Scal$ in \eqref{equ:FDGS0}. For a target FDR level $q$, the realized FDR level of the BH procedure is approximately $\pi_0 q$, where $\pi_0=|\Scal_0|/|\Scal|$ is the null proportion. The conservatism motivates us to estimate the null proportion to improve power; see Section~\ref{sec:prop}. Theorem~\ref{rmk:power} rigorously proves that our procedure is more powerful than the one using primary statistics alone.
} 

} 

\begin{remark}
\rm
\cite{Shen2002} and \cite{Huang2021} considered the case where $\mu$ has a sparse wavelet representation. {One of the goal is to detect the significant wavelet coefficients while controlling the FDR.} 
Observing that the wavelet coefficients within each scale and across different scales are related, they screened the wavelet coefficients based on the largest adjacent wavelet coefficients to gain more power. 
The generalized degrees of freedom determine the number of locations for the subsequent FDR-controlling procedure. Our approach can be potentially applied to their settings by exploring the signal structure encoded in the neighboring wavelet coefficients.
\end{remark}

\subsection{Estimating the null proportion}\label{sec:prop}
It is well known that when the number of signals is a substantial proportion of the total number of hypotheses, the BH procedure will be overly conservative. 
We develop a modification of Storey's approach \citep{Storey2002,Storey2004} to incorporate the estimation of the null proportion. As a motivation, we assume that $T_{2}(s)$ approximately follows the mixture model $T_{2}(s)\sim\pi_{0}\Ncal(\mu_0(s),1) + (1-\pi_{0}) \Ncal(\mu_1(s),1)$,
where $\mu_0(s)\leq 0$, $\mu_1(s)>0$, and $\pi_{0}$ is the prior probability that $s\in\Scal_0$. 
Let $\Phi$ be the cumulative distribution function of $\Ncal(0,1)$. 
Fixing some $\lambda\in \Rb$, we have
$$
\begin{aligned}
	\Pb\{T_{2}(s)<\lambda \}
	&= \pi_{0} \Pb\{\Ncal(\mu_0(s),1)< \lambda\} + (1-\pi_{0})\Pb\{\Ncal(\mu_1(s),1)< \lambda\}\\
	&\geq \pi_{0}\Pb\{\Ncal(\mu_0(s),1)< \lambda \}
 \geq \pi_{0}\Phi(\lambda), 
\end{aligned}
$$
\noindent where the first inequality in the second line is tighter if $(1-\pi_{0}) \Pb \{\Ncal(\mu_1(s),1)<\lambda\}$ is closer to zero and the second inequality becomes an equality when $\mu_0(s)=0$. 
The above derivation suggests a conservative estimator for $\pi_0$ given by
\begin{equation}\label{equ:est_pi0}
	\widehat{\pi}_0 
	:=\frac{\sum_{s\in\Scal} \boldsymbol{1}\{T_{2}(s)<\lambda\}}{\left|\Scal\right|\Phi(\lambda)}\approx \frac{\Pb\{T_{2}(s)<\lambda\}}{ \Phi(\lambda)}\geq \pi_0.
\end{equation}

\subsection{A feasible procedure}\label{sec:f_method}

\begin{figure}[!h]
     \centering
     \includegraphics[width=0.9\textwidth]{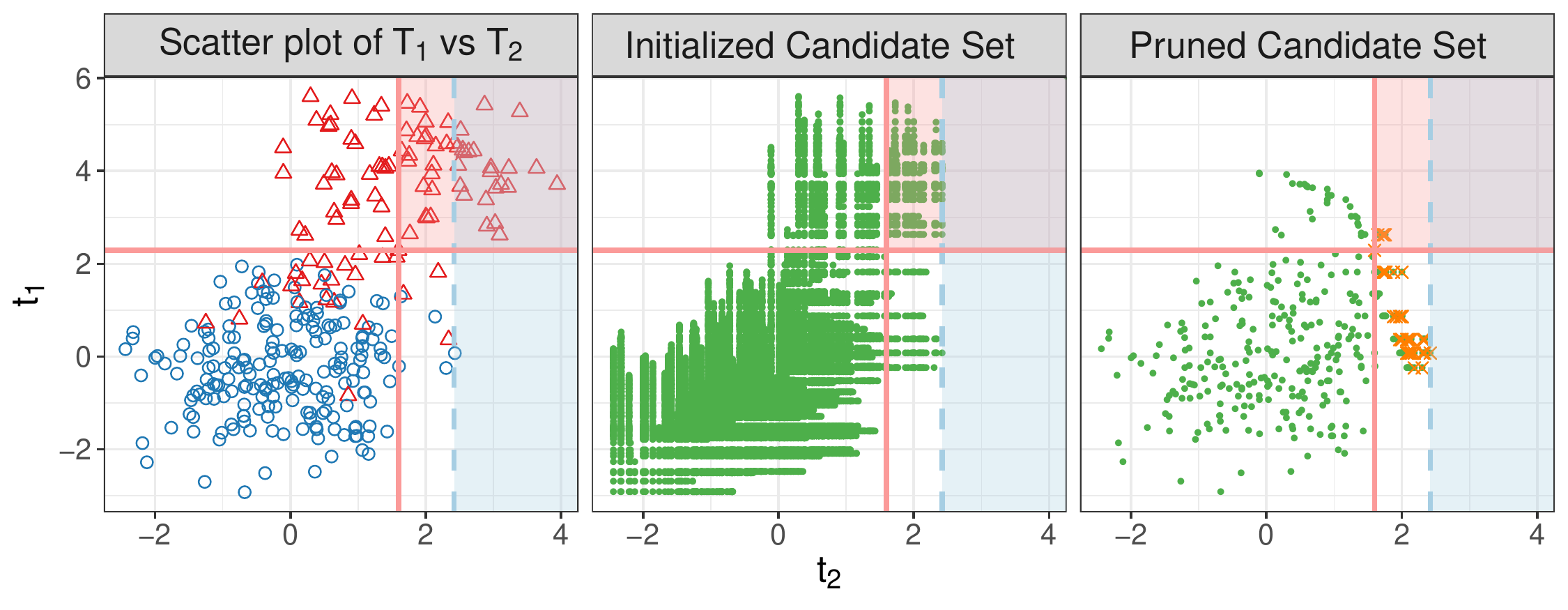}
        \caption{
        An illustration of 2d-SMT and the fast searching algorithm. 
        The spatial domain $\mathcal{S}$ includes 300 points. The red triangle ($\triangle$) and blue circle ($\circ$) in the left panel denote non-null and null locations, respectively. The solid and dashed lines are the cutoffs for the 2d-SMT procedure and the BH procedure, respectively. The orange cross ($\times$) in the right panel corresponds to the cutoff whose corresponding FDP is less than 0.1. 
        The set $\mathcal{T}'$ in the middle panel contains 14,534 points, whereas the set after our proposed pruning steps in the right panel significantly reduces the number of candidates to 345. 
        }\label{fig:Select}
\end{figure}

So far we have assumed that the spatial covariance function $k(s,s')=\text{cov}\{\epsilon(s),\epsilon(s')\}$ of the error process is known. 
In practice, we need to estimate the spatial covariance function $\text{cov}\{\epsilon(s),\epsilon(s')\}$, which has been widely investigated in the literature \citep{Sang2012, 
Katzfuss2021}; see Section~\ref{sec:cov_est} of the supplement for more details. Given the estimated covariance function, we let $\whT_{1}(s)$ and $\whT_{2}(s)$ respectively denote the feasible statistics of $T_1(s)$ and $T_2(s)$ by replacing $(\sigma(s),\tau(s),\rho(s))$ with their estimates $(\widehat{\sigma}(s),\widehat{\tau}(s),\widehat{\rho}(s))$. 
Let $\widehat{G}_{\tScal}(u)$ denote the nonparametric empirical Bayes estimate of $G_{\Scal}$ based on $\{\whT_1(s):s\in\widetilde{\mathcal{S}}\}$.
We propose the following FDP estimate which accounts for the null proportion using the idea in Section \ref{sec:prop}, 
\begin{equation}\label{equ:def_tfdp}
   \widehat{\FDP}_{\lambda,\tScal}(t_{1}, t_{2}):=\frac{\sum_{s\in\Scal} \boldsymbol{1}\{\whT_{2}(s)<\lambda\}}{m\Phi(\lambda)}
   \frac{\sum_{s\in\Scal} \int L\left\{ t_{1}, t_{2}, x, \whrho(s)\right\} d \whG_{\tScal}(x) }{1\vee\whR(t_1,t_2)},
\end{equation}
\noindent where $\whR(t_1,t_2)=\sum_{s\in \mathcal{S}} \1f \big\{\whT_{1}(s)\geq t_{1}, \whT_{2}(s)\geq t_{2} \big\}$. Thus, given the desired FDR level $q\in(0,1)$, the optimal rejection cutoffs are defined as
\begin{equation}\label{eq-opt}
(\widetilde{t}_{1}^{\star},\widetilde{t}_{2}^\star)=\argmax_{(t_1,t_2)\in \Fcal_q}\whR(t_1,t_2),
\end{equation}
\noindent where $\Fcal_q=\{(t_1,t_2)\in\mathbb{R}^2:\widehat{\FDP}_{\lambda,\tScal}(t_1,t_2)\leq q \}$. The left panel in Figure~\ref{fig:Select} exemplifies the cutoffs for the BH and 2d-SMT procedures with the target FDR level at 10\%. 
Compared to the BH, the 2d-SMT realizes
a lower cutoff for $T_2(s)$ (the red vertical line) as it excludes locations exhibiting weak neighboring signals with a cutoff for $T_1(s)$ (the red horizontal line). The lower cutoff for $T_2(s)$ in 2d-SMT leads to more true rejections in this example. 

\subsection{Spatial varying null proportions and cutoffs}\label{sec:vary_null}
Our proposed 2d-SMT is a flexible framework that can accommodate weighted BH procedures (wBH), a broad class of multiple testing procedures. The wBH method leverages the hypothesis heterogeneity by assigning location-specific cutoffs.
According to wBH, $\Hcal_{0,s}$ will be rejected with the rule $p(s):=1-\Phi(\whT_2(s))\leq \min\{\tau,w(s)t\}$, $\tau\in(0,1]$, where $\tau$ is the censoring level for all p-values and $w(s)$ is a location-specific weight that encodes external information for location $s$.
Apparently, the rejection rule is equivalent to assigning location-specific cutoffs to the primary statistic, i.e., $\whT_2(s)\geq \Phi^{-1}(1-\min\{\tau,w(s)t\})$. 
Inspired by the idea behind wBH, we extend  
2d-SMT by allowing location-specific cutoffs to further incorporate 
external information on the prior null probability and signal distribution.
Specifically, we reject $\mathcal{H}_{0,s}$ whenever $\whT_1(s)\geq c(t_1,s)$ and $\whT_2(s)\geq c(t_2,s)$,
where $c(t,s)= \Phi^{-1}(1-\min\{\tau,w(s)t\})$. 
With some abuse of notation, we let 
\begin{equation}\label{equ:def_tfdp2}
   \widehat{\FDP}_{\lambda,\tScal}(t_{1}, t_{2}):=
   \frac{\sum_{s\in\Scal}\widehat{\pi}_0(s)\int L\left\{ c(t_{1},s), c(t_{2},s), x, \whrho(s)\right\} d \whG_{\tScal}(x) }{1\vee\whR(t_1,t_2)},
\end{equation} 
\noindent where $\widehat{\pi}_0(s)$ is an estimate of the null proportion $\pi_0(s)=\Pb\{\theta(s)=0\}$ at location $s$. We reject $\mathcal{H}_{0,s}$ if $T_1(s)\geq c(\widetilde{t}_{1}^{\star},s)$ and $T_2(s)\geq c(\widetilde{t}_{2}^{\star},s)$, where
$(\widetilde{t}_{1}^{\star},\widetilde{t}_{2}^\star)$ is the solution to \eqref{eq-opt} with the FDP estimate given in (\ref{equ:def_tfdp2}).

There have been extensive recent studies on the choice of $w(s)$ in wBH. Examples include GBH \citep{hu2010}, IHW \citep{Ignatiadis2016,Ignatiadis2021}, SABHA \citep{Ang2018} and LAWS \citep{Cai2021}. The weights in these examples are either proportional to $1/\pi_0(s)$ or $\{1-\pi_0(s)\}/\pi_0(s)$, where $\pi_0(s)$ can be estimated using various approaches. 
In the spatial setting, we often use the location associated with each hypothesis as the external covariate to estimate $\pi_0(s)$. 

{ 
If additional types of covariate information are available, the proposed 2d-SMT framework is able to leverage both covariate and spatial information. 
Some covariate-adaptive FDR procedures can be utilized by assigning weights based on an estimation of covariate-specific null proportions; and the spatial information is, again, captured by the auxiliary test statistics.
In Section~\ref{sec:covarSpa} of the supplement, we present a simulation experiment that uses the group information as the covariate. 
The simulation results show that integrating both the group covariate and the spatial information can further improve the detection power.
}

\section{Implementation Details}\label{sec:imp_detail}
In this section, we discuss a few crucial points for the implementation of our method.
First of all, the auxiliary statistic $\whT_1(s)$ requires the specification of a pre-chosen neighborhood $\mathcal{N}(s)$ for each location. In our implementation, we let $\mathcal{N}(s)$ be the set of the $\kappa$-nearest neighbors around location $s$, and find that 2d-SMT is not quite sensitive to the choice of $\kappa$ and shows satisfactory power improvement provided $2\leq \kappa\leq 7$; see Section~\ref{sec:rob_nei_num} of the supplement for more details. Alternatively, one can also choose the locations within a certain distance from the location $s$ of interest as its neighbors. 
Furthermore, the neighborhood set of location $s$ can also be adaptively determined when external information that is independent of $\{X(s):s\in\mathcal{S}\}$ is provided; see Section~\ref{sec:adaneigh} of the supplement for more details.

Second, the auxiliary statistics used in NPEB estimation, $\{\whT_1(s):s\in\tScal\}$, should be from far enough spatial locations to weaken their spatial dependency. In practice, we choose $\tScal\subset\Scal$ whose neighbors have no overlaps.

Third, when the signal is sparse, the estimation of the number of false rejections in the 2d-SMT procedure may be unstable. Inspired by the idea of Knockoff+, we add a small offset to improve the selection stability. 
More precisely, we replace 
$\sum_{s\in\Scal} \int L\left\{ t_{1}, t_{2}, x, \whrho(s)\right\} d \whG_{\tScal}(x)$ in \eqref{equ:def_tfdp} by
$\sum_{s\in\Scal} \int L\left\{ t_{1}, t_{2}, x, \whrho(s)\right\} d \whG_{\tScal}(x)+q$,
where $q$ is the target FDR level. This replacement improves the selection stability for sparse signals but does not influence the power and FDR control for dense signals.


Finally, finding the optimal cutoffs in 2d-SMT requires solving the discrete constrained optimization problem \eqref{eq-opt}. 
Due to the discrete nature of \eqref{eq-opt}, the solution can be obtained if we replace $\mathcal{F}_q$ by $\{(t_1,t_2)\in\mathcal{T}:\widehat{\FDP}_{\lambda,\tScal}(t_1,t_2)\leq q \}$,
where $\mathcal{T}=\{(\whT_1(s),\whT_2(s')): s,s'\in \mathcal{S}\}$ is the set of all candidate cutoffs. A naive grid search algorithm would evaluate $\widehat{\FDP}_{\lambda,\tScal}$ at $|\mathcal{S}|^2$ values, which is computationally prohibitive for a large number of spatial locations. 
To overcome the computational bottleneck, we propose a fast algorithm to utilize the specific structure of \eqref{eq-opt} through the following three steps. We briefly introduce the basic idea below 
and defer the comprehensive discussion to Section~\ref{ref:searching_alg} of the supplement.

\begin{algorithm}[h!]
	\small
		\caption{Fast Searching Algorithm.	\label{alg:fast_alg}} 
		\hspace*{0.02in} {\bf Input:} 
		Test Statistics $\left\{\left(\whT_1(s), \whT_{2}(s)\right):s\in\Scal\right\}$; target FDR level $q$;
		
		\hspace*{0.02in} {\bf Initialization:} 
		\begin{algorithmic}[1]
			\State Initialize $(\widetilde{t}_{1}^\star,\widetilde{t}_{2}^\star,R_{\max},\FDP_{\min})=\left(-\infty,\widetilde{t}_2^{\#},\whR(-\infty,\widetilde{t}_2^{\#}),\widehat{\FDP}_{\lambda,\tScal}(-\infty,\widetilde{t}_2^{\#})\right)$
		\end{algorithmic}
		\hspace*{0.02in} {\bf Search Step:} 
		\begin{algorithmic}[1] 
			\For{$i = 1,2,\ldots,\breve{m}$}
			\State Calculate $R=\whR(t_{1,i,1},t_{2,i})$;
			\State Set $j=1+\max(0,R_{\max}-R)$;
			\While{$j \leq m_i$}
			\State Calculate $R=\whR(t_{1,i,j},t_{2,i})$ and $\widehat{\FDP}=\widehat{\FDP}_{\lambda,\tScal}(t_{1,i,j},t_{2,i})$ according to \eqref{equ:def_tfdp};
			\If{$R=R_{\max}$ and $\widehat{\FDP} <\FDP_{\min}$ or $R>R_{\max}$ and $\widehat{\FDP} \leq q$}
			\State $(\widetilde{t}_{1}^\star,\widetilde{t}_{2}^\star,R_{\max},\FDP_{\min})=(t_{1,i,j},t_{2,i},R,\widehat{\FDP})$;
			\State Update $j=j+1$;
			\Else
			\State Calculate $R_{\text{req}}=\lceil \widehat{\FDP}\times R /q\rceil$;
			\State Update $j=j+\max(1,R_{\text{req}}-R)$;
			\EndIf
			\EndWhile
			\EndFor
		\end{algorithmic}
	\hspace*{0.02in} {\bf Output:} 
	Rejection cutoff $(\widetilde{t}_{1}^\star,\widetilde{t}_{2}^\star)$.
\end{algorithm}


\noindent \textbf{Step 1.} We maintain the cutoffs that achieve the minimum FDP among all the cutoffs realizing the same rejection sets. The derivation in Section \ref{ref:searching_alg} of the supplement suggests that we only need to consider the following set of
candidate cutoffs: 
$$
 \Tcal'=
 \left\{ (\whT_{1}(s_l),\whT_{2}(s_k))
    :\whT_{1}(s_l)\leq \whT_{1}(s_k) \mbox{ and } l\leq k, \, k=1,2,\ldots,m\right\} \cup\left\{(\infty,\infty)\right\}.
$$
\noindent \textbf{Step 2.} Let 
$\widetilde{t}_2^{\#}$ be the minimum value satisfying $\widehat{\FDP}_{\lambda,\tScal}(-\infty,t_2)\leq q$. It is not hard to see that the optimal cutoff for the primary statistic in 2d-SMT is no more than $\widetilde{t}_2^{\#}$. Hence we can reduce the set of candidate cutoffs to
$\mathcal{T}''=\left\{(t_1,t_2)\in\Tcal':t_2\leq \widetilde{t}_2^{\#}\right\}$. 
\\ \textbf{Step 3.} Denote the elements in $\mathcal{T}''$ by $(t_{1,i,j},t_{2,i})$ for $i=1,2\dots,\breve{m}$ and $j =1,\dots, m_i$, where $\breve{m}$ is the number of $\widehat{T}_2(s)$'s that are smaller than or equal to $\widetilde{t}_2^{\#}$. 
Suppose the points are sorted in the following way: (1) $t_{2,1}>t_{2,2}>\cdots>t_{2,\breve{m}}$; (2) 
$t_{1,i,1}>t_{1,i,2}>\cdots>t_{1,i,m_i}$ for all $1\leq i\leq \breve{m}$.
Our algorithm involves two loops. In the outer loop, we search the cutoff for the primary statistic, while in the inner loop, we search the cutoff for the auxiliary statistic. The key idea here is to skip those cutoffs in the inner loop that are impossible to procedure a value of $\widehat{\FDP}$ equal to or below the level $q$.
For example, consider a cutoff $(t_{1,i,j},t_{2,i})$ which induces $R$ rejections with $\widehat{\FDP}_{\lambda,\tScal}(t_{1,i,j},t_{2,i})=2q$. Then the next cutoff, denoted as $(t_{1,i,j^\prime},t_{2,i})$, needs to induce at least $2R$ rejections to ensure that $\widehat{\FDP}_{\lambda,\tScal}(t_{1,i,j^\prime},t_{2,i})\leq q$. When there is no tie, increasing $j$ by $k$ brings exactly $k$ more rejections. Therefore, the next possible cutoff to be examined is $(t_{1,i,j+R},t_{2,i})$. 
The middle and left panels in Figure~\ref{fig:Select} illustrate how the set of candidate cutoff values can be reduced by Step 3.


\section{Asymptotic Results}\label{sec:theory}
In this section, we investigate the asymptotic property of the 2d-SMT procedure. We observe $\{X(s):s\in \Scal_m \}$, where $\Scal_m=\{s_1,s_2,\cdots,s_m\} \subseteq \mathcal{U} \subseteq \Rb^K$, $\mathcal{U}_0 = \{s\in\mathcal{U}_0:\theta(s)=0\}$ and $\mathcal{U}_1=\{s\in\Scal:\theta(s)=1\}$. 
We denote by $\Scal_{0,m}=\Scal_m\cap \mathcal{U}_0$ with $m_0=|\Scal_{0,m}|$ be the set of null locations and let $\Scal_{1,m}=\Scal_m\cap  \mathcal{U}_1$ with $m_1=|\Scal_{1,m}|$ be the set of non-null locations. We further let $\tScal_m\subseteq \Scal_m$ with $|\tScal_m|=\tm$ be the set of randomly selected locations for implementing the NPEB. Our asymptotic analysis requires the following regularity conditions. 

\begin{assumption}[Spatial Domain]\label{ass:incr_dom}
The spatial domain { $\mathcal{U}\subset\Rb^K$ is infinitely countable.} There exist $0 < \Delta_{l} < \Delta_{u} < \infty$, such that for every element in $\mathcal{U}$, the distance to its nearest neighbor is bounded from below and above respectively by $\Delta_l$ and $\Delta_u$, i.e., $\Delta_{l}\leq\inf_{s^\prime \neq s}\dist(s,s^\prime)\leq\Delta_{u},$ for all $s\in\mathcal{U}$,
where $\dist(\cdot, \cdot)$ is the Euclidean distance of two points.
\end{assumption}

\begin{assumption}[Neighborhood]\label{ass:neigh}
For each location $s$ in $\mathcal{U}$, its neighborhood $\Ncal(s)\subset \Scal\backslash\{s\}$ used in 2d-SMT is its nearest neighbors with size uniformly upper bounded by some positive integer $N_{nei} \in\Nb^+$, i.e., $0< |\Ncal(s) | < N_{nei}, ~ \forall s \in \mathcal{U}.$
\end{assumption}
Assumption~\ref{ass:incr_dom} states that the distance between any location in $\mathcal{U}$ and its nearest neighbor is moderately uniform \citep{Cressie1993Spatial, Jenish2012}. One example satisfying Assumption~\ref{ass:incr_dom} is the lattice $\mathcal{U}=\Zb^K$ where $\Delta_l=\Delta_u=1$.
Particularly, the lower bound ensures $\max_{s,s^\prime\in\Scal_m} \dist(s,s^\prime)$ tend to infinity as $m$ increases. 
Assumption \ref{ass:neigh} specifies the neighborhood of each location and requires the number of neighbors to be bounded for each location. 

\begin{assumption}[Second-order Structure]\label{ass:cnvrg_cov}
The variance and covariance of the error process $\epsilon(s)$ satisfy: 
\subassumption \label{ass:cnvrg_cov:a} 
There exist positive constants $B_{ud,\sigma}$, $B_{up,\sigma}$, $B_{ud,\tau}$, and $B_{up,\tau}$, such that $B_{ud,\sigma} \leq \inf_{s\in\mathcal{U}}\sigma(s)\leq\sup_{s\in\mathcal{U}}\sigma(s)<B_{up,\sigma}$ and $B_{ud,\tau} \leq \inf_{s\in\mathcal{U}}\tau(s)\leq\sup_{s\in\mathcal{U}}\tau(s)<B_{up,\tau}$ for all $s\in \mathcal{U}$;
\subassumption \label{ass:cnvrg_cov:c} The estimated covariance of $X$ are uniformly weakly consistent with a polynomial rate, i.e., $\sup_{s,s^\prime\in\Scal_m}|\widehat{\cov}\{\epsilon(s),\epsilon(s^\prime)\}-\cov\{\epsilon(s),\epsilon(s^\prime)\}|=\op(m^{-q})$ for some $q>0$.
\end{assumption}
Assumptions~\ref{ass:cnvrg_cov:a} requires the variance of the error process to be bounded away from zero and infinity. 
Assumption~\ref{ass:cnvrg_cov:c} imposes condition on the convergence rate of the covariance estimate, which is satisfied by many commonly-used estimators. For example, the maximum likelihood estimator achieves the desired convergence rate with $q=1/2$ when the parametric covariance function is locally Lipschitz continuous \citep{Mardia1984}. 



To describe the spatial dependence structure, we adopt the near epoch dependency (NED), which has been extensively studied in time series analysis \citep{James1994} and first introduced to the spatial analysis by \cite{Jenish2012}. The NED is satisfied by many classical models in spatial statistics, e.g., spatial autoregression models \citep{Cliff1981}.
Denote by $\Vcal$ a spatial domain such that $\mathcal{U}\subseteq\Vcal\subset\Rb^K$. 
Let $\left\{Y(v),v\in\Vcal\right\}$ be a random field and set $\mathfrak{F}(S)=\sigma\left\{Y(s),s\in S\right\}$ as the $\sigma$-field generated by $Y(s)$ for $s\in S\subset\Vcal$. { In reality, $\mathcal{V}$ represents a physical spatial domain in consideration, which might encompass a wide geographical area; $\mathcal{S}$, on the other hand, is a set of locations to collect the data.}

\begin{defn}[NED]\label{def:near_epoch}
    Let $X=\{X(s), s\in\Scal_m\}$ be a random field with $\|X(s)\|_{p}<\infty$ where $\|X(s)\|_p=(\Eb|X(s)|^p)^{1/p}$, $p\geq 1$. Define $Y=\{Y(s),s\in \Vcal_m\}$ as a random field where $\Scal_m\subseteq \Vcal_m$ 
    and let $\mathbf{d}=\{d_{m}(s), s\in\Scal_m\}$ be a set of finite positive constants. Then, $X$ is said to be $L_p(\mathbf{d})$-near-epoch dependent on $Y$ if 
    $$\left\|X(s)-\Eb \left\{X(s)\mid \mathfrak{F}(S)\right\} \right\|_p\leq d_{m}(s) \psi(r),$$
    \noindent where $S\subset \Vcal_m$, $r=\max\left\{t\geq 0:B(s;t)\subseteq S\right\}$ with $B(s;t)$ being a ball centered around $s$ with radius $t$, and $\psi(r)\geq 0$ is some non-increasing sequence with $\lim_{r\rightarrow\infty}\psi(r) = 0$. Here $\psi(r)$ and $d_{m}(s)$ are called NED coefficients and NED scaling factors, respectively. 
    We say $X$ is $L_p(\mathbf{d})$-NED of size $-\lambda$ if $\psi(r)=O(r^{-\mu})$ for some $\mu>\lambda>0$. Furthermore, if $\sup_{m}\sup_{s\in\Scal_m} d_{m}(s)<\infty$, we say $X$ is uniformly $L_p$-NED on $Y$.
\end{defn}

\begin{assumption}[Uniform NED]\label{ass:NED} 
    $X=\{X(s), s\in \Scal_m\}$ is uniformly $L_p$-NED on $Y=\{Y(s):s\in\Vcal_m\}$ of size $-\lambda$, for $\lambda>0$, $p\geq 2$, where $Y(s)$ are independently distributed.
\end{assumption}

In Definition~\ref{def:near_epoch}, $B(s;t)$ is the maximum ball that is completely included in $S$. 
The quantity $\|X(s)-\Eb\{X(s)\mid\mathfrak{F}(S)\}\|_p$ satisfies a 
generalized non-decreasing property with respect to $S$. 
Specifically, if $\|X(s)-\Eb \{X(s)\mid \mathfrak{F}(S)\}\|_p \leq d_m(s) \psi(r)$, we then have $\|X(s)-\Eb\{X(s)\mid \mathfrak{F}(V)\}\|_p\leq d_m(s) \psi(r)$ for any $S\subseteq V\subseteq \Scal$. 
We can choose $d_m(s)=2\|X(s)\|_p$ as the scaling factor such that $0\leq \psi(r)\leq 1$. In addition, if $X$ is $L_p$-NED on $Y$, then $X$ is also $L_q$-NED on $Y$ with the same $\{d_m(s)\}$ and $\psi(r)$ for any $q\leq p$. The condition $p\geq2$ in Assumption~\ref{ass:NED} guarantees that the variance of $\Eb \left\{X(s)\mid \mathfrak{F}(S)\right\}$ converges to that of $X(s)$ as $S$ expands. 
\cite{Jenish2012} demonstrated that the NED property can be validated in examples from both \cite{gorodetskii1978strong} and \cite{Andrews1984}. The former showed that strong mixing might fail in linear processes with normal innovations and slowly declining coefficients, while the latter illustrated the absence of $\alpha$-mixing in a simple first-order autoregressive process with independent Bernoulli innovations. 
Furthermore, for infinity-order moving average random fields, verifying the NED property involves checking the smoothness of the functional form and the absolute summability of coefficients, which is an easier task compared to verifying mixing conditions.



Under the NED assumption, we can approximate $X(s)$ with $\Eb[X(s)\mid\mathfrak{F}\{B(s;r)\}]$ and in turn approximate $T_1(s)$ with
\begin{equation}\label{eqn:tStarDef}
    T_1^*(s;r)=\Eb\left[T_1(s)\mid \mathfrak{F}\left\{\cup_{v\in \Ncal(s)}B(v;r)\right\}\right]/\zeta_r(s),
\end{equation}
\noindent 
where $\zeta_r^2(s)=\Var\left(\Eb\left[T_1(s)\mid \mathfrak{F}\left\{\cup_{v\in \Ncal(s)}B(v;r)\right\}\right]\right)$. We further require the conditional statistics to be normally distributed as required by the theory for NPEB.

\begin{assumption}[Normality]\label{ass:cond_norm}
For any $S\subset\Vcal$, $\Eb\left\{X(s)\mid \mathfrak{F}(S)\right\}$ is normal.
\end{assumption} 
Spatial linear autoregression models with Gaussian white noise satisfy Assumptions~\ref{ass:NED} and \ref{ass:cond_norm} \citep{Jenish2012nonpara}. 
Under Assumptions~\ref{ass:NED} and \ref{ass:cond_norm}, $\{T^*_1(s;r):s\in\tScal_m\}$ defined in \eqref{eqn:tStarDef} are independent and normally distributed random variables with unit variance if $r>0$ satisfies
\begin{equation}\label{equ:req_ind}
    \cup_{v\in \Ncal(s)}B(v;r)\cap \cup_{v\in \Ncal(s^\prime)}B(v;r)=\emptyset,\quad s,s^\prime\in \tScal_m.
\end{equation}
\noindent
Under Assumptions \ref{ass:incr_dom} and \ref{ass:neigh}, setting $r=\widetilde{\Delta}_{l,m}/2-N_{nei}\Delta_u$ indicates \eqref{equ:req_ind}, where
\begin{equation}\label{eqn:npebDist}
\widetilde{\Delta}_{l,m}=\inf_{s,s^\prime\in\tScal_m}\dist(s,s^\prime).
\end{equation}

\begin{assumption}[Size of Subset]\label{ass:num_tSm} 
The subset for implementing NPEB, $\tScal_m\subset\Scal_m$, satisfies $
    \tm^{1/(\lambda p)}\{\log(\tm)\}^{-1/(2\lambda )}=o(\widetilde{\Delta}_{l,m})$ and $\tm\rightarrow\infty$ as $m\rightarrow\infty$, where $\widetilde{\Delta}_{l,m}$ is defined as in \eqref{eqn:npebDist}.
\end{assumption}

Let $G_{m}(u)=m^{-1}\sumsSm\boldsymbol{1}\{\xi(s)\leq u\}$ and $G_{\tm}(u)=\tm^{-1}\sumstSm\boldsymbol{1}\{\xi(s)\leq u\}$ be the empirical distributions of the unknown means for $s\in\Scal_m$ and $s\in\tScal_m$, respectively.
\begin{assumption}[Limiting Distribution]\label{ass:emp_bayes} There exists a limiting distribution $G_0$ on $[-\nu_0,\nu_0]$ for some positive constant $\nu_0$ such that $d_{H}(f_{G_{m}},f_{G_0})=\op(1)$, as $m\rightarrow\infty$, and the subset $\tScal_m$ satisfies $d_{H}(f_{G_{\tm}},f_{G_0})=\op(1)$, as $\tm\rightarrow\infty$, where $f_G(x)=\int\phi(x-u)dG(u)$ and $d^2_H(f,g)=2^{-1}\int \big\{\sqrt{f(x)}-\sqrt{g(x)} \big\}^2  dx$ is the squared Hellinger distance between densities $f$ and $g$.
\end{assumption}

The condition about $\tm$ and $\widetilde{\Delta}_{l,m}$ in Assumption~\ref{ass:num_tSm} trades off the number of $\tScal_m$ and the distance between them, which becomes milder when decreasing the strength of near epoch dependency, i.e., increasing $\lambda$ or $p$ 
. In fact, we can choose $\tScal_m$ satisfying Assumption~\ref{ass:num_tSm} under Assumptions \ref{ass:incr_dom} and \ref{ass:neigh}; see Section~\ref{sec:relation_tm_r} of the supplement for more details. 
{Assumption~\ref{ass:emp_bayes} states that the empirical distributions of $\xi(s)$ using $s\in\Scal_m$ and $s\in\tScal_m$ have the same limiting distribution $G_0$, as the number of spatial samples goes to infinity.} 

\begin{assumption}[Null Proportion]\label{ass:pi0}
	$m_0/m\rightarrow \pi_0 \in (0,1)$, as $m\rightarrow\infty$.
\end{assumption}
\begin{assumption}[{Asymptotic True/False Rejection Proportion}]\label{ass:cnvrg_true}
	As $m_0$ and $m_1$ tend to infinity, for every $\left(t_1,t_2\right)\in\Rb\times\Rb$, we have
$$
 \begin{aligned}
    &\frac{\sum_{s\in\Scal_{0,m}} \Pb\left\{T_{1}(s)\geq t_{1},T_{2}(s) \geq t_{2}\right\}}{m_0}:=\Eb\left\{\frac{V_m(t_1,t_2)}{m_0}\right\}
 {\rightarrow} K_{0}\left(t_{1}, t_{2}\right)
 ,\\
 &\frac{\sum_{s\in\Scal_{1,m}} \Pb\left\{T_{1}(s) \geq t_{1},T_{2}(s) \geq t_{2}\right\}}{m_1}:=\Eb\left\{\frac{S_m(t_1,t_2)}{m_1}\right\} {\rightarrow} K_{1}\left(t_{1}, t_{2}\right), 
 \end{aligned}
	$$
\vskip -0.5em
\noindent
where 
$K_{0}\left(t_{1}, t_{2}\right)
      \leq\lim_{m\rightarrow\infty} \sum_{s\in\Scal_{m}}\int L\{t_1,t_2,x,\rho(s)\} \, d G_0(x)/m$, 
and both $K_{0}\left(t_{1}, t_{2}\right)$ and $K_{1}\left(t_{1}, t_{2}\right)$ are non-negative continuous functions of $\left(t_{1}, t_{2}\right)$.
\end{assumption}
{Let $K_1(-\infty,t_2)$ be the limit of $\sum_{s\in\Scal_{1,m}} \Pb\left\{T_{2}(s) \geq t_2\right\}/m_1$ as $m_1$ goes to infinity.} 
Define
\begin{equation}\label{equ:def_F_FDP_inf}
 \FDP_{\lambda}^\infty\left(t_{1}, t_{2}\right):=\lim_{m\rightarrow\infty}\frac{F(\lambda)\sum_{s\in\Scal_m}\int L(t_1,t_2,x,\rho(s))d G_0(x)}{m\Phi(\lambda)K(t_1,t_2)},
\end{equation}
\noindent where $K(t_1,t_2)=\pi_0K_0(t_1,t_2)+(1-\pi_0)K_1(t_1,t_2)$ and $F(\lambda)=\sum_{s\in\Scal_m}\pi_{0}\Phi(\lambda)+(1-\pi_{0})\{1-K_{1}(-\infty, \lambda)\}$.

\begin{assumption}[Existence of Cutoffs]\label{ass:pre_control}
    There exist $t_{1}^{\star}$ and $t_{2}^{\star}$ such that $\FDP_{\lambda}^{\infty}\left(t_{1}^{\star}, 0\right)<q$, $\FDP_{\lambda}^{\infty}\left(0, t_{2}^{\star}\right)<q$, and $K\left(t_{1}^{\star}, t_{2}^{\star}\right)>0$.
\end{assumption}
Assumption~\ref{ass:pi0} requires the asymptotic null proportion to be strictly between zero and one, which rules out the case of very sparse signals. Assumption~\ref{ass:cnvrg_true} characterizes the expected proportions of true and false rejections among the alternative and null hypotheses, respectively, as functions with respect to threshold $(t_1,t_2)$.
The upper bound condition of $K_0(t_1,t_2)$ appeared in Assumption~\ref{ass:cnvrg_true} is valid in many examples. For instance, it is fulfilled when (a) the signal strength around the null locations is weaker than that around the non-null locations on average; and (b) the error process is stationary, $\Scal_m$ is observed at the lattice $\Scal=\Zb^K$, and $\Ncal(s)$ is selected as the $\kappa$-nearest neighbors, so that $\rho(s)\equiv\rho$ for some $\rho\in(0,1)$. 
Assumption~\ref{ass:pre_control} reduces the searching region for the optimal cutoff to a rectangle. { Assumptions~\ref{ass:pi0}, \ref{ass:cnvrg_true}, and \ref{ass:pre_control} align with existing conditions for proving asymptotic FDR control \citep{Storey2004, Ferreira2006, Benjamini2007}, in which just the primary statistic is considered.}

{These assumptions are in general mild conditions for asymptotic FDR control in spatial multiple testing.
Assumptions~\ref{ass:incr_dom}, \ref{ass:cnvrg_cov}, \ref{ass:pi0}, \ref{ass:cnvrg_true}, and \ref{ass:pre_control} are widely used in spatial multiple testing \citep{Cressie1993Spatial,Storey2004,Ferreira2006,Benjamini2007,Jenish2012}; and Assumptions~\ref{ass:neigh}, \ref{ass:NED}, \ref{ass:cond_norm}, \ref{ass:num_tSm}, and \ref{ass:emp_bayes} are specifically required for the 2d-SMT method. 
In particular, the condition on distances between locations in Assumption~\ref{ass:incr_dom} is requisite for increasing domain asymptotics \citep{Cressie1993Spatial}. 
Assumptions~\ref{ass:neigh} and \ref{ass:num_tSm} regularize the sizes of subsets for constructing auxiliary statistics and implementing NPEB.
When a practitioner implements the 2d-SMT method, these two assumptions can be directly satisfied by choosing suitable neighbors and a subset used for NPEB.
Assumption~\ref{ass:cnvrg_cov} imposes regular conditions on the moment of the error process and requires the covariance of the error process can be consistently estimated \citep{Sun2015}.
Assumptions~\ref{ass:NED}, \ref{ass:cond_norm}, and \ref{ass:emp_bayes} are technical conditions for showing the convergence of the GMLE under certain dependencies: Assumption~\ref{ass:NED} imposes some specific weak dependence on the random field; Assumption~\ref{ass:cond_norm} is crucial as we utilize the theory of NPEB developed for Gaussian location model; Assumption~\ref{ass:emp_bayes} requires that the empirical distribution of means possesses a limit. Assumptions~\ref{ass:pi0}, \ref{ass:cnvrg_true}, and \ref{ass:pre_control} naturally generalize the conditions for asymptotic FDR control in the Storey's procedure \citep{Storey2004, Ferreira2006,Benjamini2007}, where the rejection regions transit from one-dimensional to two-dimensional.} 

The FDR of the 2d-SMT procedure is given by 
\begin{equation}\label{equ:FDR}
    \widetilde{\FDR}_{m} =
\Eb\left\{\FDP\left(\wtt_{1}^{\star}, \wtt_{2}^{\star}\right)\right\},
\quad\text{where}\quad\FDP(t_1,t_2)=\frac{\whV_{m}(t_1,t_2)}{\whV_{m}(t_1,t_2)+\whS_{m}(t_1,t_2)},
\end{equation}
\noindent $(\wtt_{1}^{\star}, \wtt_{2}^{\star})$ is defined in \eqref{eq-opt}, and $\whV_{m}\left(t_{1}, t_{2}\right)=\sum_{s\in\Scal_{0,m}} \1f\{\whT_{1}(s) \geq t_{1},\whT_{2}(s) \geq t_{2}\}$ and $\whS_{m}\left(t_{1}, t_{2}\right)=\sum_{s\in\Scal_{1,m}} \1f\{\whT_{1}(s) \geq t_{1},\whT_{2}(s) \geq t_{2}\}$ are respectively the numbers of false and true rejections. 
\begin{thm}\label{thm:cnvrg_fdr}
Under Assumptions~\ref{ass:incr_dom}--\ref{ass:pre_control}, we have $\limsup_{m\rightarrow\infty}\widetilde{\FDR}_{m}\leq q$.
\end{thm}
Theorem~\ref{thm:cnvrg_fdr} states that 2d-SMT procedure asymptotically controls the FDR. The proof of Theorem~\ref{thm:cnvrg_fdr}, which is deferred to Section~\ref{app:proof} of the supplement, relies on two facts: {(i) \eqref{equ:def_tfdp} uniformly converges to \eqref{equ:def_F_FDP_inf} and $\FDP(t_1,t_2)$ in \eqref{equ:FDR} satisfies the uniform law of large numbers over the rectangle encompassed by $|t_1|\leq t_{1}^\star$ and $|t_2|\leq t_{2}^\star$; (ii) \eqref{equ:def_F_FDP_inf} is asymptotically larger than $\FDP(t_1,t_2)$ in \eqref{equ:FDR}. We make two technical innovations.
First, we establish $d_{H}\left(f_{\whG_{\tm}}, f_{G_{0}}\right)=\op(1)$ in Lemma~\ref{lemma:cnvrg_emp_bayes_cp}, which is the first result, to our knowledge, for the convergence of GMLE estimated from dependent observations. Second, we address the challenge posed by the non-Lipschitz nature of $\1f\{T_1(s)\geq t_1,T_2(s)\geq t_2\}$ {in Lemma~\ref{lemma:cnvrg_true}}, which expands the applications of the NED-based law of large numbers. 

We next formalize the power improvement of 2d-SMT over 1d-SMT (setting $t_1$ as $-\infty$) and defer the proof to Section~\ref{sec:pow_proof} of the supplement. Denote the thresholds for 1d- and 2d-SMT as $t^{1d}_2=\argmax_{t_2\in\Fcal^{1d}_{q,\infty}} K(-\infty,t_2)$ and $(t^{2d}_1,t^{2d}_2)= \argmax_{(t_1,t_2)\in\Fcal^{2d}_{q,\infty}} K(t_1,t_2)$, respectively, where $ \Fcal^{1d}_{q,\infty}=\{t_2:\FDP_{\lambda}^\infty(-\infty,t_2)\leq q\}$ and $\Fcal^{2d}_{q,\infty}=\{(t_1,t_2):\FDP_{\lambda}^\infty(t_1,t_2)\leq q\}$. 
The corresponding percentages of true discoveries in the asymptotic sense are respectively defined as $\PTD^{1d}=K_1(-\infty,t^{1d}_2)$ and $\PTD^{2d}=K_1(t^{2d}_1,t^{2d}_2)$. 
\begin{thm}\rm\label{rmk:power} Under Assumptions~\ref{ass:pi0} and \ref{ass:cnvrg_true}, we have (i) $\PTD^{2d}\geq\PTD^{1d}$ and (ii) $\PTD^{2d}>\PTD^{1d}$ if additionally $K_{0}\left(t^{2d}_{1}, t^{2d}_{2}\right)<\lim_{m\rightarrow\infty} \sum_{s\in\Scal_{m}}\int L\{t^{2d}_{1}, t^{2d}_{2},x,\rho(s)\} \, d G_0(x)/m$.   
\end{thm}
Theorem~\ref{rmk:power} shows that 2d-SMT offers superior power compared to 1d-SMT. Interestingly, 2d-SMT can achieve a strictly higher power than 1d-SMT, even when both methods yield the same total number of discoveries, i.e., $K(t_1^{2d},t_2^{2d})=K(-\infty,t_2^{1d})$. This power boost, as discussed alongside Assumption~\ref{ass:cnvrg_true}, stems not only from more discoveries but also because signals near null locations tend to be weaker than those near non-null locations.}

\setcounter{equation}{0} 
\section{Simulation Studies}\label{sec:simu}
We conduct extensive simulations to evaluate the performance of the proposed 2d-SMT procedure. We consider various simulation settings, including (1) 1d and 2d spatial domains; (2) known and unknown covariance structures; (3) different signal shapes; and (4) simple and composite nulls. 
We investigate the specificity and sensitivity of different methods under different settings by varying the signal intensities, magnitudes, and degrees of dependency. 


We compare the 2d procedure with the following competing methods: Storey's procedure \citep[][ST]{Storey2002}; Independent hypothesis weighting \citep[][IHW]{Ignatiadis2016,Ignatiadis2021}; Structure adaptive BH procedure with the stepwise constant weights 
\citep[][SABHA]{Ang2018}; Locally adaptive weighting and screening \citep[][LAWS]{Cai2021}; Adaptive p-value thresholding procedure \citep[][AdaPT]{Lei2018}; Dependence-adjusted BH procedure \citep[][dBH]{Fithian2020}.
As discussed in Section \ref{sec:vary_null}, our idea can combine with the ST, SABHA, and IHW methods to further enhance their power by borrowing neighboring information. We denote the corresponding procedures by 2D (ST), 2D (SA), and 2D (IHW), respectively, and include them in the numerical comparisons.

Throughout, we focus on testing the one-sided hypotheses $\mathcal{H}_{0,s}:\mu(s)\leq 0$ versus $\mathcal{H}_{a,s}:\mu(s)>0$. We set the target FDR level at $q=0.1$ and report the FDP and power (the number of true discoveries divided by the total number of signals) by averaging over 100 simulation runs.
For the set of neighbors $\mathcal{N}(s)$ of the location $s$, we use the $\kappa$-nearest neighbors for each $s$.
A sensitivity analysis of $\kappa$ is conducted in Section \ref{sec:rob_nei_num} of the supplement and empirically suggests $\kappa$ to be an integer between $2$ and $7$. 
In this section, we use $\kappa=4$.

We consider the process $X(s)=\mu(s)+\epsilon(s)$ defined on the one-dimensional domain $\mathcal{S}=[0,30]$. We observe the process $X(s)$ at 900 locations that are evenly distributed over $\mathcal{S}$. We introduce three \textit{data generating mechanisms} for the signal process $\mu(s)$ and consider three \textit{signal sparsity levels} within each mechanism.

\begin{itemize}[topsep=0pt,itemsep=0pt,parsep=0pt]
\item \textbf{Setup \RNum{1}}: $\mu(s)=\gamma\mu_0(s)$, where $\gamma$ determines the magnitude and $\mu_0(s)$ is generated from B-spline basis functions to control the signal densities and locations. Three different shapes of $\mu_0(s)$ are considered,  
which correspond to the sparse, medium, and dense signal cases, respectively.

\item \textbf{Setup \RNum{2}}: $\mu(s)= \gamma\delta(s)$ with $\delta(s)\sim \operatorname{Bernoulli}(\bar{\pi}_0(s))$. The non-null probability functions 
$\bar{\pi}_0(s)$ exhibit similar patterns as those of $\mu_0(s)$ described in Setup \RNum{1}.

\item \textbf{Setup \RNum{3}}: $\mu(s)=\gamma G(s)$, where $G(s)$ follows a Gaussian process with a constant mean $\bar{\mu}$ and the covariance function $k_\mu(s,s^\prime)=\sigma_\mu^2\exp\{-\left(\|s-s^\prime\|/\rho_\mu\right)^k\}$ with $k=1$, $\sigma^2_\mu=3$ and $\rho_\mu=0.3$. We set $\bar{\mu}=-2.5,-2,-1$ for the sparse, medium, and dense signal cases, respectively. Their non-null proportions are nearly $4\%$, $8\%$, and $25\%$. 
\end{itemize} 

The shapes of $\mu_0(s)$ in Setup \RNum{1}, the generated signals $\delta(s)$ in Setup \RNum{2}, and the simulated signals $G(s)$ in Setup \RNum{3} are depicted in Figures~\ref{fig:1D_sig_spline}--\ref{fig:1D_sig_mvnorm} of the supplement, respectively. 
For the magnitude $\gamma$, we considered $\gamma \in \{2,3,4\}$ in Setups \RNum{1} and \RNum{3}, and $\gamma \in \{1,1.5,2\}$ in Setup \RNum{2}.
We generated the noise process $\epsilon(s)$ from a mean-zero Gaussian process with the covariance function
$k_{\epsilon}(s,s^\prime; r, k, \rho_\epsilon)=(1-r)\boldsymbol{1}\{s=s^\prime\}+r\exp\{-\left(\left\|s-s^\prime\right\|/\rho_\epsilon\right)^k\}$. Here, $r$ determines the relative percentage of nugget effect, $\rho_\epsilon$ measures the strength of dependency, and $k$ controls the decay rate of dependence. We demonstrated three different degrees of spatial dependence through the following choices of $(r,k,\rho_\epsilon)$: (1) $r=0.5$, $k=1$, $\rho_\epsilon=0.05$ (exponential kernel); (2) $r=0.8$, $k=1$, $\rho_\epsilon=0.1$ (exponential kernel); and (3) $r=0.6$, $k=2$, $\rho_\epsilon=0.2$ (Gaussian kernel). 
The above (1)--(3) combinations of 
$(r,k,\rho_\epsilon)$ represent the weak, medium, and strong correlation among locations, respectively; see Figure~\ref{fig:1D_Cov} of the supplement. 
Here we assume only one observation is available at each location and the covariance is known.

\begin{figure}[!h]
     \centering
    \includegraphics[width=0.9\textwidth]{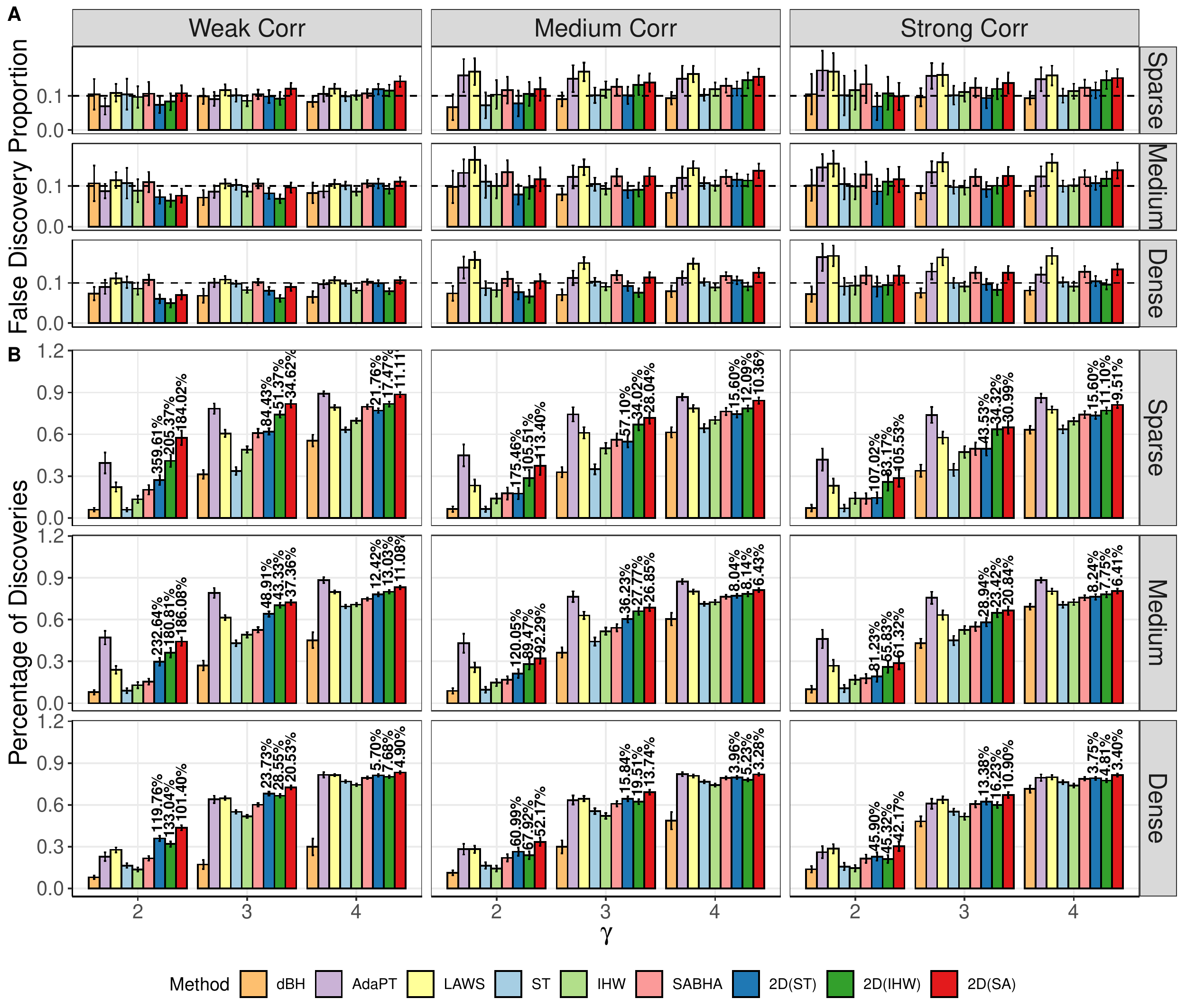}
        \caption{The mean and ($1.96$ multiple of) the standard error of FDP (Panel A) and power (Panel B) under Setup \RNum{1} with $\gamma\in\{2,3,4\}$. The percentages on the top of bars represent the power improvement of 2d procedures compared to their 1d counterparts. 
        }
        \label{fig:Spline_Res}
        \vspace{-1em}
\end{figure}

\begin{figure}[!h]
     \centering
     \includegraphics[width=0.9\textwidth]{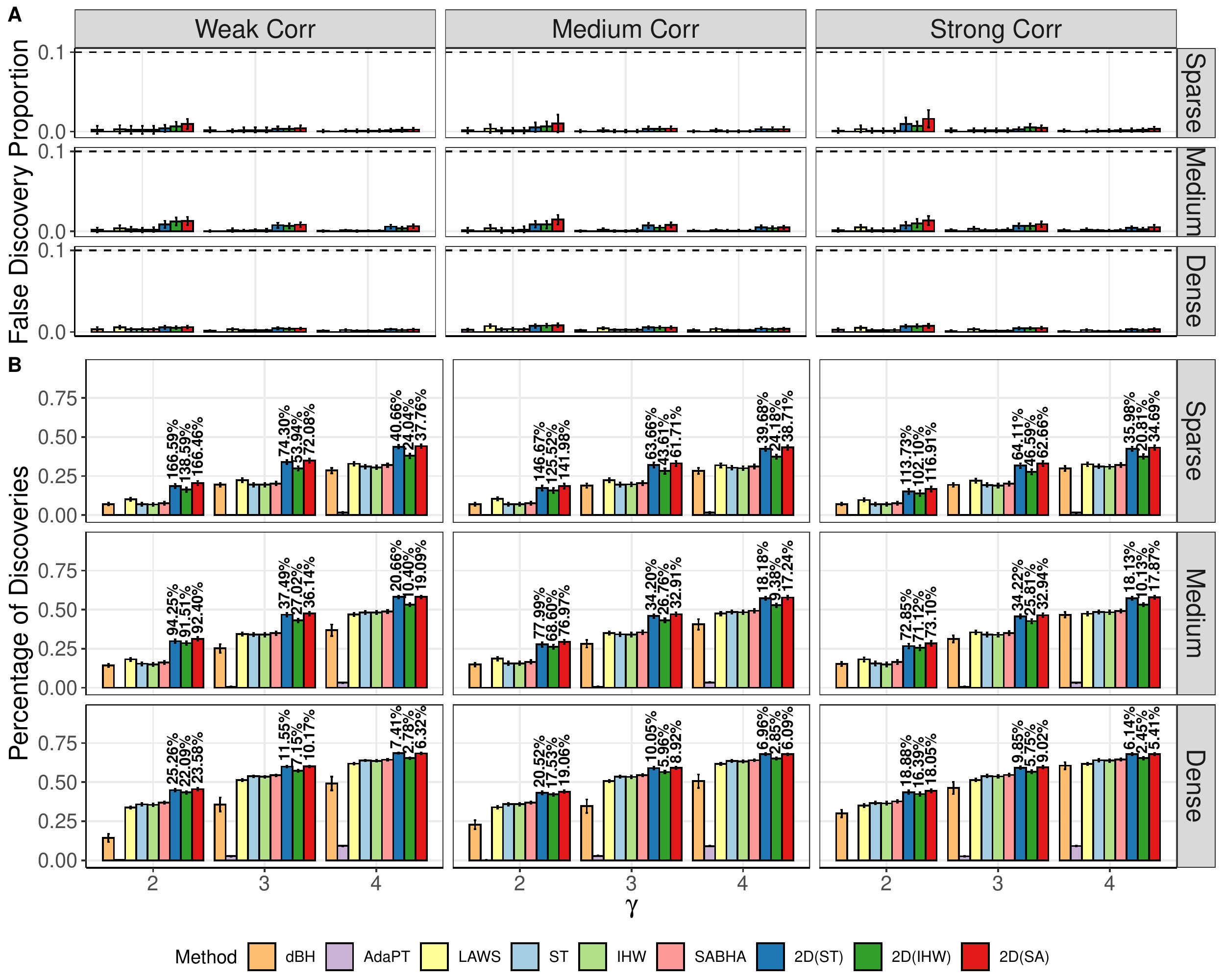}
        \caption{The mean and ($1.96$ multiple of) the standard error of FDP (Panel A) and power (Panel B) under Setup \RNum{3} with $\gamma\in\{2,3,4\}$. The percentages on the top of bars represent the power improvement of 2d procedures compared to their 1d counterparts. 
        }
        \label{fig:Mv_Res}
        \vspace{-1em}
\end{figure}

We applied the competing methods to the generated datasets, and the empirical FDP and power under Setups I and III are summarized in Figures \ref{fig:Spline_Res} and \ref{fig:Mv_Res} based on 100 replicates. Under Setup I, ST, IHW, SABHA, 2D (ST), and dBH controlled the FDR reasonably well across all cases. LAWS, AdaPT, 2D (IHW), and 2D (SA) were inflated for the medium and strong correlation cases, with LAWS being the worst. 2D (SA) and AdaPT were generally more powerful than the other methods, while dBH was quite conservative. 
As expected, the 2d procedures outperformed their 1d counterparts in terms of power. The results for Setup \RNum{2} were displayed in Figure~\ref{fig:Unif_Res} of the supplement and generally similar to those in Setup \RNum{1}. Under Setup \RNum{3}, the empirical FDPs were close to zero, indicating all methods were conservative due to the composite null effect. The 2d-SMT procedures provided the highest power compared to the other approaches. 
Overall, the 2d-SMT procedures achieved remarkable improvements in power for either the weak correlation, the sparse signal, or the feeble magnitude cases. 
In Section~\ref{sec:increase_m} of the supplement, we also performed simulations where the location sizes are increased to $m=2000$. With larger location sizes, the 2d-SMT procedures exhibited more reliable FDR control under Setups \RNum{1}--\RNum{2} and demonstrated a significant improvement in power under Setup \RNum{3}. 
In Section \ref{sec:Simu_TwoD} of the supplement, we further considered a spatial process defined on a two-dimensional domain with multiple observations at each location to estimate the unknown covariance. The 2d-SMT procedures generally provided the best trade-off between FDR and power, especially when the correlation was weak. 
LAWS achieved the highest power at the cost of higher FDR. 
AdaPT provided reliable FDR control in all cases but their power were dominated by 2D (ST), 2D (IHW), and 2D (SA) for sparse signals and weak correlation. Additionally, 2d-SMT appeared to be robust to covariance function misspecifications, as we used an exponential kernel to estimate the covariance which was indeed generated from a Gaussian kernel. 

\setcounter{equation}{0} 

\section{Ozone Data Analysis}\label{sec:real_data}
Ozone has double-edged effects on human health: 
{ozone} in the upper atmosphere (stratospheric ozone) shields humans from harmful ultraviolet (UV) radiation; 
while ozone at ground level (tropospheric ozone) triggers a variety of adverse health effects on human, sensitive vegetation and ecosystems \citep{Weinhold2008,Liu2022}.
The US Environmental Protection Agency (EPA) formulates regulations to 
reduce tropospheric ozone levels in outdoor air. The majority of tropospheric ozone occurs through the reaction of nitrogen oxides (\NOx{}), carbon monoxide (CO), and volatile organic compounds (VOCs) in the atmosphere when exposed to sunlight, particularly under the UV spectrum \citep{Warneck2000}. 


We applied our proposed 2d procedures and their 1d alternatives to identify the locations where the decreasing trend is below a pre-specified level for the Contiguous United States from 2010 to 2021. The data were the annual averages of the fourth-highest daily maximum 8-hour ozone concentrations
(see \href{http://www.epa.gov/airexplorer/index.htm}{http://www.epa.gov/airexplorer/index.htm}). 
To facilitate the analysis, we retained 697 stations (\rNum{1}) having a single site, 
(\rNum{2}) having full records across the years, and (\rNum{3}) being recorded by the World Geodetic
System (WGS84). 
{A regression model with mean-zero stationary Gaussian process of error has been widely used to analyze spatial data, e.g., for temperature \citep{French2013} and for ozone \citep{Sun2015}. We followed the model in \cite{Sun2015} to obtain the test statistics.} In particular, we first fitted the following linear model for each location
\begin{equation}\label{equ:ozone_model}
\Xf(s)=\mu_0(s)+\beta(s) \tf +\sigma_\epsilon(s)\bepsilon(s),
\end{equation}
\noindent
where $\Xf(s)=(X_{2010}(s),\cdots,X_{2021}(s))^{\top}$ was the observed ozone level measured in parts per billion (ppb), $\tf=(2010,2011,\cdots,2021)^\top$ was the predictor capturing the time trend, $\beta(s)$ was the slope at site $s$, $\bepsilon(s)=(\epsilon_{2010}(s),\cdots,\epsilon_{2021}(s))^{\top}$ was assumed to follow a mean-zero Gaussian process with the exponential kernel function $k_{\epsilon}(\cdot,\cdot;r, 1, \rho_\epsilon)$, and $\sigma_\epsilon(s)$ was the standard deviation of noise at site $s$. 
For each $\beta_0\in \{0.1,0.2,0.3,0.4,0.5\}$, we were interested in testing whether the ozone level declined more than $\beta_0$ ppb per year at each site, i.e., $\Hcal_{0,s}:\beta(s)\geq-\beta_0$ versus $\Hcal_{1,s}:\beta(s)<-\beta_0$. We first conducted simple linear regression and obtained the OLS estimates of $(\mu_0(s),\beta(s),\sigma_\epsilon(s))$, denoted as $(\whmu_0(s),\whbeta(s),\whsig_\epsilon(s))$.
Then, we obtained $(\whr,\whrho_\epsilon)$ by fitting the kernel function to the residuals $\widehat{\bepsilon}(s):=\{\Xf(s)-\whbeta(s)\tf-\whmu_0(s)\}/\whsig_\epsilon(s)$. Finally, {the proposed 2d procedures are conducted with the target FDR level at 10\% using the primary test statistic calculated as $\whT_2(s)=\{\whbeta(s)+\beta_0\}/\whsig_{\whbeta}(s)$, 
and the auxiliary test statistic given by $\whT_{1}(s)=\sum_{v\in \mathcal{N}(s)}\{\whbeta(v)+\beta_0\}/\whtau(s)$, where $\Ncal(s)$ was the set containing the two-nearest neighbors of $s$ and $\whtau(s)=\sum_{v,v^\prime\in \mathcal{N}(s)}\widehat{\cov}\{\whbeta(v),\whbeta(v^\prime)\}$.}

\begin{figure}[!h]
     \centering
    \subfigure[$\beta_0 =0.2$, ST versus 2D (ST)]{
    \includegraphics[width=0.45\textwidth]{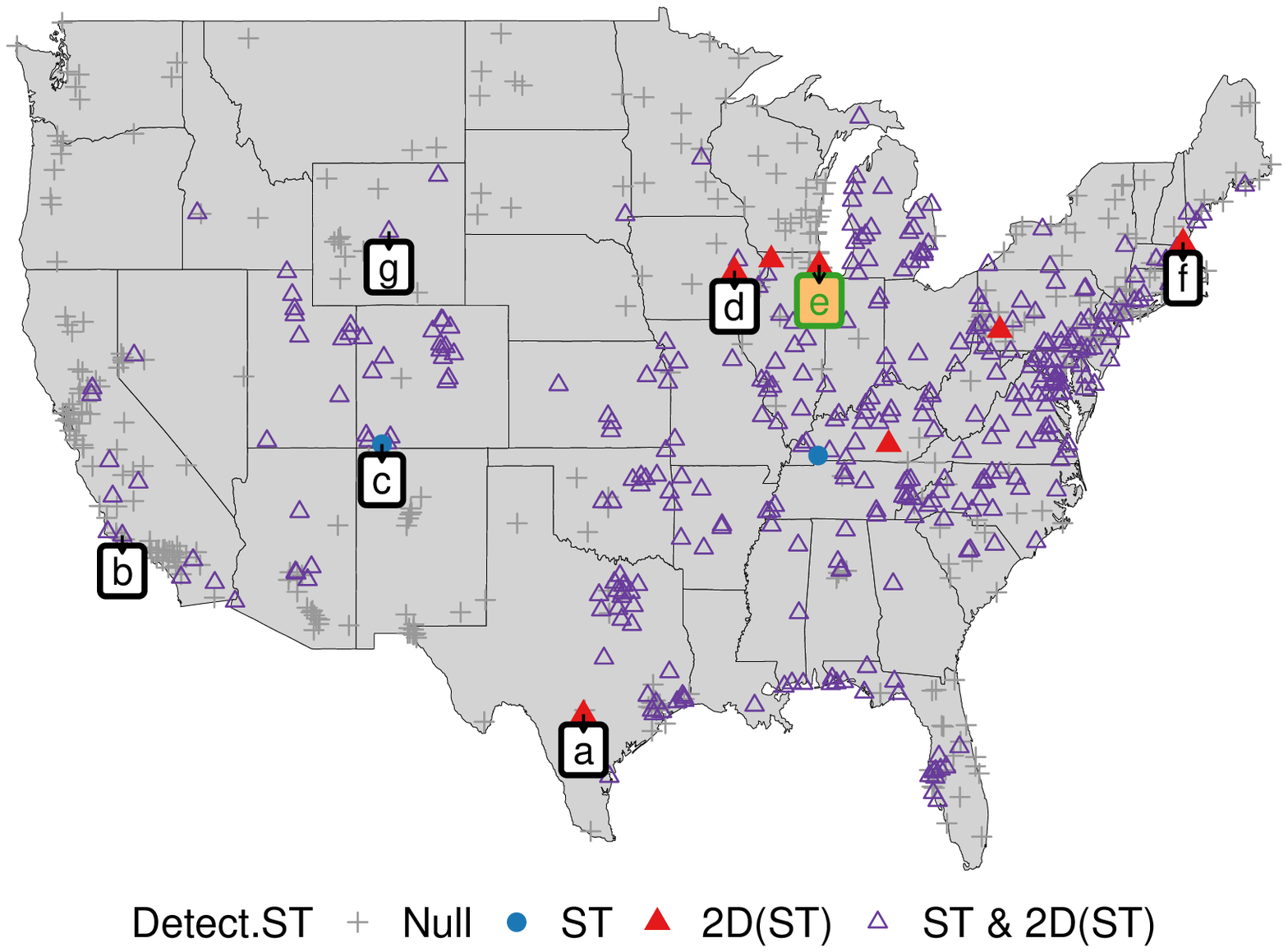}}
    \subfigure[$\beta_0 =0.3$, ST versus 2D (ST)]{
    \includegraphics[width=0.45\textwidth]{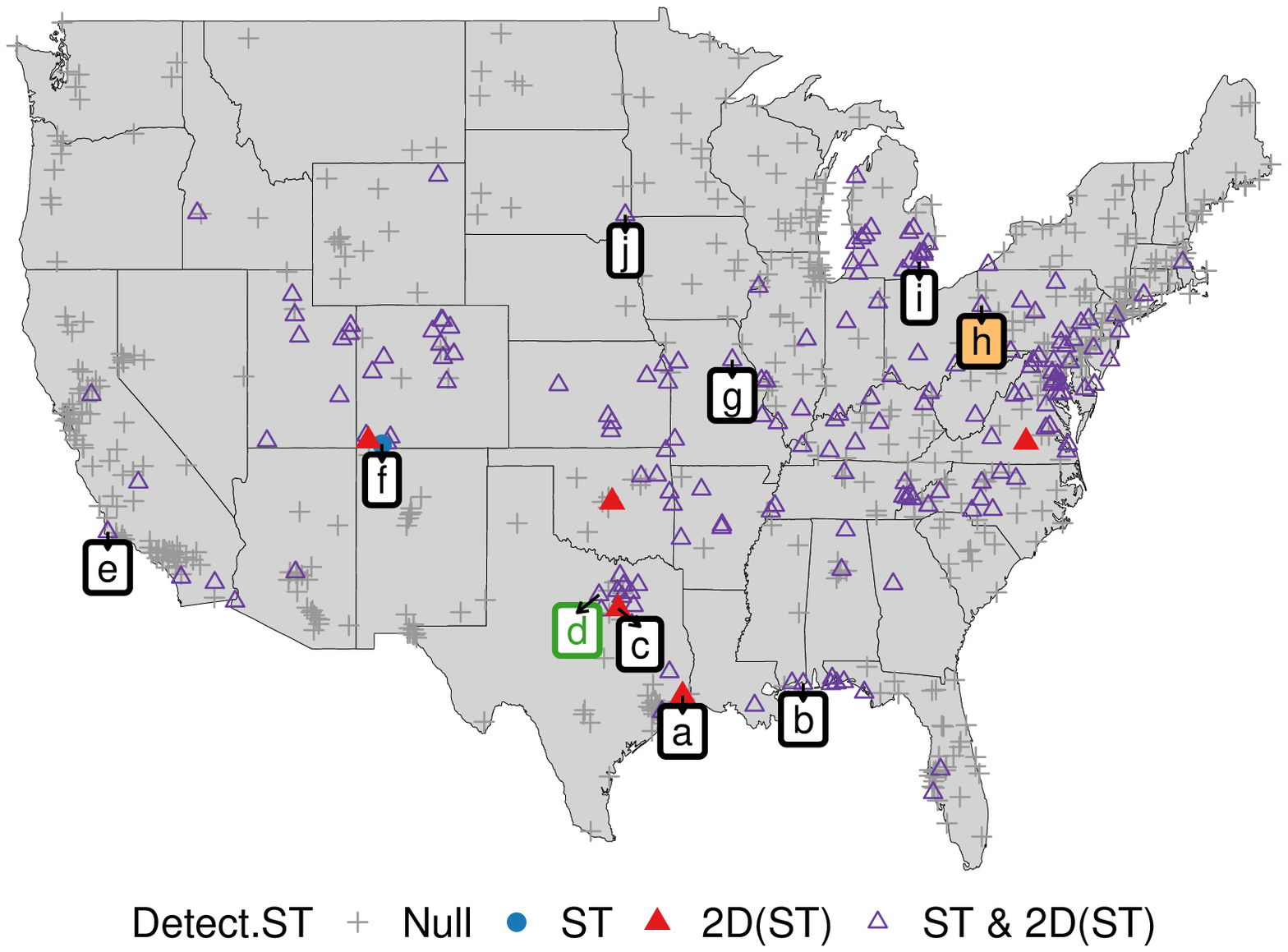}}
    \subfigure[$\beta_0 =0.3$, SABHA versus 2D (SA)]{\label{fig:ozone_analysis:c}
    \includegraphics[width=0.45\textwidth]{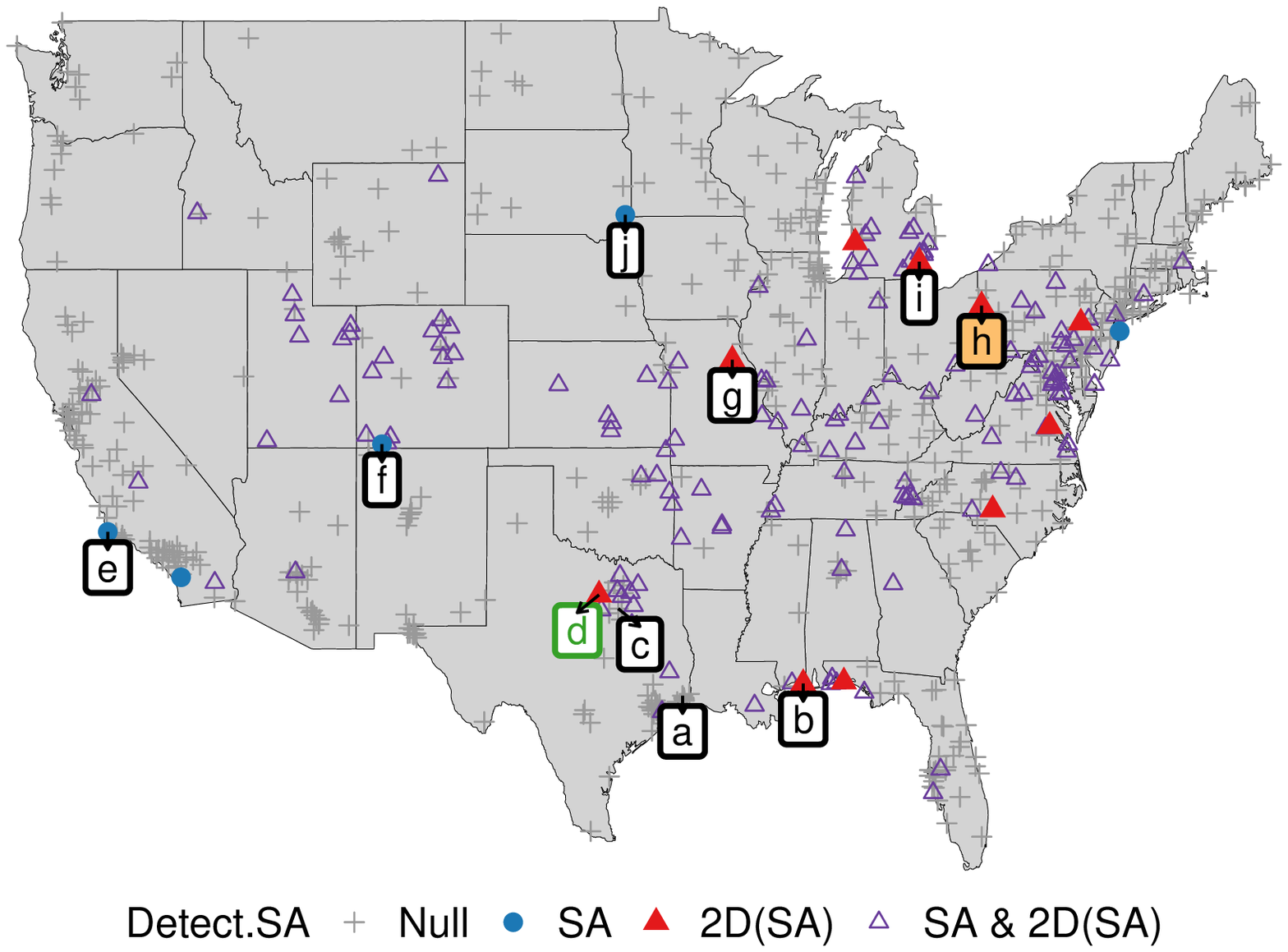}}
    \subfigure[$\beta_0 =0.5$, SABHA versus 2D (SA)]{\label{fig:ozone_analysis:d}
    \includegraphics[width=0.45\textwidth]{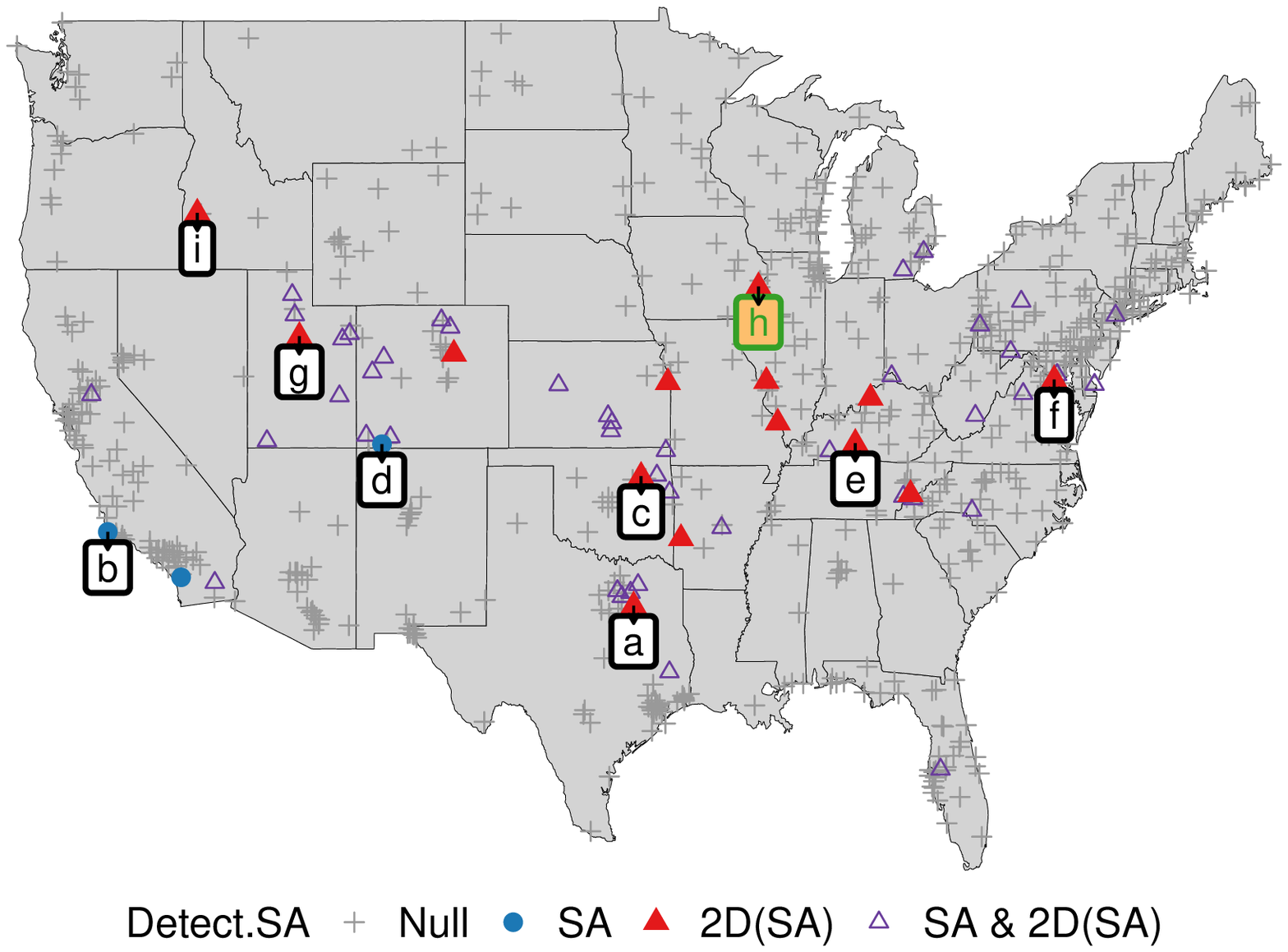}}
     \setlength{\belowcaptionskip}{-5pt}
     \caption{Results for ozone data analysis. 
     The red triangles ($\blacktriangle$), blue circles ($\bullet$), purple triangles ($\triangle$), and grey plus ($+$) signs represent the locations detected by the 2d procedure only, the 1d procedure only, both procedures, and neither one of the procedures, respectively. The labelled locations: (\rNum{1}) are detected by either 1d or 2d procedures but not both; and (\rNum{2}) possess either CO or \NOtwo{} records across 2010 to 2021. {In each sub-figure, the location with an orange background/green typeface indicates the greatest decline in CO/\NOtwo{} among the labelled locations.}
     }
     \label{fig:ozone_analysis} 
\end{figure}

\textit{Locations with significant ozone level decline.} 
We applied 2D (ST), 2D (IHW), 2D (SA), and their 1d alternatives to identify the non-null locations. 
We trisected the ranges of the latitude and longitude, which divided the whole area into nine different regions and allocated each site a categorical variable; see Figure~\ref{fig:ozone_reg} of the supplement for the division. 
Our analysis employed the categorical variable as the covariate in IHW and as the group indicator in SABHA. 
As shown in Figure~\ref{fig:ozone_analysis}, 2d procedures generally discovered more locations with significant decreasing ozone levels than their 1d counterparts did.

\textit{Ozone precursor.} 
The EPA has been making efforts to reduce tropospheric ozone by executing air pollution control strategies, including formulating vehicle and transportation standards, regional haze and visibility rules, and regularly reviewing the National Ambient Air Quality Standards. 
The universal ozone precursors (\NOx{}, CO, and VOCs) first respond to these strategies and then influence the ozone levels.
Indeed, some studies found the emissions of \NOtwo{} and CO account for the increase in background ozone levels \citep{Chin1994, Vingarzan2004, Han2011}. 
Motivated by these findings, we collected the contemporaneous CO and \NOtwo{} data from EPA and focused on the locations detected by either the 1d procedures or the 2d procedures but not both. 
We aimed to scrutinize the trends of CO and \NOtwo{} at these locations and explain our findings.

To this end, we regressed the CO and \NOtwo{} levels on $\mathbf{t}$ separately and recorded the slopes 
to understand the increasing/decreasing trends of the CO and \NOtwo{} levels. We summarized the major findings in Figure~\ref{fig:ozone_analysis} and Tables~\ref{tab:prec_descent}--\ref{tab:avr_prec} (of the supplement).
First, the locations detected only by the 2d procedures always included the ones with the most significant decline in the CO or \NOtwo{} levels (i.e., the locations with the orange background or green typeface labels in Figure~\ref{fig:ozone_analysis});   
see Table~\ref{tab:prec_descent} of the supplement. 
Second, the average decline (measured by the average of the standardized slopes) of the locations detected only by the 2d procedures was larger than that of the locations detected only by the 1d procedures.
Take Figure~\ref{fig:ozone_analysis:d} as an example, where we had \NOtwo{} records at locations a, b, c, d, f, g, h and i, and CO records at locations b, c, e, f and h. SABHA detected the locations b and d, while 2D (SA) identified the locations a, c, e, f, g, h, and i. The average decline in the \NOtwo{} levels was -3.63 for the locations detected by SABHA as compared to -4.68 for the locations detected by 2D (SA). As for CO, the average standardized slope was -0.96 for locations detected by SABHA in comparison to -1.40 for the locations identified by 2D (SA). In general, the CO and \NOtwo{} levels tended to decrease more rapidly for those locations detected by 2D (SA) except for the case with $\beta_0=0.3$; 
see Table~\ref{tab:avr_prec} of the supplement.

\textit{Ozone data simulation.}
To further validate our findings and to demonstrate the effectiveness of the 2d procedures, we conducted a simulation where we generated data mimicking the structure of the original data. Specifically, we generated the ozone level data from 2010 to 2021 through \eqref{equ:ozone_model} by setting $\beta(s)=\whbeta(s)$, $\mu(s)=\whmu(s)$, $\sigma_\epsilon(s)=\whsig_\epsilon(s)$, $r=\whr$, and $\rho_\epsilon=\whrho_{\epsilon}$. 
We processed the data and conducted multiple testing in the same way as discussed before. 
Table~\ref{tab:ozone_simu} shows that the 2d procedures achieved equal or higher power compared to the 1d alternatives {while controlling FDR under 10\%}. {To assess the robustness of our method, we followed \cite{Sun2015} to conduct additional simulations by using Gaussian kernel and empirical covariance matrix to generate synthetic data based on \eqref{equ:ozone_model} and remained to use the exponential kernel to analyze the synthetic data. The results are summarized in Section~\ref{sec:o3_data_more} of the supplement, which is consistent with the findings reported in Table~\ref{tab:ozone_simu}.}
\begin{table}[!h]\small
  \centering
  \caption{Mean and standard deviation of {FDPs and percentage of true discoveries (PTDs) for the simulated ozone data.} The results are based on 100 simulation runs.\label{tab:ozone_simu}}
  
\resizebox{\linewidth}{!}{
    \begin{tabular}{rrrrrrrr}
    \toprule
    \toprule
Criterion &$\beta_0$ & \multicolumn{1}{l}{ST} & \multicolumn{1}{l}{IHW} & \multicolumn{1}{l}{SABHA} & \multicolumn{1}{l}{2D (ST)} & \multicolumn{1}{l}{2D (IHW)} & \multicolumn{1}{l}{2D (SA)} \\
    \midrule
    \multicolumn{1}{c}{\multirow{5}[4]{*}{FDP}}& 0.5  &0.021(0.028)&0.021(0.027)&0.038(0.036)&0.022(0.029)&0.021(0.028)&0.052(0.040)\\
    &0.4  &0.019(0.021)&0.018(0.019)&0.024(0.020)&0.021(0.022)&0.018(0.019)&0.028(0.020)\\
    &0.3  &0.014(0.015)&0.012(0.013)&0.012(0.012)&0.015(0.015)&0.012(0.013)&0.013(0.012)\\
    &0.2  & 0.014(0.012)&0.008(0.009)&0.010(0.009)&0.015(0.013)&0.008(0.009)&0.010(0.009)\\
   & 0.1  &0.014(0.012)&0.006(0.007)&0.007(0.007)&0.014(0.012)&0.007(0.007)&0.007(0.007)\\
    \midrule
    \multicolumn{1}{c}{\multirow{5}[4]{*}{PTD}}& 0.5  &0.287(0.123)&0.287(0.120)&0.416(0.150)&0.292(0.125)&0.287(0.120)&\textbf{0.474(0.151)}\\
    &0.4  &0.409(0.153)&0.393(0.139)&0.474(0.111)&0.414(0.154)&0.393(0.139)&\textbf{0.506(0.106)}\\
    &0.3  &0.545(0.150)&0.508(0.137)&0.520(0.125)&\textbf{0.552(0.151)}&0.507(0.137)&0.532(0.121)\\
    &0.2  & 0.691(0.125)&0.627(0.112)&0.632(0.114)&\textbf{0.697(0.125)}&0.627(0.112)&0.638(0.113)\\
   & 0.1  &0.796(0.090)&0.714(0.087)&0.721(0.088)&\textbf{0.799(0.090)}&0.714(0.087)&0.725(0.087)\\
    \bottomrule
    \end{tabular}%
    }\vspace{-5.5mm}%
\end{table}%

\setcounter{equation}{0} 

\section{Discussion}\label{sec:discussion}
This paper proposes a new FDR-controlling procedure, 2d-SMT, to improve the signal detection power by incorporating the spatial information encoded in neighboring observations. It provides a unique perspective on utilizing spatial information, which is fundamentally different from the existing covariate and structural adaptive multiple testing procedures. The spatial information is gathered through an auxiliary statistic, which is used to screen out the noise. A primary statistic from the location of interest is then used to determine the existence of the signal. 2d-SMT is particularly effective when the signals exhibit in clusters. 
{We demonstrate the usefulness of 2d-SMT through simulation studies and the analysis of an ozone data set. We recommend 2D (ST) among different variants of 2d-SMT because it provides the most stable FDR control performance in numerical experiments and enjoys proven asymptotic FDR control under weak spatial dependence. }


To conclude, we point out a few future research directions. First, as discussed in Section~\ref{sec:vary_null}, the 2d-SMT is flexible to combine with various weighted BH procedures. 
One challenge is, however, to establish a rigorous FDR control theory for the resulting weighted 2d-SMT procedures.
Second, 
we use the $\kappa$-nearest neighbors to construct the auxiliary statistic in implementation. A more delicate strategy is to apply a weighting scheme to pool sufficient information from nearby locations. Third, extending the idea in 2d-SMT to other statistical problems, such as mediation analysis in causal inference, is of interest. 
\section*{Supplementary Material}
The online Supplementary Material contains our proofs of Theorems~\ref{thm:cnvrg_fdr} and \ref{rmk:power}, additional numerical results, some discussions about the estimation for the covariance of noises, the details of Algorithm \ref{alg:fast_alg}, and the address to download the reproducible code of this work.
\vskip 14pt

\bibliographystyle{asa}
\bibliography{mybib}

\end{document}


\title{Supplementary Material for ``Powerful Spatial Multiple Testing via Borrowing Neighboring Information"}
  \author{Linsui Deng$^{1}$,
  Kejun He$^{1}$\thanks{Correspondence: \href{mailto:kejunhe@ruc.edu.cn}{kejunhe@ruc.edu.cn} and \href{mailto:zhangxiany@stat.tamu.edu}{zhangxiany@stat.tamu.edu}.}~,
  Xianyang Zhang$^{2}$\footnotemark[1] 
  \\
$^{1}$The Center for Applied Statistics, Institute of Statistics and Big Data, \\ Renmin University of China, Beijing, China \\
$^{2}$Department of Statistics, Texas A\&M University, College Station, USA\\
}

  \maketitle

The supplement is organized as follows.
Section~\ref{sec:add_exp_details} includes some details on the simulation studies and the additional results of the ozone data analysis.
Section~\ref{app:proof} presents two main lemmas and the proof of Theorem~\ref{thm:cnvrg_fdr}. 
The two main lemmas are proved in Section~\ref{sec:prf_lemma_12} with some preliminary lemmas and discussions. 
Section~\ref{sec:prf_lemmas} provides detailed proofs of the preliminary lemmas. Section~\ref{sec:pow} proves the theory of the power improvement of the 2d-SMT compared to the 1d-SMT and provides a concrete example to investigate the power improvement.
We discuss the estimation for the covariance of noises in Section~\ref{sec:cov_est}. 
Section~\ref{ref:searching_alg} thoroughly describes the details of Algorithm \ref{alg:fast_alg} in Section~\ref{sec:imp_detail} of the main paper which overcomes the computational bottleneck of a naive grid search. 
The R package and reproducible code of this work are accessible at  \hyperlink{https://github.com/denglinsui/TwoDSMT}{https://github.com/denglinsui/TwoDSMT} and \hyperlink{https://github.com/denglinsui/2dSMT-manuscript-sourcecode}{https://github.com/denglinsui/2dSMT-manuscript-sourcecode}, respectively.

\section{Additional Experimental Results}\label{sec:add_exp_details}
This section provides additional results for the simulations and the real data analysis with some implementation details.
In particular, Section~\ref{sec:simu_set} visualizes the simulation settings in Section~\ref{sec:simu} of the main paper. Section~\ref{sec:increase_m} first reports the FDPs and powers under Setup II with location size $m=900$, and then presents the results of location size $m=2000$ under Setups I--III.
Section~\ref{sec:Simu_TwoD} describes and reports the simulations studies for a two-dimensional domain where the covariance is unknown and needs to be estimated.
In Section~\ref{sec:rob_nei_num}, we investigate the sensitivity of 2d-SMT to the number of observations in the nearest neighbors. 
Section~\ref{sec:adaneigh} explores a data-adaptive approach to determine the number of neighbors for each location. 
In Section~\ref{sec:covarSpa}, we examine the integration of covariate and spatial information within the framework of the 2d-SMT method. 
Section~\ref{sec:o3_data_more} depicts the partition of the Contiguous United States into nine regions, and provides analytical and numerical findings related to the ozone data.

\subsection{Simulation Settings}\label{sec:simu_set}
\begin{figure}
    \centering
    \subfigure[Setup \RNum{1}. \label{fig:1D_sig_spline}]{ \includegraphics[width=0.9\textwidth]{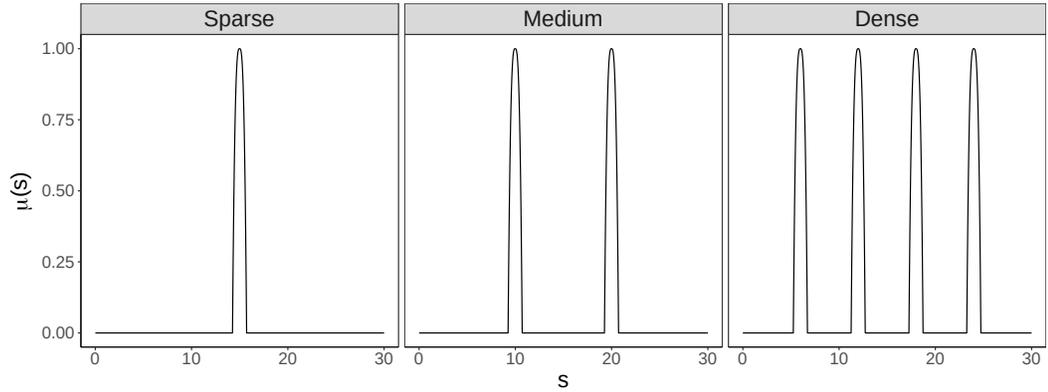}}
    \subfigure[Setup \RNum{2}. \label{fig:1D_sig_ber} ]{ \includegraphics[width=0.9\textwidth]{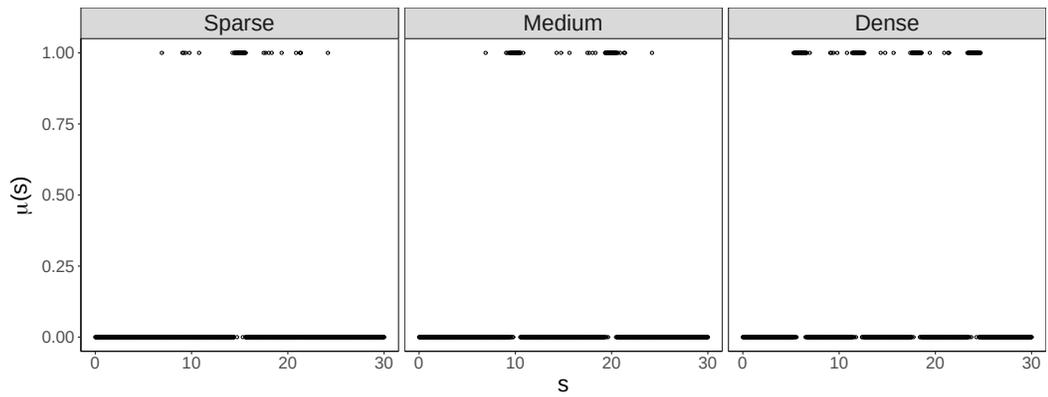}}
    \subfigure[Setup \RNum{3}. \label{fig:1D_sig_mvnorm} ]{ \includegraphics[width=0.9\textwidth]{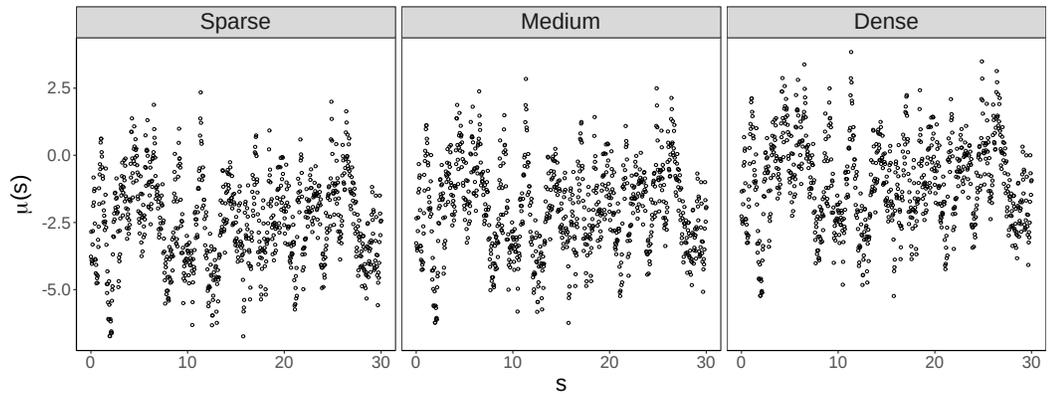}}
    \caption{The one-dimensional signal process $\mu(s)$ ($s\in[0,30]$) with $\gamma=1$ in Section \ref{sec:simu} of the main paper. The top to bottom panels correspond to Setups I--III of our simulation settings respectively.}
    \label{fig:1D_Sig}
\end{figure}

\begin{figure}
    \centering
    \includegraphics[width=0.95\textwidth]{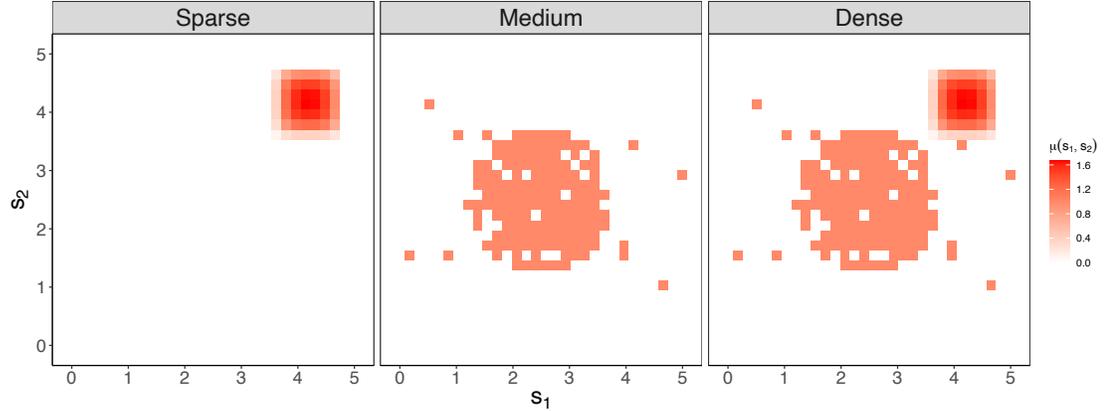}
    \caption{The two-dimensional signal process $\mu(s_1,s_2)$ ($(s_1,s_2)\in[0,5]^2$) with $\gamma=1$ in Section \ref{sec:Simu_TwoD}. The left to right panels correspond to sparse, medium, and dense signals of our simulation settings respectively.}
    \label{fig:2D_Sig}
\end{figure}

\begin{figure}
    \centering
    \includegraphics[width=0.95\textwidth]{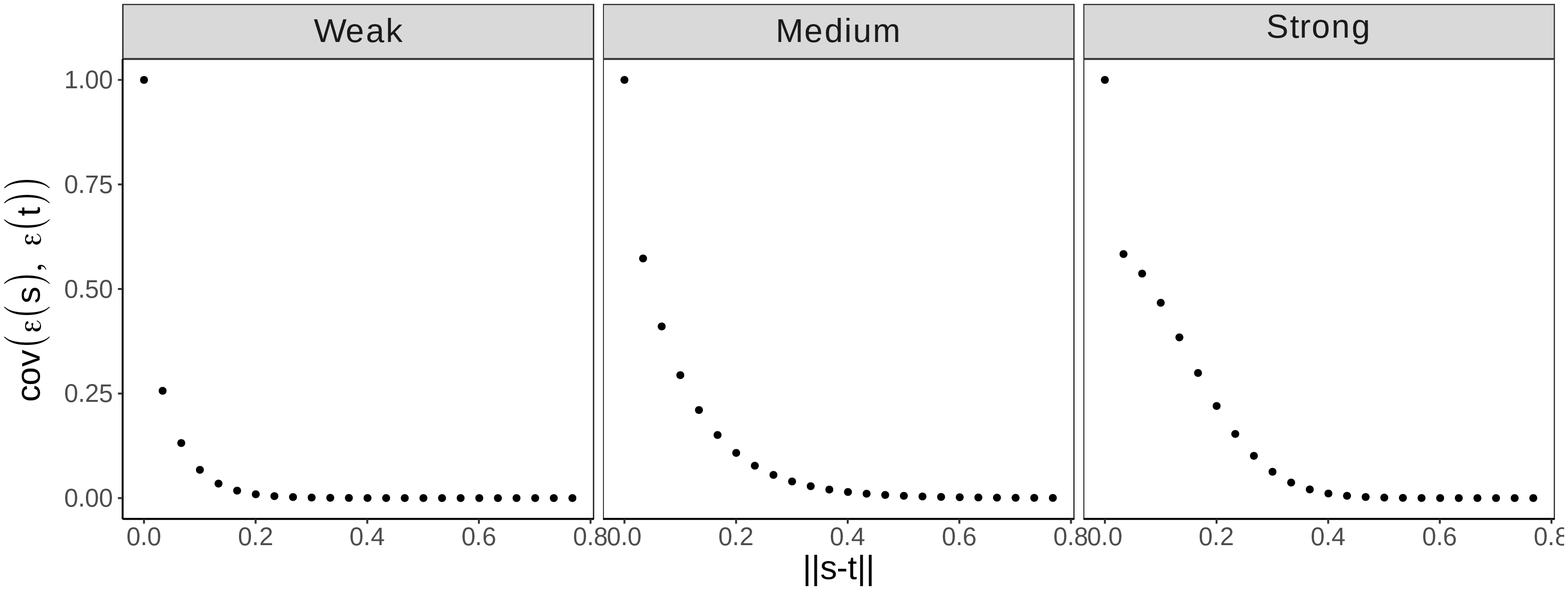}
    \caption{The covariance structure of the noise process $\epsilon(s)$ used in our simulation studies. The $x$-axis is the distance between $s$ and $t$, denoted as $\left\|s-t\right\|$, and $y$-axis is the covariance between $\epsilon(s)$ and $\epsilon(t)$, denoted as $\mathrm{cov}\{\epsilon(s),\epsilon(t)\}$. The choices of $(r,\rho_\epsilon,k)$ for the weak, medium, and strong dependence strengths (from left to right) are: (1) $r=0.5$, $\rho_\epsilon=0.05$, $k=1$ (exponential kernel), (2) $r=0.8$, $\rho_\epsilon=0.1$, $k=1$ (exponential kernel), and (3) $r=0.6$, $\rho_\epsilon=0.2$, $k=2$ (Gaussian kernel), respectively.
    }
    \label{fig:1D_Cov}
\end{figure}

\begin{itemize}
\item Figure~\ref{fig:1D_Sig} depicts the one-dimensional signal process $\mu(s)$ for Setups I--III as described in Section \ref{sec:simu} of the main paper. The plots are depicted with magnitude $\gamma=1$.
\item Figure~\ref{fig:2D_Sig} shows the two-dimensional signal process for three signal sparsity levels in Section \ref{sec:Simu_TwoD}. The plots are depicted with magnitude $\gamma=1$.
\item Figure~\ref{fig:1D_Cov} displays the spatial covariances of the noise process $\epsilon(s)$ versus the spatial distance for three dependency strengths as described in Sections \ref{sec:simu}. 
\end{itemize}

\subsection{Additional Simulation Studies for a One-Dimensional Domain}\label{sec:increase_m}
\begin{figure}[!h]
     \centering
    \includegraphics[width=\textwidth]{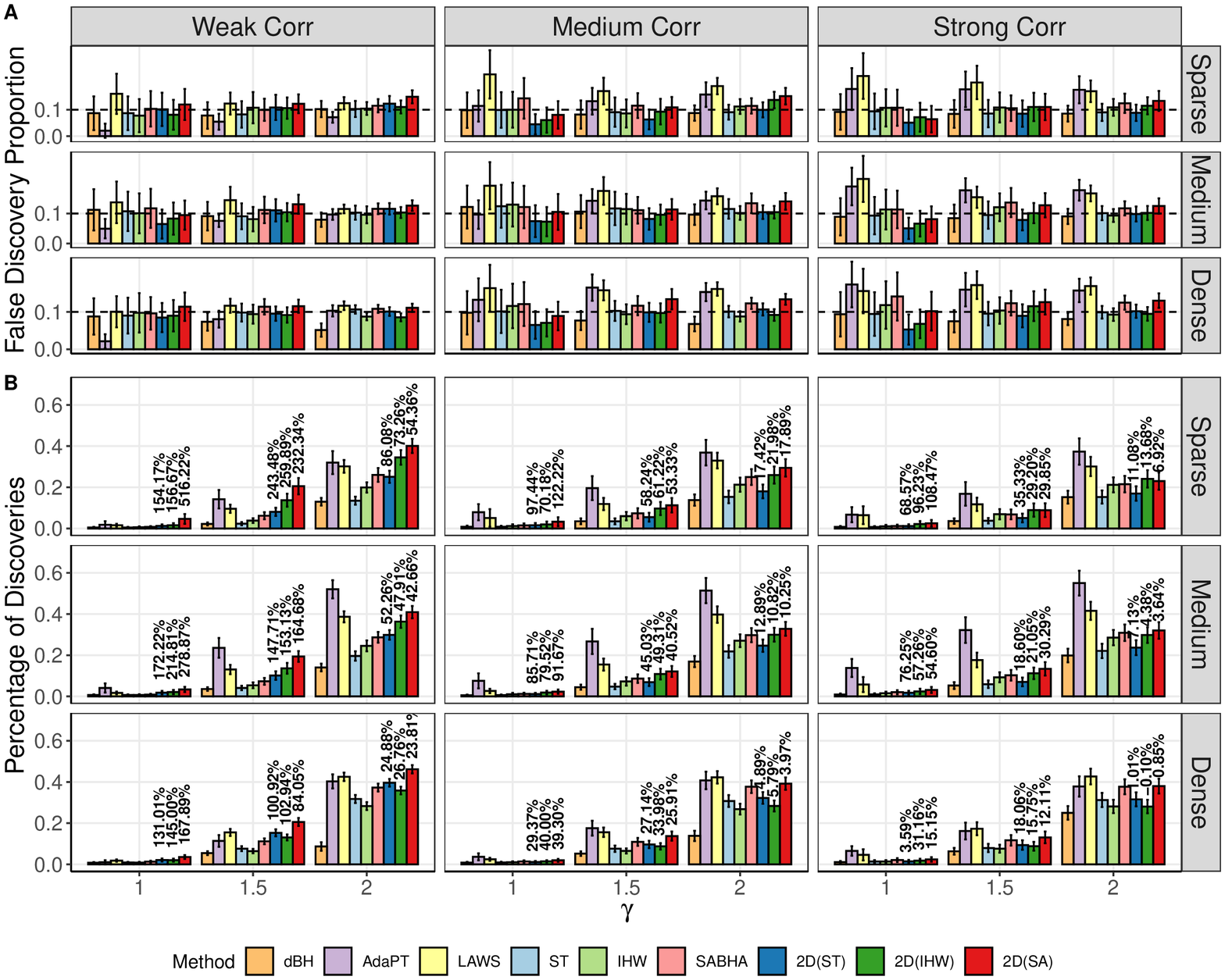}
        \caption{The mean and ($1.96$ multiple of) the standard error of FDP (Panel A) and power (Panel B) under Setup \RNum{2} with $\gamma\in\{1,1.5,2\}$. 
        The percentages on the top of bars represent the power improvement of 2d procedures compared to their 1d counterparts.}
        \label{fig:Unif_Res}
\end{figure}

In this section, we demonstrate the performance of the 2d procedures under the Setup II (see Section~\ref{sec:simu} of the main paper) with $m=900$ and under Setups I--III with $m=2000$. The results for Setup \RNum{2} were displayed in Figure~\ref{fig:Unif_Res} and generally similar to those in Setup \RNum{1} of our main paper. 

\begin{figure}[!h]
     \centering
    \includegraphics[width=\textwidth]{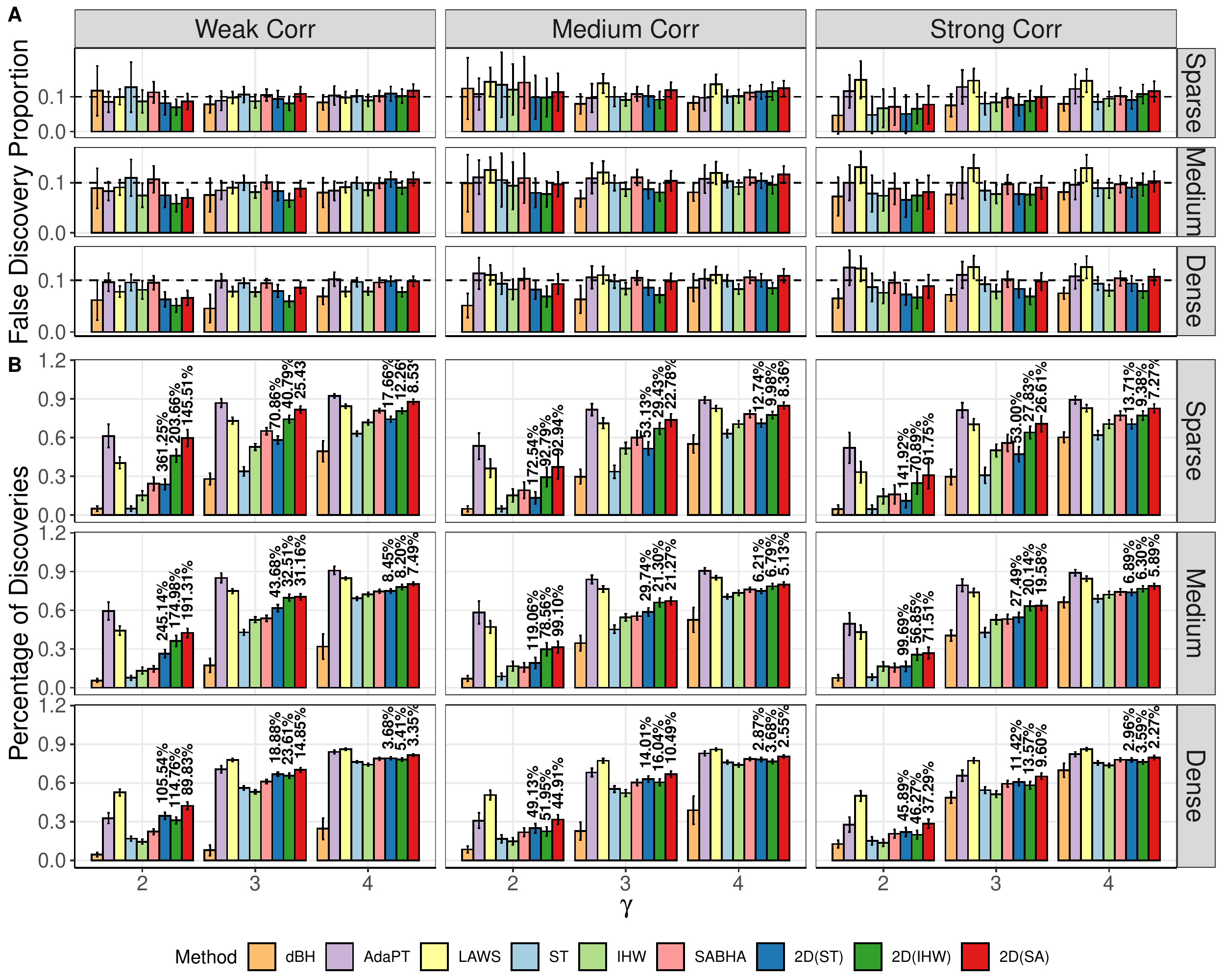}
        \caption{The mean and ($1.96$ multiple of) the standard error of FDP (Panel A) and power (Panel B) under Setup \RNum{1} with $\gamma\in\{2,3,4\}$ and $m=2000$. The percentages on the top of bars represent the power improvement of 2d procedures compared to their 1d counterparts. 
        }
        \label{fig:Spline_m2000_Res}
\end{figure}

\begin{figure}
     \centering
    \includegraphics[width=\textwidth]{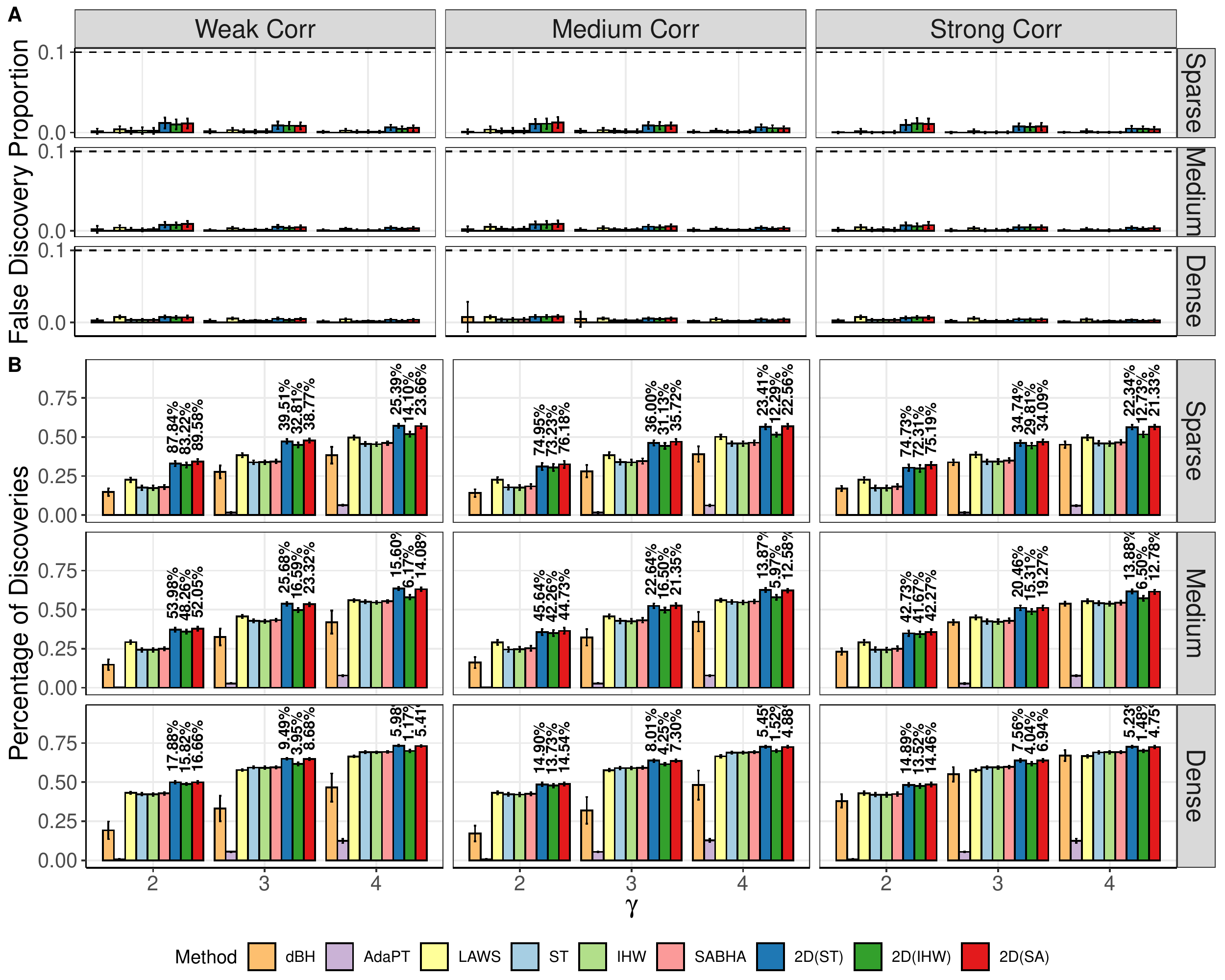}
        \caption{The mean and ($1.96$ multiple of) the standard error {of FDP (Panel A) and power (Panel B)} under Setup \RNum{3} with $\gamma\in\{2,3,4\}$ and $m=2000$. The percentages on the top of bars represent the power improvement of 2d procedures compared to their 1d counterparts. 
        }
        \label{fig:Mv_m2000_Res}
\end{figure}

Then, we demonstrate the performance of the 2d procedures for one-dimensional domain with increased location sizes. 
The simulation setting is the same as Section~\ref{sec:simu} of the main paper but the process $X(s)$ was observed at $m=2000$ locations evenly distributed over the domain $\Scal = [0, 60]$. 
The simulation results are reported in Figures~\ref{fig:Spline_m2000_Res} and \ref{fig:Mv_m2000_Res}. It can be seen that under Setups I--II, 2D (ST), 2D (IHW) and 2D (SA) satisfactorily controlled FDR in all cases. 
Compared to the results of $m=900$, the performance of LAWS and AdaPT with $m=2000$ became better in terms of FDR control. However, they were still more likely to have FDR inflation in contrast with the proposed 2d procedures. 
Under Setup III, the 2d procedures were much more powerful than other methods. 

\begin{figure}[!h]
     \centering
    \includegraphics[width=\textwidth]{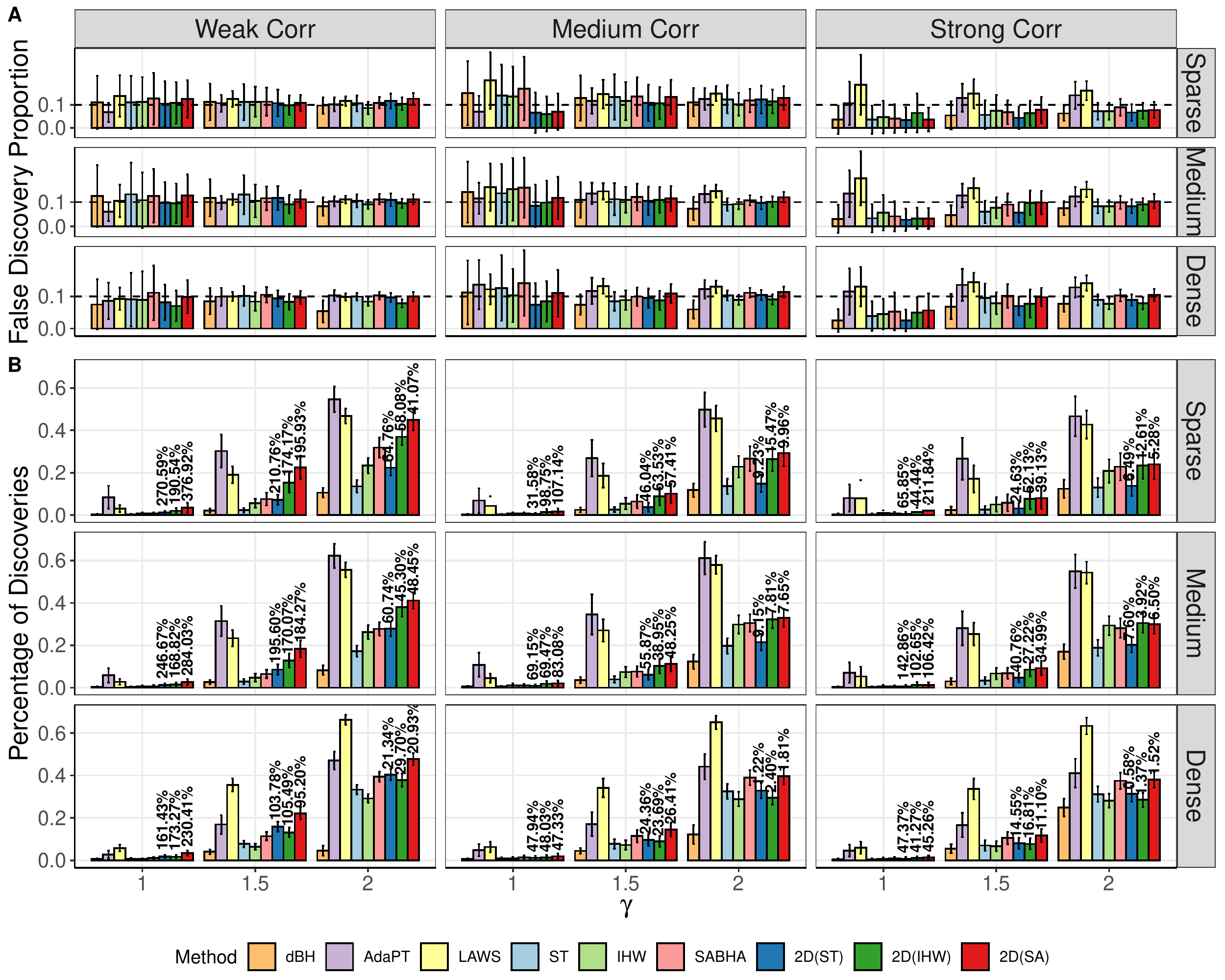}
        \caption{The mean and ($1.96$ multiple of) the standard error {of FDP (Panel A) and power (Panel B)} under Setup \RNum{2} with $\gamma\in\{1,1.5,2\}$ and $m=2000$. 
        The percentages on the top of bars represent the power improvement of 2d procedures compared to their 1d counterparts. 
        }
        \label{fig:Unif_m2000_Res}
\end{figure}

\subsection{Simulation Studies for a Two-Dimensional Domain with Unknown Covariance}\label{sec:Simu_TwoD}
In this section, we consider a spatial process $X(s)=\mu(s)+\epsilon(s)$ defined on the unit square $\Scal=[0,5]^2$. We observe the process on a $30\times 30$ lattice within the unit square. 
The noise $\epsilon(s)$ was generated from a mean-zero Gaussian process defined on $[0,5]^2$ with the same covariance function as described in Section \ref{sec:simu} of the main paper based on the distances between locations. 
Unlike Section~\ref{sec:simu} of the main paper, we now assume that the covariance structure is unknown, but three replications are available at each location. Given multiple realizations at each location, we could employ the maximum likelihood estimation with a pre-specified family of covariance functions to estimate the spatial covariance structure. The details are provided in Section~\ref{sec:cov_est} of the supplement. We consider two structures for the signal process $\mu(s)$ with $s=(s_1,s_e)$.

\begin{itemize}[topsep=0pt,itemsep=0pt,parsep=0pt]
    \item \textbf{Setup \RNum{4} (smooth signal)}: $\mu_{\text{sm}}(s)=\gamma \mu_0(s),$ where $\mu_0(s)=f_1(s_1)f_2(s_2)$ and $\{f_i\}_{i=1,2}$ being generated using B-spline functions as in Setup \RNum{1}.
    \item \textbf{Setup \RNum{5} (clustered signal)}: $\mu_{\text{cl}}(s)=\gamma\mu_0(s)$, where $\mu_0(s)\sim \operatorname{Bernoulli}(\bar{\pi}_0(s))$ with $\bar{\pi}_0(s)=
0.9$, if $(s_1-1/2)^2+(s_2-1/2)^2\leq(1/4)^2$; and $0.01$ otherwise. 
\end{itemize}
Three \textit{signal sparsity levels} based on Setups IV and V were investigated: (1) sparse signal: $\mu_0(s)=\mu_{\text{cl}}(s)$; (2) medium signal: $\mu_0(s)=\mu_{\text{sm}}(s)$; and (3) dense signal: $\mu_0(s)=\mu_{\text{sm}}(s)+\mu_{\text{cl}}(s)$. 
The realized signal processes associated with these sparsity levels are shown in Figure~\ref{fig:2D_Sig} of the supplement, and their percentages of the non-null locations are 5\%, 17\%, and 23\%, respectively. 
We set the magnitude $\gamma$ at values $\{0.5,1,1.5\}$ in both setups.

\begin{figure}[!h]
     \centering
     \includegraphics[width=\textwidth]{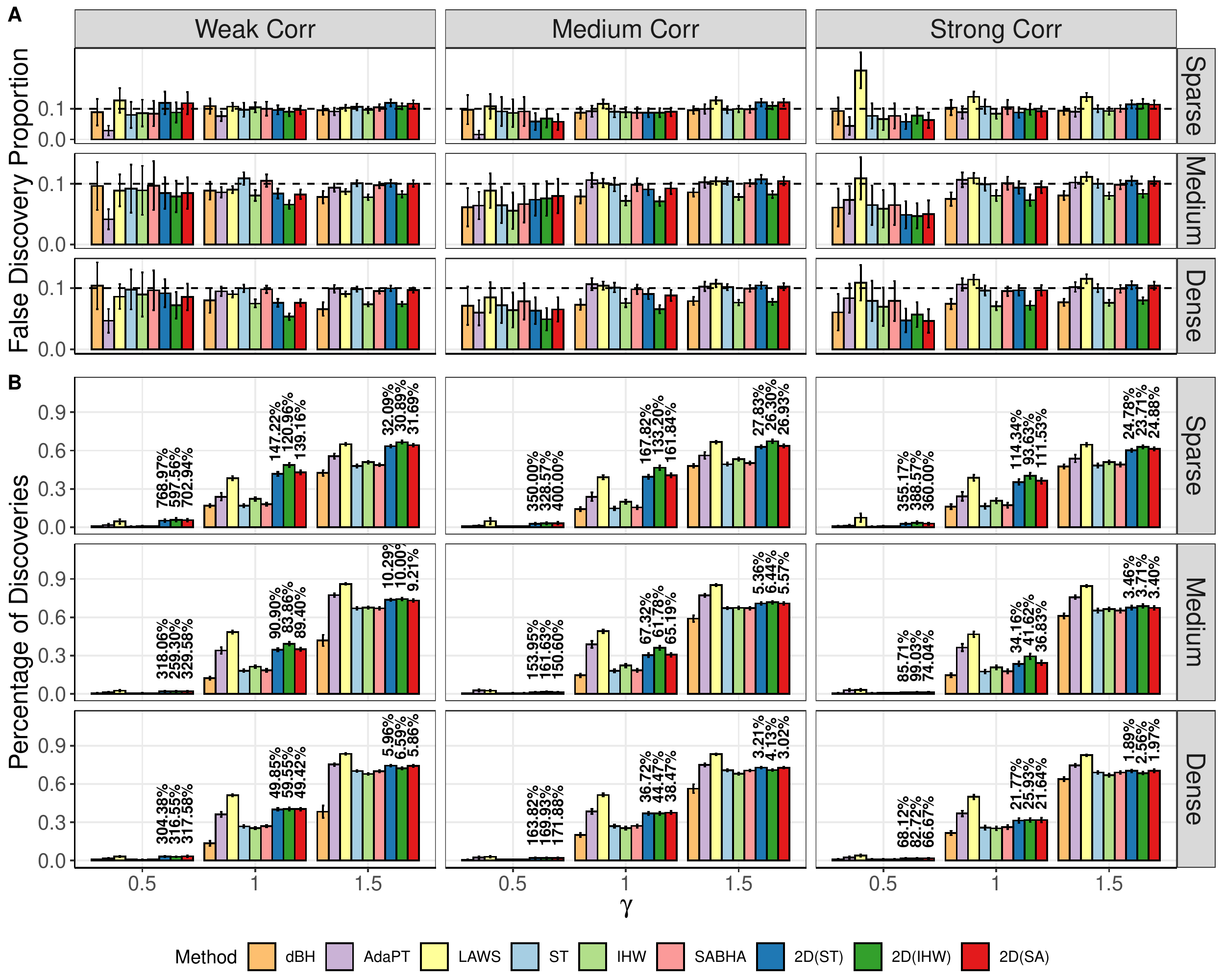}
        \caption{The mean and ($1.96$ multiple of) the standard error {of FDP (Panel A) and power (Panel B)} on the two-dimensional domain with $\gamma\in\{0.5,1,1.5\}$. The percentages on the top of bars represent the power improvement of 2d procedures compared to their 1d counterparts.
        }
        \label{fig:2D_Res}
\end{figure}

We report the numerical results of the FDP and power of competing methods based on 100 simulation runs in Figure~\ref{fig:2D_Res}. 
Generally speaking, the 2d procedures had the best FDR and power trade-off. LAWS showed higher power at the expense of FDR inflation. AdaPT provided reliable FDR control in all cases but their power were dominated by 2D (ST), 2D (IHW), and 2D (SA) for the sparse signal and weak correlation structures.
The power improvements from the 2d procedures were most significant when the correlation was weak. It is also worth mentioning that in the case of strong correlation, the underlying covariance function was based on the Gaussian kernel while we estimated the covariance structure using the exponential kernel. 
The 2d procedures appeared robust to the misspecification subject to the parametric family of covariance functions.

\subsection{Sensitivity to the Number of Nearest Neighbors}\label{sec:rob_nei_num}
\begin{figure}
    \centering
    \includegraphics[width=0.95\textwidth]{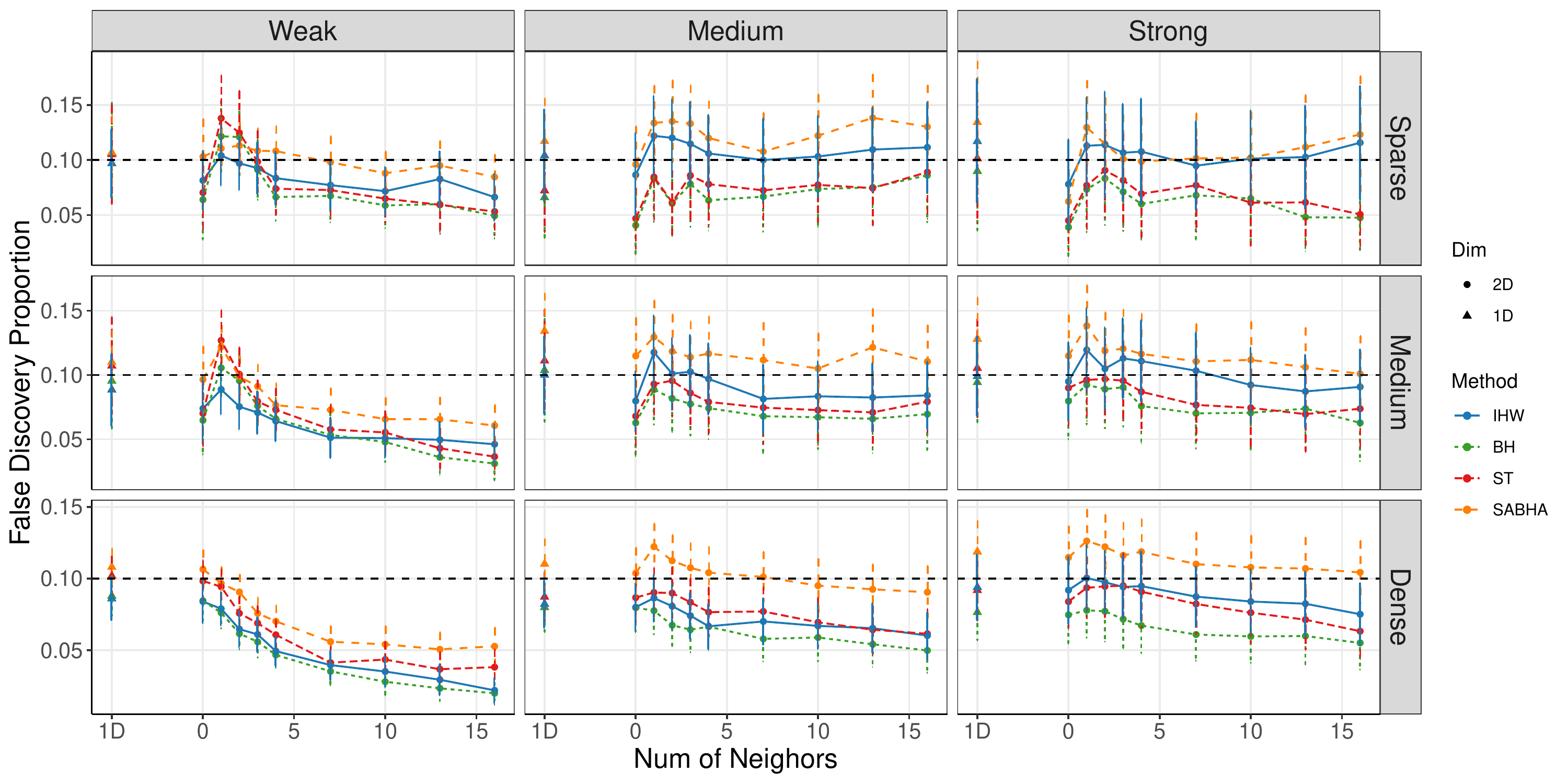}
    \caption{The mean and ($1.96$ multiple of) the standard error of FDP under Setup \RNum{1} with $\gamma=2$. Each color represents one particular 1d procedure (the circle at the leftmost column of each plot) and its 2d counterpart (the triangles).}
    \label{fig:spline_fdp_h}
\end{figure}

\begin{figure}
    \centering
    \includegraphics[width=0.95\textwidth]{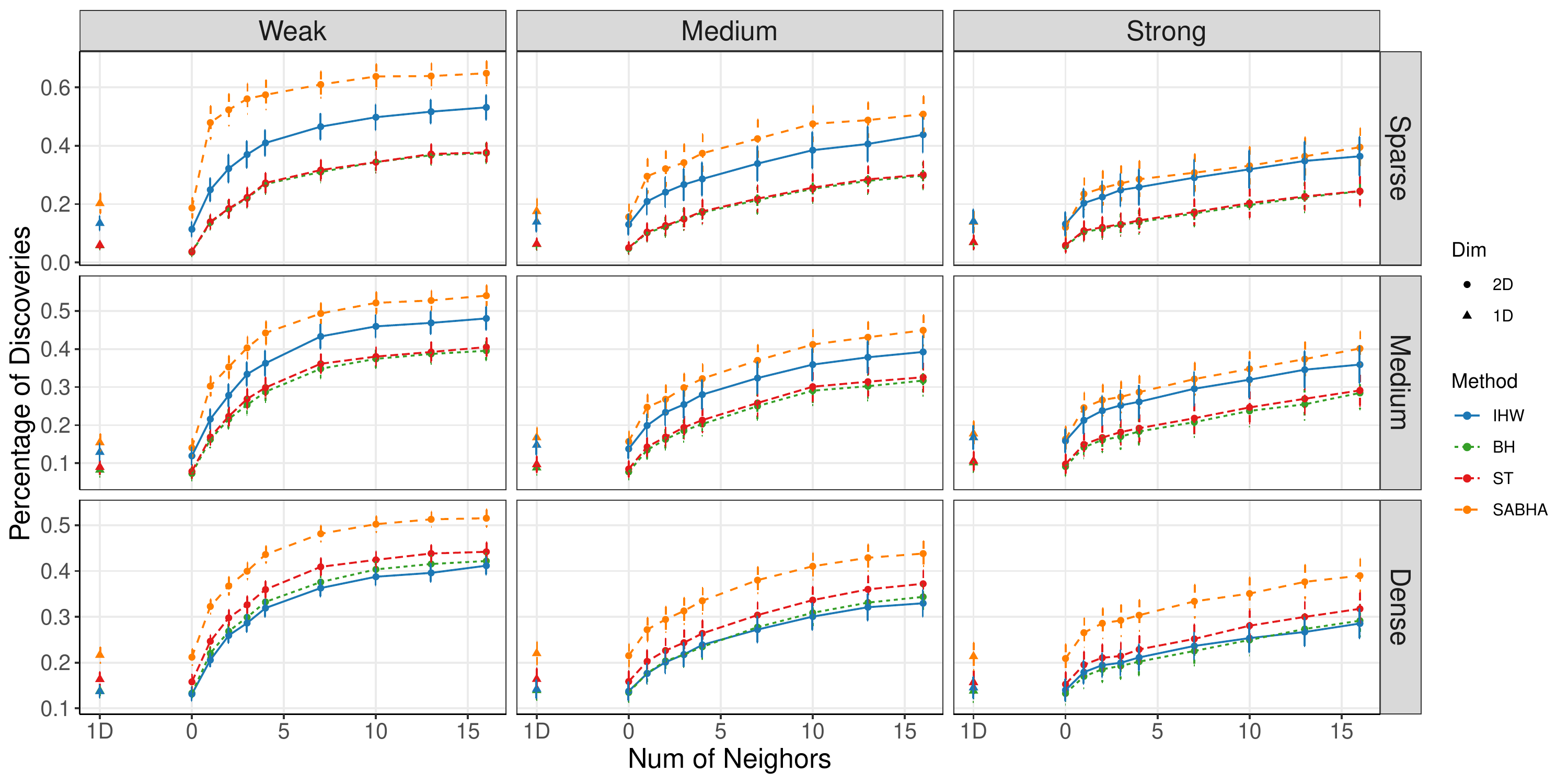}
    \caption{The mean and ($1.96$ multiple of) the standard error of power under Setup \RNum{1} with $\gamma=2$. Each color represents one particular 1d procedure (the circle at the leftmost column of each plot) and its 2d counterpart (the triangles).}
    \label{fig:spline_power_h}
\end{figure}

\begin{figure}
    \centering
    \includegraphics[width=0.95\textwidth]{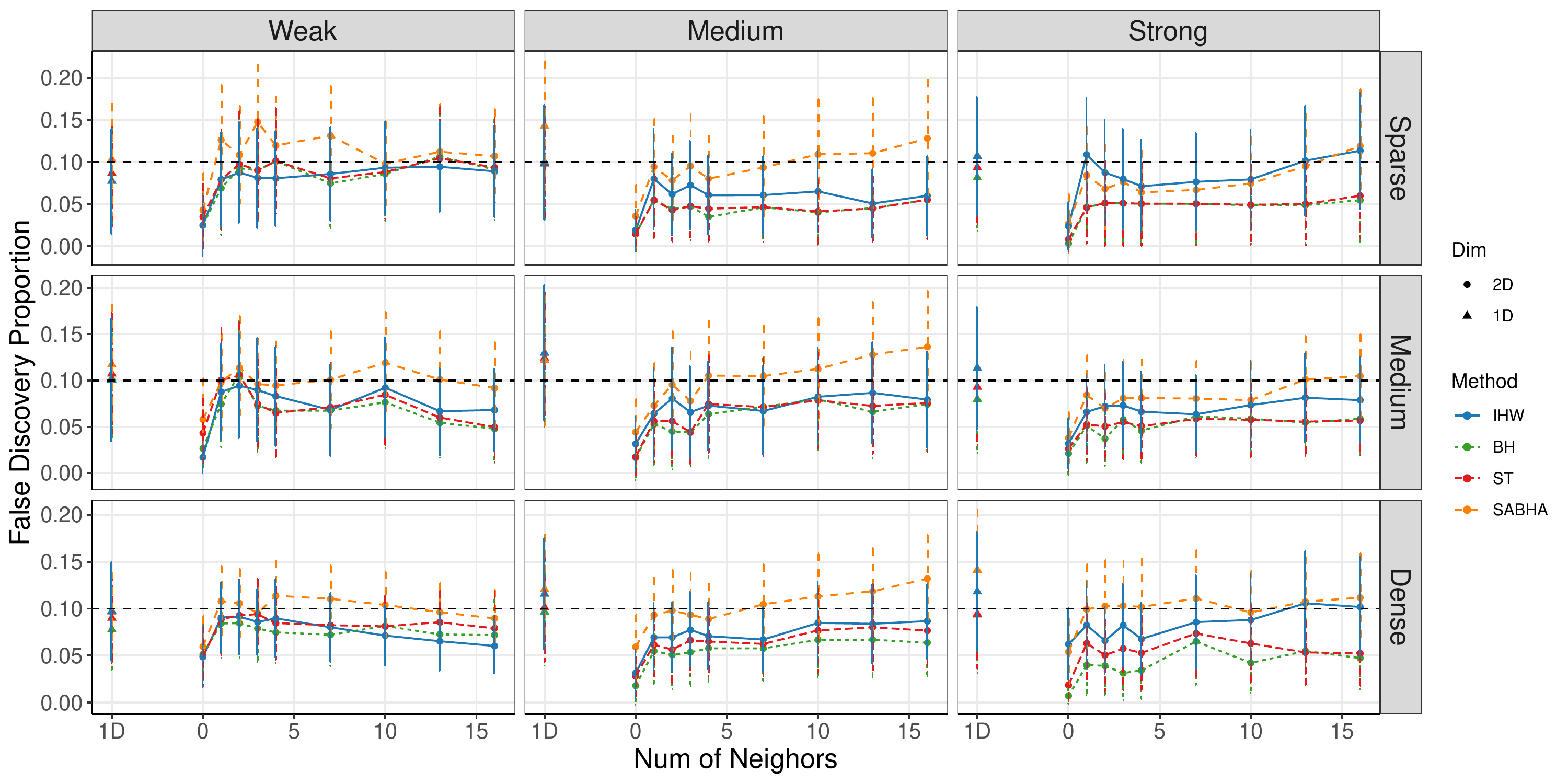}
    \caption{The mean and ($1.96$ multiple of) the standard error of FDP under Setup \RNum{2} with $\gamma=1$. Each color represents one particular 1d procedure (the circle at the leftmost column of each plot) and its 2d counterpart (the triangles).}
    \label{fig:unif_fdp_h}
\end{figure}

\begin{figure}
    \centering
    \includegraphics[width=0.95\textwidth]{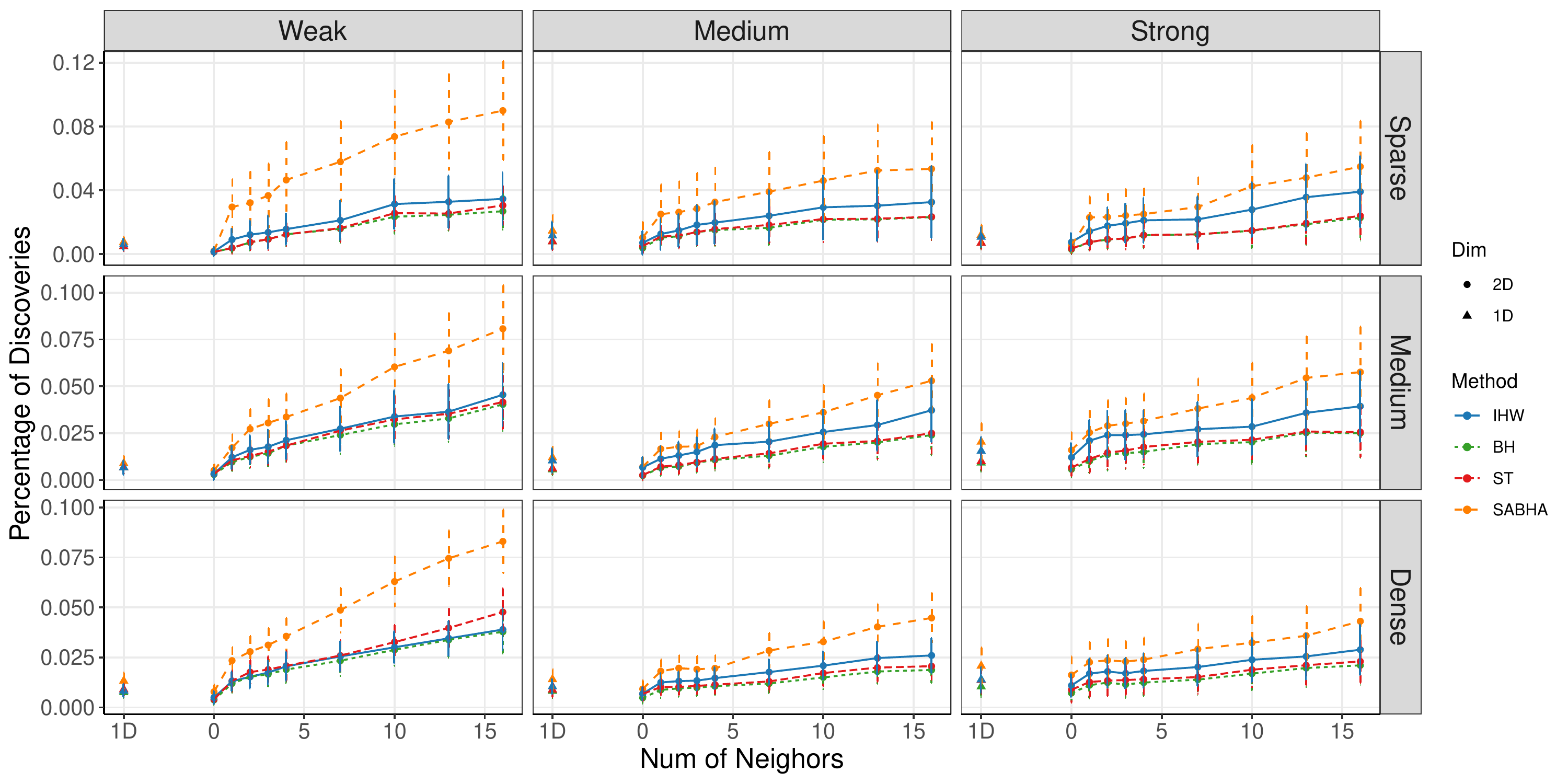}
    \caption{The mean and ($1.96$ multiple of) the standard error of power under Setup \RNum{2} with $\gamma=1$. Each color represents one particular 1d procedure (the circle at the leftmost column of each plot) and its 2d counterpart (the triangles).}
    \label{fig:unif_power_h}
\end{figure}

\begin{figure}
    \centering
    \includegraphics[width=0.95\textwidth]{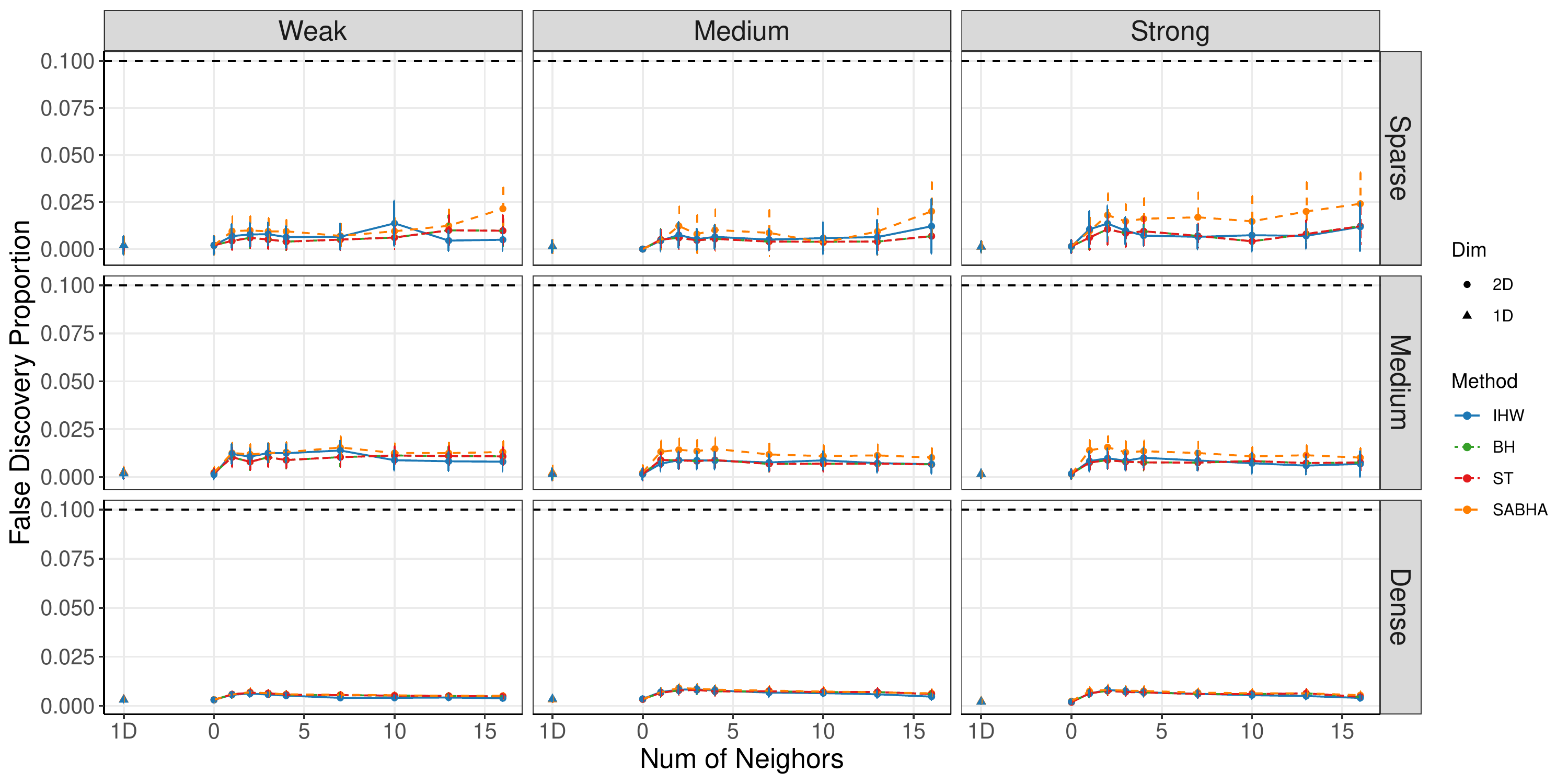}
    \caption{The mean and ($1.96$ multiple of) the standard error of FDP under Setup \RNum{3} with $\gamma=2$. Each color represents one particular 1d procedure (the circle at the leftmost column of each plot) and its 2d counterpart (the triangles).}
    \label{fig:mv_fdp_h}
\end{figure}

\begin{figure}
    \centering
    \includegraphics[width=0.95\textwidth]{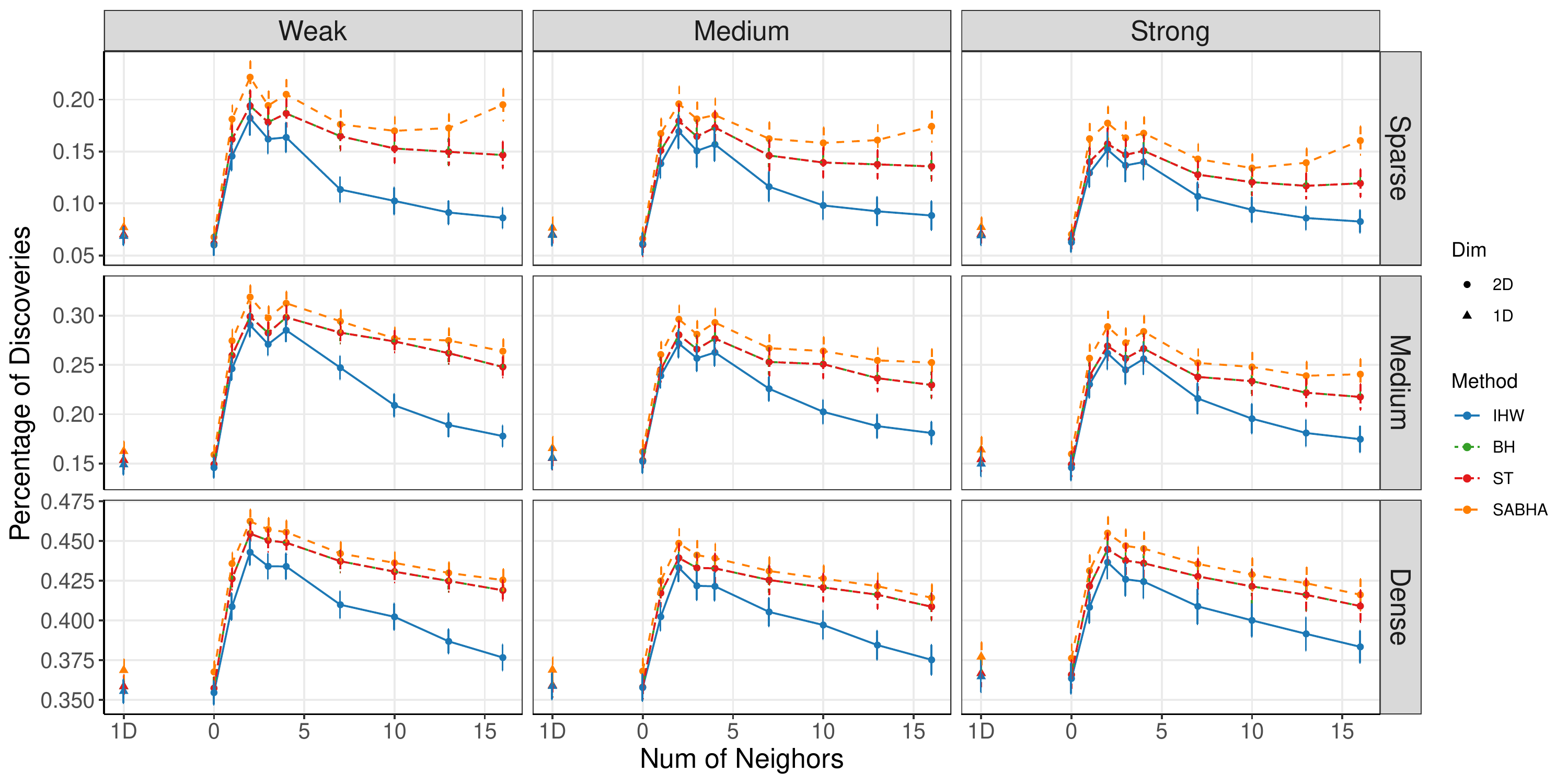}
    \caption{The mean and ($1.96$ multiple of) the standard error of power under Setup \RNum{3} with $\gamma=2$. Each color represents one particular 1d procedure (the circle at the leftmost column of each plot) and its 2d counterpart (the triangles).}
    \label{fig:mv_power_h}
\end{figure}

In this section, we conduct a simulation study to investigate the sensitivity of 2d-SMT to the number of observations in $\mathcal{N}(s)$.
To be concrete, we let $\mathcal{N}(s)$ be the $\kappa$-nearest neighbors for each location $s$.  
We focus on the settings in Setups I--III and let $\kappa\in\{1,2,3,4,7,10,13,16\}$. 
The false discovery proportions (FDPs) and powers of BH, ST, SABHA, and IHW (with the global null proportion estimate) as well as their corresponding 2d versions are summarized in Figures~\ref{fig:spline_fdp_h}--\ref{fig:mv_power_h}. 
The difference between the 2d procedures with zero neighbor and their 1d counterparts is that the 2d procedures add a small offset $q$ (i.e., the FDR level) to the estimate of the number of false rejections (see Section~\ref{sec:imp_detail} of the main paper).

Figures~\ref{fig:spline_fdp_h}, \ref{fig:unif_fdp_h}, and \ref{fig:mv_fdp_h} show that the FDP increased when incorporating the first neighbor, then gradually decreased and maintained stability afterwards as more neighbors were included for the three setups.
As seen from Figures~\ref{fig:spline_power_h} and \ref{fig:unif_power_h}, including more neighbors improved the detection power for all the 2d procedures, which was significantly higher than that of the corresponding 1d counterpart under all setups of Setups I and II. 
In Figure \ref{fig:mv_power_h}, although the detection power of the 2d procedure remained larger than that of the corresponding 1d counterpart, the different 2d procedures behaved differently for varying $\kappa$ under Setup III. 
In particular, the powers of 2D (IHW), 2D (ST), and 2D (BH) gradually decreased when $\kappa$ was larger than 7. For 2D (SABHA), when the signal was sparse, its detection power first decreased and then increased when more neighbors were included. For the medium and dense signal cases, its behavior was similar to the other 2d procedures. These findings empirically suggest that one may choose $\kappa$ between 2 and 7 in 2d-SMT to control the FDP and improve the detection power.

\subsection{Data-Adaptive Neighborhood}\label{sec:adaneigh}
So far, each location has been assigned an equal number of neighbors. In this section, we present a simple strategy to determine the number of neighbors, and then assess its effectiveness through numerical simulations. 
The idea of this strategy is to adaptively enlarge the neighborhood size of locations that are more likely to be within a large cluster of spatial signals.
We achieve this by continuously including neighbors for location $s$ from its nearest neighbors (as candidates) until a primary statistic below zero is found from a candidate, i.e., we use the sign of the primary statistic as an initial criterion to distinguish the nulls and the alternatives in candidates. 
To be specific, let $\Ncal_\kappa(s)$ denote the set of $\kappa$-nearest neighbors of location $s$. The number of neighbors for location $s$ is  determined as follows
\begin{equation}\label{equ:kappas}
    \kappa_s =
\left\{ 
    \begin{array}{l}
        4,  \qquad \qquad \qquad \qquad \qquad \qquad  T_2(v)<0 \text{ for some }v\in \Ncal_2(s) \\
        \max\left\{2\leq \kappa\leq 7:T_2 (v)>0\text{ for all }v\in \Ncal_\kappa(s)
\right\}, ~~\text{otherwise,}\\
    \end{array}
\right.
\end{equation}
where a value of four is used in Section~\ref{sec:simu} of the main paper and the range of $\kappa$ is suggested in Section~\ref{sec:rob_nei_num}. 
In this experiment, we considered two variants to implement the above strategy: 1) utilized a separate dataset specifically for the neighbor selection process, and 2) used the same dataset for both the neighbor selection process and the subsequent inference.

We generated the synthetic datasets according to Setups I--III in Section~\ref{sec:simu} of the main paper, equipped with the medium signal and weak correlation. 
Four different methods were compared: the standard ST and its three 2d variants. The first variant, 2D$_{\mathrm{Fix}}$(ST), assigned 4-nearest neighbors for all locations.  
The second variant, 2D$_{\mathrm{Ada},1}$(ST), and the third variant, 2D$_{\mathrm{Ada},2}$(ST), determined the number of neighbors according to \eqref{equ:kappas} using a separate dataset and the inference dataset, respectively. 

\begin{figure}
    \centering
    \includegraphics[width=\textwidth]{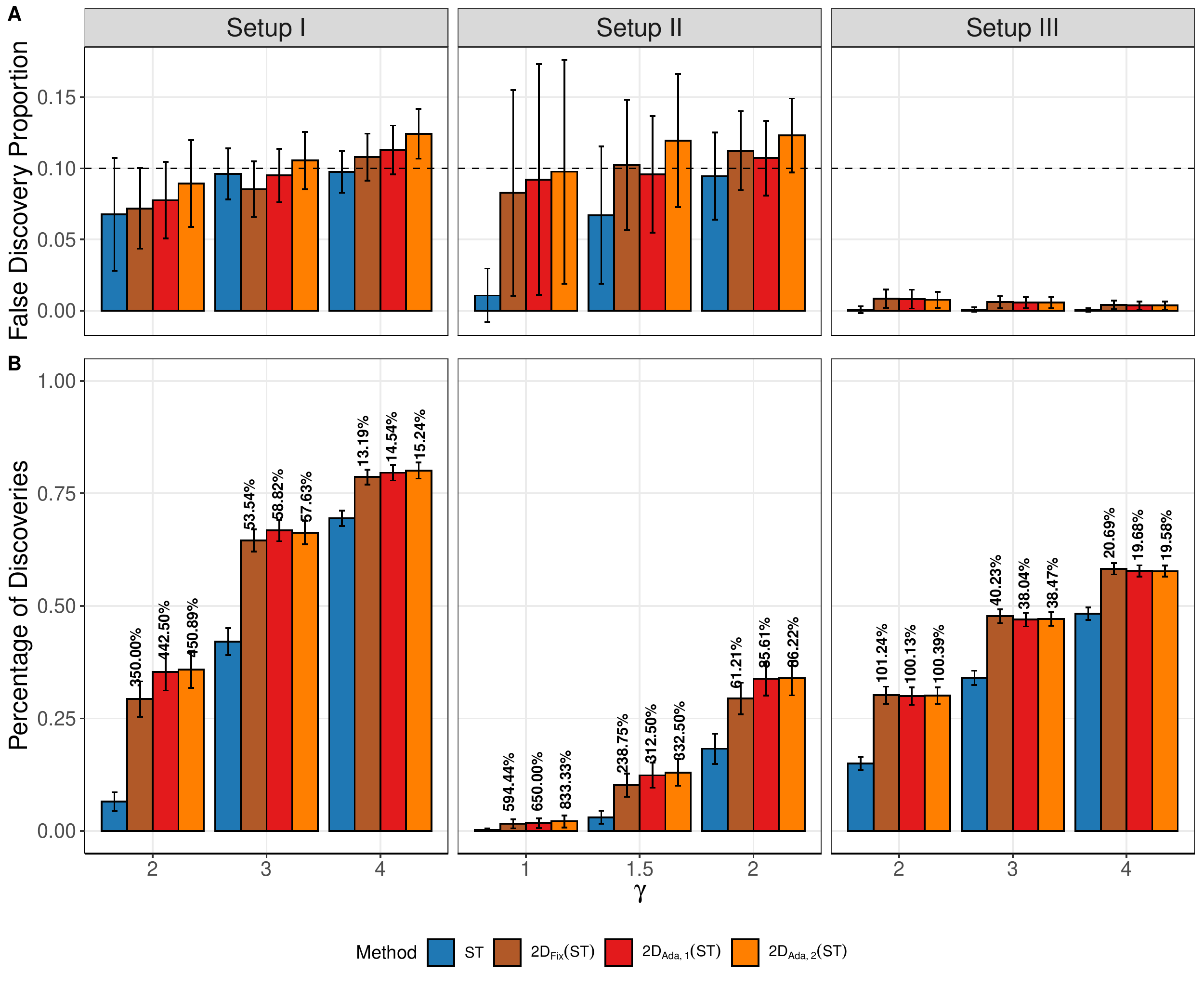}
    \caption{The mean and (1.96 multiple of) the standard error of FDP (Panel A) and power (Panel B) under Setup I with $\gamma\in\{2,3,4\}$, Setup II with $\gamma\in\{1,1.5,2\}$, and Setup III with $\gamma\in\{2,3,4\}$. The percentages on the top of bars represent the power improvement of 2D$_{\mathrm{Fix}}$(ST), 2D$_{\mathrm{Ada},1}$(ST) and 2D$_{\mathrm{Ada},2}$(ST) compared to ST.}
    \label{fig:1D_Neigh}
\end{figure}

We report the numerical results of the empirical FDP and power of the competing methods based on 100 simulation runs in Figure~\ref{fig:1D_Neigh}. All 2d procedures exhibited an enhanced power relative to the standard ST procedure. Our neighbor selection strategy, as detailed in \eqref{equ:kappas}, boosted the power in Setups I and II. Notably, the 2D$_{\mathrm{Ada},2}$(ST) approach encountered difficulties in controlling FDR, particularly in Setup I with $\gamma=4$. Its performance highlights the necessity for caution in data reuse, especially when no additional dataset for neighbor selection is available. 

\begin{figure}
    \centering
    \includegraphics[width=\textwidth]{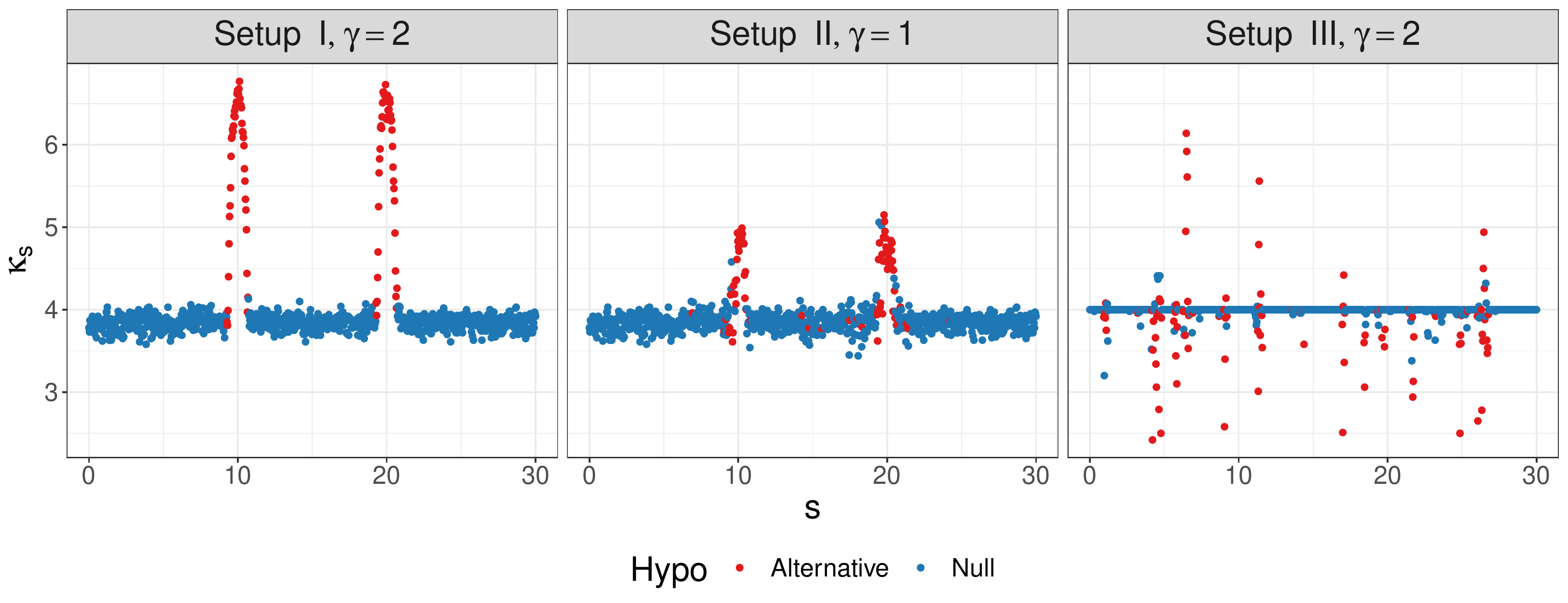}
    \caption{The average number of neighbors for 900 locations based on 100 simulation runs.}
    \label{fig:Neigh}
\end{figure}

Figure~\ref{fig:Neigh} displays the average number of neighbors for all locations based on 100 simulation runs. The strategy \eqref{equ:kappas} effectively allocated a larger set of neighbors to locations under the alternative in Setups I and II, thereby improving power through adaptive neighbor selection. As illustrated in Figure~\ref{fig:1D_sig_mvnorm}, given that locations under the alternative were dispersed into several small clusters in Setup III, determining neighbors became more complex.  In conclusion, while adaptive selection of neighbors can improve detection efficiency, further investigation is required to ensure both the safety and efficacy of the neighbor selection strategy.

\subsection{Combining Covariate and Spatial Information}\label{sec:covarSpa}
Section~\ref{sec:vary_null} of the main paper introduces the concept of weighted thresholds, highlighting the potential to further exploit spatial information. When both covariate and spatial information are available, we can utilize the covariate information through weights and spatial information through auxiliary statistics. In this section, we conduct simulations to investigate the ability of the 2d-SMT procedure to simultaneously use spatial and covariate information. 

\begin{figure}
    \centering
    \includegraphics[width=0.4\textwidth]{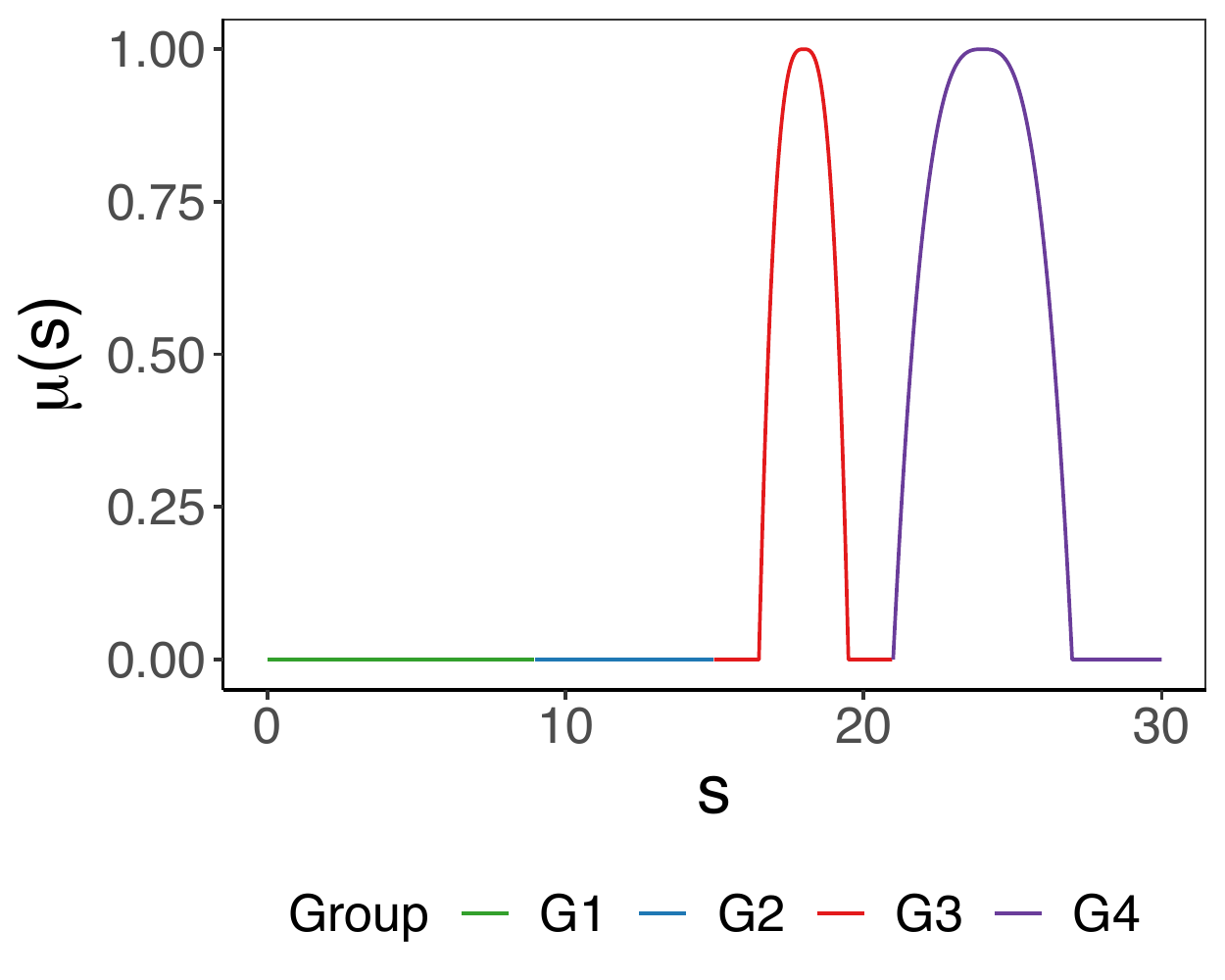}
    \caption{The one-dimensional signal process $\mu(s)$ ($s\in[0,30]$) with $\gamma=1$ that possesses both spatial and group information.}
    \label{fig:1D_Sig_Spa_Grp}
\end{figure}

In a variant of Setup I, as illustrated in Figure~\ref{fig:1D_Sig_Spa_Grp}, we partitioned the locations into four groups: the first two groups were all under the null, whereas the third and fourth groups comprised approximately $1/2$ and $2/3$ of locations under the alternative, respectively. In this setup, the signal pattern exhibits both grouping designs and spatial trends. We considered the three degrees of spatial dependence as shown in Figure~\ref{fig:1D_Cov}. We evaluated four types of weights, deriving eight methods by considering both the 1d and 2d approaches. The BH and 2D (BH) used uniform weights $w(s)\equiv 1$. The FDP estimator of 2D (ST) is detailed in Section~\ref{sec:f_method} of the main paper, while the FDP estimator of ST sets the threshold for primary statistics $t_1$ to be $-\infty$. To incorporate group information, we utilized the null proportion estimators as described in Section~\ref{sec:prop} for each group, denoted as groupwise null proportion estimates $\widehat{\pi}_0(s)$. Based on these estimates, $\mathrm{SA}_\mathrm{Grp}$ and 2D ($\mathrm{SA}_\mathrm{Grp}$) assigned weights inversely proportional to $\widehat{\pi}_0(s)$, and $\mathrm{LAWS}_\mathrm{Grp}$ and 2D ($\mathrm{LAWS}_\mathrm{Grp}$) assigned weights proportional to $\{1-\widehat{\pi}_0(s)\}/\widehat{\pi}_0(s)$.

\begin{figure}
     \centering
    \includegraphics[width=\textwidth]{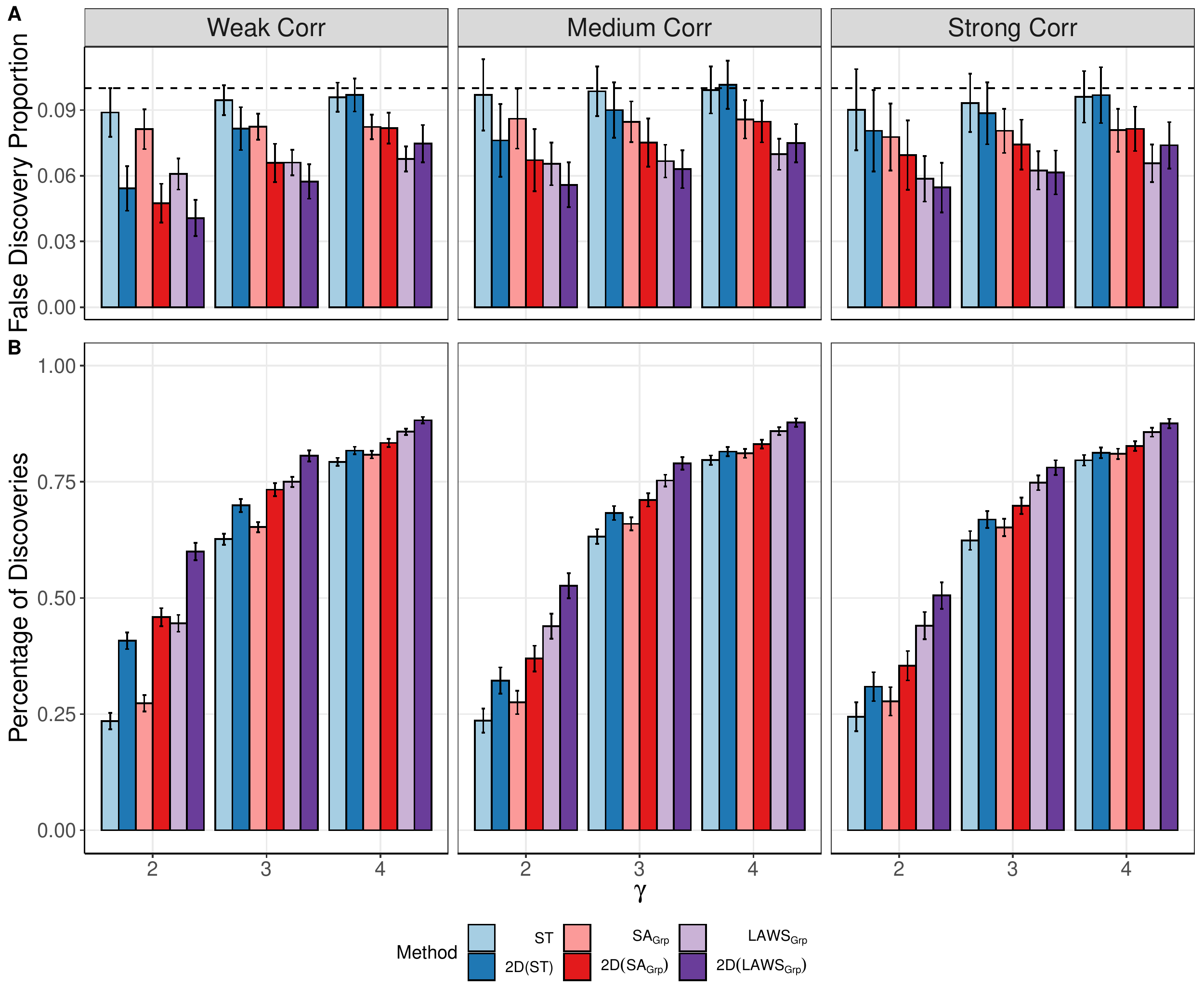}
        \caption{The mean and ($1.96$ multiple of) the standard error {of FDP (Panel A) and power (Panel B)} under the setup that possesses both spatial pattern and group information with $\gamma\in\{2,3,4\}$ and $m=900$. The percentages on the top of bars represent the power improvement of 2d procedures compared to their 1d counterparts. 
        }
        \label{fig:Spa_grp_m900_Res}
\end{figure}

We report the numerical results of the empirical FDP and power of the eight methods based on
100 simulation runs in Figure~\ref{fig:Spa_grp_m900_Res}. All procedures controlled the $\FDR$ fairly well. The performance of the four 1d procedures indicated that incorporating null proportion estimates can significantly enhance detection power, and including groupwise estimates can provide further improvements. Furthermore, $\mathrm{LAWS}_\mathrm{Grp}$, which employs a weighting function that uses group information more aggressively, detected more signals. The 2d procedures outperformed their 1d counterparts by incorporating spatial information. In general, 2D ($\mathrm{LAWS}_\mathrm{Grp}$) achieved the highest power while controlling the FDR under the nominal level.

\subsection{Additional Results for the Ozone Data Analysis}
\label{sec:o3_data_more}

\begin{figure}
    \centering
    \includegraphics[width=0.5\textwidth]{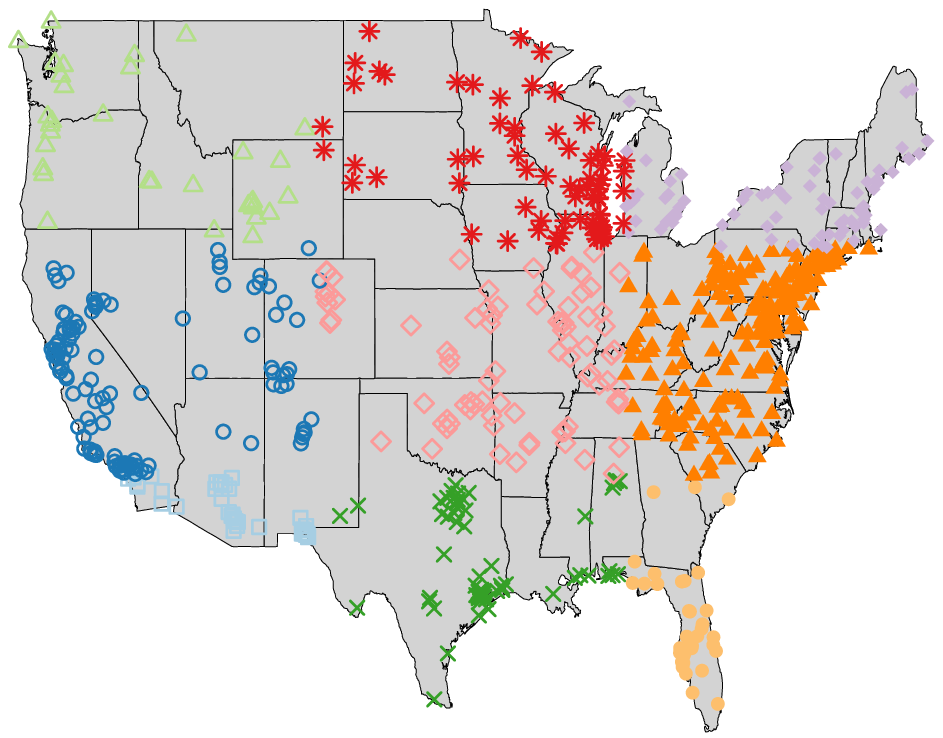}
    \caption{The Contiguous United States is divided into nine different regions based on latitude and longitude, i.e., each station belongs to a region coded between one and nine. The stations in the same region are depicted using the same shape and color.}
    \label{fig:ozone_reg}
\end{figure}

Figure~\ref{fig:ozone_reg} uses distinct shapes and colors to illustrate the partition of the Contiguous United States into nine different regions as discussed in Section \ref{sec:real_data} of the main paper. The region to which each location belongs is treated as a categorical variable and is used as the covariate in IHW and the group indicator in SABHA.

Table~\ref{tab:prec_descent} shows the locations with the most significant decline in CO or \NOtwo{} concentration levels and whether the competing methods detected them in the study of the decline of the ozone level for various $\beta_0$. 
Table~\ref{tab:avr_prec} presents the average standardized slopes of CO and \NOtwo{} at the locations detected by the 2d procedure or the 1d procedure, but not both. It shows that the \NOtwo{} concentration level at the locations detected by the 2d procedures decreased faster than their 1d counterparts on average. 
For CO, when the null hypothesis was $\beta_0 = 0.2$ or $\beta_0=0.5$, the average standardized slope of its concentration level at the locations detected by the 2d procedures remained smaller than their 1d counterparts. 
The only exception was when $\beta_0=3$, the locations detected by the 1d procedure (SABHA) had a smaller average standardized slope of CO level than those detected by 2D (SA).

\begin{table}[htbp]
  \centering
  \caption{The locations with the most significant decline in the CO or \NOtwo{} levels. Reported are the locations where at least one of the 2d procedures gives different decisions to their corresponding 1d counterparts. The column ``Stand. Coeff.'' represents the value of the standardized coefficient.} 
  \resizebox{\textwidth}{!}{
    \begin{tabular}{crrrrcccccc}
    \toprule
    \toprule
    \multirow{2}[2]{*}{\makecell[c]{Ozone\\precursor}} & \multicolumn{1}{c}{\multirow{2}[2]{*}{$\beta_0$}} & \multicolumn{1}{c}{\multirow{2}[2]{*}{Lat.}} & \multicolumn{1}{c}{\multirow{2}[2]{*}{Lon.}} & \multicolumn{1}{c}{\multirow{2}[2]{*}{\makecell[c]{Stand.\\ Coeff.}}} & \multicolumn{6}{c}{Methodology} \\
  \cmidrule{6-11}         &       &       &       &       & ST    & 2D (ST) & IHW   & 2D (IHW) & SABHA & 2D (SA) \\
    \midrule
    \multirow{5}[2]{*}{CO} & 0.1  & 40.63  & -75.34  & -0.56  & \xmark & \cmark & \xmark & \xmark & \xmark & \xmark \\
          & 0.2  & 42.14  & -87.80  & -0.94  & \xmark & \cmark & \xmark & \xmark & \xmark & \xmark \\
          & 0.3  & 41.00  & -80.35  & -2.19  & \cmark & \cmark & \cmark & \cmark & \xmark & \cmark \\
          & 0.4  & 36.20  & -95.98  & -0.85  & \cmark & \cmark & \cmark & \cmark & \xmark & \cmark \\
          & 0.5  & 41.53  & -90.59  & -3.46  & \xmark & \xmark & \xmark & \xmark & \xmark & \cmark \\
    \midrule
    \multirow{5}[2]{*}{NO2} & 0.1  & 40.63  & -75.34  & -3.49  & \xmark & \cmark & \xmark & \xmark & \xmark & \xmark \\
          & 0.2  & 42.14  & -87.80  & -4.60  & \xmark & \cmark & \xmark & \xmark & \xmark & \xmark \\
          & 0.3  & 32.87  & -97.91  & -43.10  & \cmark & \cmark & \cmark & \cmark & \xmark & \cmark \\
          & 0.4  & 35.41  & -94.52  & -7.27  & \xmark & \xmark & \xmark & \xmark & \xmark & \cmark \\
          & 0.5  & 41.53  & -90.59  & -9.16  & \xmark & \xmark & \xmark & \xmark & \xmark & \cmark \\
    \bottomrule
    \end{tabular}%
    }
  \label{tab:prec_descent}%
\end{table}%

\begin{table}[htbp]
  \centering
  \caption{The average standardized slopes of CO and \NOtwo{} at the locations where at least one of the 2d procedures gives different decisions to their corresponding 1d counterparts. Each letter refers to the specific location depicted in Figure~\ref{fig:ozone_analysis} of the main paper.}
    \begin{tabular}{clcccccc}
    \toprule
    \toprule
    \multicolumn{1}{c}{\multirow{3}[3]{*}{\makecell[c]{Ozone\\precursor}}} & \multicolumn{1}{c}{\multirow{3}[3]{*}{Methodology}} & \multicolumn{6}{c}{$\beta_0$} \\ \cmidrule(lr){3-8}&& \multicolumn{2}{c}{0.2} & \multicolumn{2}{c}{0.3} & \multicolumn{2}{c}{0.5} \\
 \cmidrule(lr){3-4}\cmidrule(lr){5-6}\cmidrule(lr){7-8}
          &       & \multicolumn{1}{c}{Loc.} & \multicolumn{1}{c}{Avr.} &\multicolumn{1}{c}{Loc.}   & \multicolumn{1}{c}{Avr.}&\multicolumn{1}{c}{Loc.}  & \multicolumn{1}{c}{Avr.} \\
    \midrule
    \multirow{4}[2]{*}{CO} & ST    &  \makecell[c]{—}        &    \makecell[c]{—}     &   \makecell[c]{—}       &    \makecell[c]{—}     &    \makecell[c]{—}      &  \makecell[c]{—}   \\
          & 2D (ST) & d,e & -0.48  &   \makecell[c]{—}       &    \makecell[c]{—}      &    \makecell[c]{—}      &  \makecell[c]{—}   \\
          & SABHA &   \makecell[c]{—}    &   \makecell[c]{—}       & e, j   & 0.11  & b     & -0.96  \\
          & 2D (SA) &   \makecell[c]{—}       &   \makecell[c]{—}       & h, i   & 0.56  & c, e, f, h & -1.40  \\
    \midrule
    \multirow{4}[2]{*}{\NOtwo{}} & ST    & c& -1.30  & f     & -1.30  &    \makecell[c]{—}      &  \makecell[c]{—}   \\
          & 2D (ST) & a, e, f & -2.73  & a, c   & -6.44  &    \makecell[c]{—}      &  \makecell[c]{—}   \\
          & SABHA &     \makecell[c]{—}     &    \makecell[c]{—}      & e, f, j & -2.81  & b, d   & -3.63  \\
          & 2D (SA) &   \makecell[c]{—}       &      \makecell[c]{—}   & b, d, g & -17.17  & a, c, f, g, h, i & -4.68  \\
    \bottomrule
    \end{tabular}%
  \label{tab:avr_prec}%
\end{table}%

Motivated by \cite{Sun2015}, we conducted further simulations to study the sensitivity of 2d-SMT to covariance misspecification. We considered two approaches to determine the covariance for simulating the ozone level data. The first approach fitted the residuals with a Gaussian kernel, and the second utilized the empirical covariance matrix to accommodate non-stationarity. We then generated the simulated data accordingly to \eqref{equ:ozone_model} of the main paper with modified covariance. The subsequent steps followed those of the ozone simulations in Section~\ref{sec:real_data}, which used the exponential kernel. Tables~\ref{tab:ozonGau} and \ref{tab:ozonEmp} demonstrate that the 2d procedures still achieved equal or higher power compared to their 1d counterparts while controlling FDR under 10\%. To summarize, our procedure is robust to model misspecification, which confirms the reasonableness of modeling ozone data with \eqref{equ:ozone_model} of the main paper.

\begin{table}[!h]\small
  \centering
  \caption{{Mean and standard deviation of FDPs and percentage of true discoveries (PTDs) for simulated ozone data using the Gaussian kernel.} The results are based on 100 simulation runs.}\label{tab:ozonGau}
  
\resizebox{\linewidth}{!}{
    \begin{tabular}{rrrrrrrr}
    \toprule
    \toprule
Criterion &$\beta_0$ & \multicolumn{1}{l}{ST} & \multicolumn{1}{l}{IHW} & \multicolumn{1}{l}{SABHA} & \multicolumn{1}{l}{2D (ST)} & \multicolumn{1}{l}{2D (IHW)} & \multicolumn{1}{l}{2D (SA)} \\
    \midrule
    \multicolumn{1}{c}{\multirow{5}[4]{*}{FDP}}
    & 0.5&0.024(0.027)&0.025(0.029)&0.042(0.030)&0.025(0.028)&0.025(0.029)&0.054(0.036)\\
     &0.4&0.022(0.018)&0.020(0.017)&0.026(0.017)&0.023(0.019)&0.020(0.017)&0.033(0.018)\\
    &0.3&0.017(0.014)&0.013(0.013)&0.014(0.012)&0.017(0.015)&0.013(0.013)&0.015(0.012)\\
    &0.2  & 0.017(0.013)&0.010(0.009)&0.012(0.010)&0.017(0.013)&0.010(0.009)&0.012(0.010)\\
   & 0.1  &0.015(0.011)&0.008(0.008)&0.009(0.008)&0.015(0.012)&0.008(0.008)&0.009(0.008)\\
    \midrule
    \multicolumn{1}{c}{\multirow{5}[4]{*}{PTD}}& 0.5  &0.277(0.097)&0.280(0.095)&0.412(0.115)&0.281(0.099)&0.280(0.095)&\textbf{0.474(0.114)}\\
    &0.4
    &0.401(0.115)&0.391(0.106)&0.481(0.079)&0.409(0.117)&0.391(0.106)&\textbf{0.517(0.081)}\\
    &0.3  &0.547(0.112)&0.508(0.099)&0.511(0.097)&\textbf{0.554(0.113)}&0.508(0.099)&0.521(0.096)\\
    &0.2  &0.695(0.088)&0.628(0.079)&0.633(0.081)&\textbf{0.702(0.088)}&0.628(0.079)&0.640(0.081)\\
   & 0.1  &0.799(0.060)&0.716(0.061)&0.725(0.060)&\textbf{0.803(0.060)}&0.716(0.061)&0.729(0.060)\\
    \bottomrule
    \end{tabular}%
    }
\end{table}%

\begin{table}[!h]\small
  \centering
  \caption{{Mean and standard deviation of FDPs and percentage of true discoveries (PTDs) for the simulated ozone data using the empirical covariance matrix.} The results are based on 100 simulation runs.}\label{tab:ozonEmp}
  
\resizebox{\linewidth}{!}{
    \begin{tabular}{rrrrrrrr}
    \toprule
    \toprule
Criterion &$\beta_0$ & \multicolumn{1}{l}{ST} & \multicolumn{1}{l}{IHW} & \multicolumn{1}{l}{SABHA} & \multicolumn{1}{l}{2D (ST)} & \multicolumn{1}{l}{2D (IHW)} & \multicolumn{1}{l}{2D (SA)} \\
    \midrule
    \multicolumn{1}{c}{\multirow{5}[4]{*}{FDP}}& 0.5  &0.018(0.036)&0.018(0.035)&0.027(0.040)&0.019(0.038)&0.018(0.036)&0.036(0.044)\\
    &0.4&0.017(0.034)&0.014(0.027)&0.018(0.029)&0.018(0.034)&0.014(0.027)&0.020(0.030)\\
    &0.3  &0.016(0.025)&0.011(0.019)&0.012(0.020)&0.017(0.026)&0.011(0.019)&0.013(0.020)\\
    &0.2  & 0.016(0.021)&0.009(0.015)&0.010(0.015)&0.016(0.022)&0.009(0.015)&0.010(0.016)\\
   & 0.1  &0.016(0.021)&0.008(0.012)&0.008(0.012)&0.016(0.021)&0.008(0.012)&0.008(0.013)\\
    \midrule
    \multicolumn{1}{c}{\multirow{5}[4]{*}{PTD}}& 0.5  &0.258(0.189)&0.264(0.182)&0.369(0.207)&0.261(0.192)&0.265(0.183)&\textbf{0.416(0.216)}\\
    &0.4&0.375(0.221)&0.363(0.205)&0.412(0.181)&0.380(0.223)&0.362(0.206)&\textbf{0.438(0.176)}\\
    &0.3  &0.500(0.229)&0.465(0.207)&0.480(0.190)&\textbf{0.506(0.229)}&0.464(0.207)&0.491(0.187)\\
    &0.2  &0.645(0.206)&0.582(0.189)&0.586(0.187)&\textbf{0.650(0.205)}&0.582(0.190)&0.592(0.186)\\
   & 0.1  &0.761(0.152)&0.677(0.146)&0.682(0.149)&\textbf{0.765(0.151)}&0.677(0.147)&0.686(0.148)\\
    \bottomrule
    \end{tabular}%
    }
\end{table}%

\section{Proof of Theorem 1 of the Main Paper}\label{app:proof}
\allowdisplaybreaks

We introduce two lemmas, which play key roles in the proof of Theorem~\ref{thm:cnvrg_fdr} of the main paper.

\begin{lemma}\label{lemma:cnvrg_discovery}
	Under Assumptions~\ref{ass:incr_dom}--\ref{ass:cnvrg_true} of the main paper, for any $t^\prime_1$, $t^\prime_2>0$, we have
{\begin{equation}\label{eqn:k0FirstStep}
\begin{aligned}
	\sup _{|t_{1}| \leq t_{1}^{\prime},|t_{2}| \leq t_{2}^{\prime}} & \bigg|\frac{1}{m}\sumsSm \bigg[ \int L\left\{ t_{1}, t_{2},x,\whrho(s)\right\} d \whG_{\tm}(x) \\
	& \qquad \qquad \qquad -
	\int L\left\{ t_{1}, t_{2},x,\rho(s)\right\} d G_{0}(x)\bigg ]\bigg|
	=\op(1),
\end{aligned}
\end{equation}
as $m\rightarrow\infty$.}
\end{lemma}

\begin{lemma}\label{lemma:cnvrg_proportion}
	Under Assumptions~\ref{ass:incr_dom}--\ref{ass:cond_norm}, \ref{ass:pi0}, and \ref{ass:cnvrg_true} of the main paper, for any $t^\prime_1$, $t^\prime_2>0$, we have
	\begin{align}
	\sup _{|t_{1}| \leq t_{1}^{\prime}, |t_{2}| \leq t_{2}^{\prime}}\left|\frac{1}{m_0} \whV_{m}\left(t_{1}, t_{2}\right) - K_{0}\left(t_{1}, t_{2}\right)\right|&=\op(1), \label{eqn:lemS2Eq1}\\
	\sup _{|t_{1}| \leq t_{1}^{\prime}, |t_{2}| \leq t_{2}^{\prime}}\left|\frac{1}{m_1} \whS_{m}\left(t_{1}, t_{2}\right)-K_{1}\left(t_{1}, t_{2}\right)\right|&=\op(1), 
	\label{eqn:lemS2Eq2}\\
 \left|\whF_{m}(\lambda)-F(\lambda)\right|&=\op(1), \label{eqn:lemS2Eq3}
	\end{align}
	as $m=m_0 + m_1 \to \infty$,
	where $\whF_{m}(\lambda)=m^{-1}\sum_{s\in\Scal_{m}} \1f\{\whT_{2}(s) \leq \lambda\}$.
\end{lemma}

Lemma~\ref{lemma:cnvrg_discovery} states that the estimated number of false discoveries in the proposed 2d-SMT procedure converges to the limiting process defined in Assumption~\ref{ass:cnvrg_true} of the main paper. 
Lemma~\ref{lemma:cnvrg_proportion} states the uniform convergence for the processes counting the numbers of false and true rejections and the empirical cumulative distribution function for $\whT_{2}(s)$. 
In the following derivations, we let $C$ be a positive constant which can be different from line to line.

\begin{proof}[Proof of Theorem~\ref{thm:cnvrg_fdr} of the main paper]

We fix $t_1^\prime=t_1^\star$ and $t_2^\prime=t_2^\star$ as defined in Assumption \ref{ass:pre_control} of the main paper. To prove Theorem~\ref{thm:cnvrg_fdr}, we first show
\begin{equation}\label{equ:cnvrg_FDP_t_infty}
\sup_{|t_{1}| \leq t_{1}^{\star}, |t_{2}| \leq t_{2}^{\star}}
\left|
\widehat{\FDP}_{\lambda,\tScal_m}\left(t_{1}, t_{2}\right)-\FDP_{\lambda}^{\infty}\left(t_{1}, t_{2}\right)\right|=\op(1),
\end{equation}
and
\begin{equation}\label{equ:cnvrg_FDP_K}
\sup _{|t_{1}| \leq t_{1}^{\star}, |t_{2}| \leq t_{2}^{\star}}
\left|
\frac{ \whV_{m}\left(t_{1}, t_{2}\right)}{ \whV_{m}\left(t_{1}, t_{2}\right)+ \whS_{m}\left(t_{1}, t_{2}\right)}-\frac{\pi_{0} K_{0}\left(t_{1}, t_{2}\right)}{K\left(t_{1}, t_{2}\right)}\right| =\op(1).
\end{equation}
To show \eqref{equ:cnvrg_FDP_t_infty}, according to Lemma~\ref{lemma:cnvrg_proportion} and Assumption \ref{ass:pi0} of the main paper, we have
$$
    \sup_{|t_{1}| \leq t_{1}^{\star}, |t_{2}| \leq t_{2}^{\star}}\left|\whK_m\left(t_{1}, t_{2}\right)-K\left(t_{1}, t_{2}\right)\right| =\op(1),
$$
where $\whK_{m}(t_{1}, t_{2})=m^{-1}\big\{ \whV_{m}(t_{1}, t_{2})+ \whS_{m} (t_{1}, t_{2})\big\}$. For any $|t_1|\leq t_1^\star$, $|t_2|\leq t_2^\star$ and large enough $m$, we get
    $$
    \left|\whK_{m}\left(t_{1}, t_{2}\right)-K\left(t_{1}, t_{2}\right)\right| \leq \frac{\left|K\left(t_{1}, t_{2}\right)\right|}{2},
    $$
which implies
\begin{equation}\label{equ:low_K}
    \left|\whK_{m}\left(t_{1}, t_{2}\right)\right| \geq \frac{\left|K\left(t_{1}, t_{2}\right)\right|}{2} \geq \frac{K\left(t_{1}^{\star}, t_{2}^{\star}\right)}{2}>0
\end{equation}
because $\inf_{|t_{1}| \leq t_{1}^{\star}, |t_{2}| \leq t_{2}^{\star}}\left|K\left(t_{1}, t_{2}\right)\right| \geq K\left(t_{1}^{\star}, t_{2}^{\star}\right)>0$. For large enough $m$, it follows that
$$\begin{aligned}
& \widehat{\FDP}_{\lambda,\tScal_m}\left(t_{1}, t_{2}\right)-\FDP_{\lambda}^{\infty}\left(t_{1}, t_{2}\right) \\
=&
\frac{m^{-1}K(t_1,t_2)F_{m}(\lambda)\sumsSm \int L\left\{ t_{1}, t_{2}, x, \whrho(s)\right\}  d \whG_{\tm}(x)
-
\whK_{m}(t_1,t_2)F(\lambda)K_0(t_1,t_2)
}
{\Phi(\lambda) \whK_{m}(t_1,t_2) K(t_1,t_2)}
\\
\leq &
\frac{2m^{-1}K(t_1,t_2)F_{m}(\lambda)\sumsSm\int L\left\{ t_{1}, t_{2}, x, \whrho(s)\right\}  d \whG_{\tm}(x)
-
\whK_{m}(t_1,t_2)F(\lambda)K_0(t_1,t_2)
}
{\Phi(\lambda) K^2(t^\star_1,t^\star_2)},
\end{aligned}
$$
where the last inequality holds by \eqref{equ:low_K}. Thus \eqref{equ:cnvrg_FDP_t_infty} follows from Lemma~\ref{lemma:cnvrg_discovery} and Lemma~\ref{lemma:cnvrg_proportion}. Similarly, we can prove \eqref{equ:cnvrg_FDP_K}.

Next we use \eqref{equ:cnvrg_FDP_t_infty} and \eqref{equ:cnvrg_FDP_K} to show $\limsup_{m\rightarrow\infty}\widetilde{\FDR}_{m}\leq q$. Due to Assumption~\ref{ass:pre_control} of the main paper, we have $\FDP_{\lambda}^{\infty}\left(t_{1}^{\star}, 0\right)<q$ and $\FDP_{\lambda}^{\infty}\left(0, t_{2}^{\star}\right)<q$. Then, for large enough $m$, we have 
$$
    \widehat{\FDP}_{\lambda,\tScal_m}\left(t^\star_{1}, 0\right)<q
    ~\mbox{ and }~
    \widehat{\FDP}_{\lambda,\tScal_m}\left(0, t^\star_{2}\right)<q,
$$
which implies that $(\wtt_1^\star,\wtt_2^\star)$ satisfies $|\wtt_1^\star|\leq t^\star_1$ and $|\wtt_2^\star|\leq t^\star_2$. Thus, we get
\begin{equation}\label{equ:mono}
\begin{aligned}
    & \widehat{\FDP}_{\lambda,\tScal_m}\left(\wtt_{1}^{\star}, \wtt_{2}^{\star}\right)
    -
    \frac{ \whV_{m}\left(\wtt_{1}^{\star}, \wtt_{2}^{\star}\right)}{ \whV_{m}\left(\wtt_{1}^{\star}, \wtt_{2}^{\star}\right)+ \whS_{m}\left(\wtt_{1}^{\star}, \wtt_{2}^{\star}\right)} \\
    & \qquad \geq \inf _{|t_{1}| \leq t_{1}^{\star}, |t_{2}| \leq t_{2}^{\star}}
    \left\{\widehat{\FDP}_{\lambda,\tScal_m}\left(t_{1}, t_{2}\right)-\frac{ \whV_{m}\left(t_{1}, t_{2}\right)}{ \whV_{m}\left(t_{1}, t_{2}\right)+ \whS_{m}\left(t_{1}, t_{2}\right)}\right\}.
\end{aligned}
\end{equation}
Rearranging the second formula, we obtain
\begin{align*}
& \inf _{|t_{1}| \leq t_{1}^{\star}, |t_{2}| \leq t_{2}^{\star}}
    \left\{\widehat{\FDP}_{\lambda,\tScal_m}\left(t_{1}, t_{2}\right)-\frac{ \whV_{m}\left(t_{1}, t_{2}\right)}{ \whV_{m}\left(t_{1}, t_{2}\right)+ \whS_{m}\left(t_{1}, t_{2}\right)}\right\}\\
    &\qquad =\inf _{|t_{1}| \leq t_{1}^{\star}, |t_{2}| 
    \leq t_{2}^{\star}}
    \left\{\widehat{\FDP}_{\lambda,\tScal_m}\left(t_{1}, t_{2}\right)
    -
    \FDP_{\lambda}^{\infty}\left(t_{1}, t_{2}\right)+\frac{\pi_{0} K_{0}\left(t_{1}, t_{2}\right)}{K\left(t_{1}, t_{2}\right)} \right.\\
    & \qquad \qquad \qquad \qquad  \quad \left. -
    \frac{ \whV_{m}\left(t_{1}, t_{2}\right)}{ \whV_{m}\left(t_{1}, t_{2}\right)+ \whS_{m}\left(t_{1}, t_{2}\right)}
    +
    \FDP_{\lambda}^{\infty}\left(t_{1}, t_{2}\right)-\frac{\pi_{0} K_{0}\left(t_{1}, t_{2}\right)}{K\left(t_{1}, t_{2}\right)}\right\},
\end{align*}
which converges in probability to 
$$
\inf _{|t_{1}| \leq t_{1}^{\star}, |t_{2}| \leq t_{2}^{\star}}
    \left\{\FDP_{\lambda}^{\infty}\left(t_{1}, t_{2}\right)-\frac{\pi_{0} K_{0}\left(t_{1}, t_{2}\right)}{K\left(t_{1}, t_{2}\right)}\right\}
$$
according to \eqref{equ:cnvrg_FDP_t_infty} and \eqref{equ:cnvrg_FDP_K}. As
\begin{align*}
    -1
    \leq 
    \inf _{|t_{1}| \leq t_{1}^{\star}, |t_{2}| \leq t_{2}^{\star}}
    \left\{\widehat{\FDP}_{\lambda,\tScal_m}\left(t_{1}, t_{2}\right)-\frac{ \whV_{m}\left(t_{1}, t_{2}\right)}{ \whV_{m}\left(t_{1}, t_{2}\right)+ \whS_{m}\left(t_{1}, t_{2}\right)}\right\}
    \leq 
    \widehat{\FDP}_{\lambda,\tScal_m}\left(\wtt_{1}^{\star}, \wtt_{2}^{\star}\right)
    \leq q,
\end{align*}
Lebesgues's dominated convergence theorem implies
\begin{align*}
    & \liminf_{m\rightarrow\infty}
    \Eb\left[
    \inf _{|t_{1}| \leq t_{1}^{\star}, |t_{2}| \leq t_{2}^{\star}}
    \left\{\widehat{\FDP}_{\lambda,\tScal_m}\left(t_{1}, t_{2}\right)-\frac{ \whV_{m}\left(t_{1}, t_{2}\right)}{ \whV_{m}\left(t_{1}, t_{2}\right)+ \whS_{m}\left(t_{1}, t_{2}\right)}\right\}
    \right]\\
    = &  \inf _{|t_{1}| \leq t_{1}^{\star}, |t_{2}| \leq t_{2}^{\star}}
    \left\{\FDP_{\lambda}^{\infty}\left(t_{1}, t_{2}\right)-\frac{\pi_{0} K_{0}\left(t_{1}, t_{2}\right)}{K\left(t_{1}, t_{2}\right)}\right\}
    \geq0,
\end{align*}
where the last inequality stands due to the fact that $F(\lambda)/\Phi(\lambda)\geq \pi_0$ and $K_{0}\left(t_{1}, t_{2}\right)
      \leq\lim_{m\rightarrow\infty} \sum_{s\in\Scal_{m}}\int L\{t_1,t_2,x,\rho(s)\} \, d G_0(x)/m$ implied by \eqref{equ:def_F_FDP_inf} in Assumption \ref{ass:cnvrg_true} of the main paper. It then yields that
\begin{align*}
    & \liminf_{m\rightarrow\infty}\Eb\left[
    \widehat{\FDP}_{\lambda,\tScal_m}\left(\wtt_{1}^{\star}, \wtt_{2}^{\star}\right)
    -
    \frac{ \whV_{m}\left(\wtt_{1}^{\star}, \wtt_{2}^{\star}\right)}{ \whV_{m}\left(\wtt_{1}^{\star}, \wtt_{2}^{\star}\right)+ \whS_{m}\left(\wtt_{1}^{\star}, \wtt_{2}^{\star}\right)}
    \right]\\
    & \qquad \geq \liminf_{m\rightarrow\infty}
    \Eb\left[
    \inf _{|t_{1}| \leq t_{1}^{\star}, |t_{2}| \leq t_{2}^{\star}}
    \left\{\widehat{\FDP}_{\lambda,\tScal_m}\left(t_{1}, t_{2}\right)-\frac{ \whV_{m}\left(t_{1}, t_{2}\right)}{ \whV_{m}\left(t_{1}, t_{2}\right)+ \whS_{m}\left(t_{1}, t_{2}\right)}\right\}
    \right]\\
    & \qquad =  \inf _{|t_{1}| \leq t_{1}^{\star}, |t_{2}| \leq t_{2}^{\star}}
    \left\{\FDP_{\lambda}^{\infty}\left(t_{1}, t_{2}\right)-\frac{\pi_{0} K_{0}\left(t_{1}, t_{2}\right)}{K\left(t_{1}, t_{2}\right)}\right\}
    \geq0,
\end{align*}
where the first inequality holds because of \eqref{equ:mono} and the monotonicity of expectation. Finally, 
we obtain
$$
    \limsup_{m\rightarrow\infty}
    \widetilde{\FDR}_{m}
    \leq
    \liminf_{m\rightarrow\infty}
    \Eb\left\{
    \widehat{\FDP}_{\lambda,\tScal_m}\left(\wtt_{1}^{\star}, \wtt_{2}^{\star}\right)
    \right\}
    \leq 
    q,
$$
which completes the proof. 
%
\end{proof}

\section{Proofs of Lemmas S.1 and S.2}\label{sec:prf_lemma_12}

In this section, we prove Lemmas~\ref{lemma:cnvrg_discovery} and \ref{lemma:cnvrg_proportion} with the help of some preliminary lemmas. 
In particular, Section~\ref{sec:conn_NPEB} presents the logic flow why the proposed estimator $\whG_{\tm}$ is able to estimate the limiting distribution of the signal process $\xi(s)$ by introducing multiple intermediate variants.
Section~\ref{sec:intro_lemmas} states some preliminary lemmas for proving Lemma~\ref{lemma:cnvrg_discovery}. In Section \ref{sec:prf_1_2}, we complete the proofs of Lemmas~\ref{lemma:cnvrg_discovery} and \ref{lemma:cnvrg_proportion}.

\subsection{Notation and Proof Sketch of Lemma \ref{lemma:cnvrg_discovery}}
\label{sec:conn_NPEB}
To better present the proof of Lemma \ref{lemma:cnvrg_discovery}, this section briefly introduces some notation and intermediate quantities to connect the proposed general maximum likelihood estimator (GMLE) with the limiting distribution $G_0$.
To begin with, we present a fact of GMLE that is useful in the following proof. There exists a discrete solution to \eqref{equ:GMLE_hat} of the main paper with no more than $|\widetilde{\mathcal{S}}|+1$ support points by the Carath{\'e}odory's theorem. 
Specifically, we can write the solution as
\begin{equation}\label{eqn:gTildeDef}
	\widetilde{G}_{\widetilde{\mathcal{S}}}(u) = \sum_{i=1}^{\tilde{l}} \widetilde{\pi}_i\boldsymbol{1}\left\{\widetilde{v}_i\leq u\right\},\quad \sum_{i=1}^{\tilde{l}} \widetilde{\pi}_i=1,
	\quad \text{and} \quad \widetilde{\pi}_i > 0,
\end{equation}
where $\left\{\widetilde{v}_i\right\}_{i=1}^{\tilde{l}}$ is the set of support points and $\tilde{l}\leq |\widetilde{\mathcal{S}}|+1$. 
The support of $\widetilde{G}_{\widetilde{\mathcal{S}}}$ is within the range of $\{T_1(s):s\in\widetilde{\mathcal{S}}\}$ due to the monotonicity of $\phi(x-u)$ in $|x-u|$, where $\phi(x)$ is the density of standard normal distribution. 

\begin{figure}
\resizebox{\textwidth}{!}{
\begin{tikzpicture}[node distance=2cm]
\tikzstyle{every node}=[font=\footnotesize]
\node (T1) [Pf_item] {$T_1(s)$};
\node (whT1) [Pf_item,left of = T1, xshift=-4cm] {$\whT_1(s)$};
\node (T1*r) [Pf_item,right of = T1, xshift=4cm] {$T^*_1(s;r)$};
\node (T1r) [Pf_item,below of = T1] {$T_1(s;r)$};
\node (wtGtmr) [Pf_item,below of = T1r] {$\wtG_{\tm,r}$};
\node (whGtm) [Pf_item,left of = wtGtmr, xshift=-4cm] {$\whG_{\tm}$};
\node (wtG*tmr) [Pf_item,right of = wtGtmr, xshift=4cm] {$\wtG^*_{\tm,r}$};
\node (Gtmr) [Pf_item,below of = wtG*tmr] {$G_{\tm,r}$};
\node (Gtm) [Pf_item,left of = Gtmr, xshift=-4cm] {$G_{\tm}$};
\node (G0) [Pf_item,below of = Gtm] {$G_{0}$};
\node (limit) [Pf_itemlong,below of = G0, xshift=4cm] {{ $\sum_{s\in\Scal_{m}} \int L\left\{ t_{1}, t_{2},x,\rho(s)\right\} d G_{0}(x)/m$}};
\node (est) [Pf_itemlong,left of = limit, xshift=-8cm] {$\sum_{s\in\Scal_m} \int L\left\{ t_{1}, t_{2},x,\whrho(s)\right\} d \whG_{\tm}(x)/m$};
\node (upwhT1) [above of=whT1, yshift=-1cm,xshift=-0.5cm] {};
\node (upT1*r) [above of=T1*r, yshift=-1cm,xshift=0.5cm] {};
\node (leftwhT1) [left of=whT1, xshift=0.5cm] {};
\node (leftdownwhT1) [below of=leftwhT1,yshift=-7cm] {};
\node (rightT1*r) [right of=T1*r, xshift=-0.5cm] {};
\node (rightdownT1*r) [below of=rightT1*r,yshift=-7cm] {};
\draw [Pf_arrow] (T1) -- node[anchor=south]  {(\rNum{1}) var. est. }(whT1);
\draw [Pf_arrow] (T1) -- node[anchor=south]  {(\rNum{2}) norm. ind. app.}(T1*r);
\draw [Pf_arrow] (T1) -- node[anchor=east] {(\rNum{3}) norm.}(T1r);
\draw [Pf_arrow] (T1r) --node[anchor=west, xshift = -15pt, yshift = -18pt,rotate=22] {(\rNum{3}) ind. app.} (T1*r);
\draw [Pf_arrow] (T1r) -- (T1*r);
\draw [Pf_arrow] (whGtm) --node[anchor=south] {(\rNum{4}) var. consis.} (wtGtmr);
\draw [Pf_arrow] (wtGtmr) -- node[anchor=south] {(\rNum{4}) ind. app.}(wtG*tmr);
\draw [Pf_arrowbig] (est) --node[anchor=south] {Lemma~\ref{lemma:cnvrg_discovery}} (limit);
\draw [Pf_arrowdash] (whT1) -- (whGtm);
\draw [Pf_arrowdash] (T1r) -- (wtGtmr);
\draw [Pf_arrowdash] (T1*r) -- (wtG*tmr);
\draw [Pf_arrow] (wtG*tmr) -- node[anchor=east] {(\rNum{5}) GMLE}node[anchor=west,yshift=0.05cm] {consis.}(Gtmr);
\draw [Pf_arrow] (Gtmr) --node[anchor=south]  {(\rNum{6}) norm. consis.} (Gtm);
\draw [Pf_arrow] (Gtm) --node[anchor=east,yshift=6pt]  {(\rNum{6}) large} node[anchor=east,yshift = -6pt]  {sample} node[anchor=west]  {consis.}  (G0);
\draw [Pf_arrowdash] (whGtm) -- (est);
\draw [Pf_arrowdash] (G0) -| (limit);
\draw [Pf_arrowsum] ( upwhT1) --node[anchor=south] {Variance consistency + Independent approximation} (upT1*r);
\draw [Pf_arrowsum] (leftwhT1) --node[anchor=west,rotate=90,xshift=-40pt,yshift=8pt] {Practical approach} (leftdownwhT1);
\draw [Pf_arrowsum] (rightT1*r) --node[anchor=west,rotate=-90,xshift=-55pt,yshift=8pt] {Theoretical analysis} (rightdownT1*r);
\end{tikzpicture}
}
\caption{
The diagram of proving Lemma \ref{lemma:cnvrg_discovery} using the intermediate variants. The overall strategies are: (\rNum{1}) replacing the unknown variance with the true variance $\tau(s)$; (\rNum{2}) replacing the correlated auxiliary statistics with its normalized independent approximation; (\rNum{3}) analyzing the normalized independent approximation by its normalization term and independent approximation term; (\rNum{4}) showing GMLE estimated from different sources are close; (\rNum{5}) showing GMLE $\wtG^*_{\tm,r}$ estimated from $T_1^*(s;r)$ converges to $G_{\tm,r}$; 
and (\rNum{6}) proving that $G_{\tm,r}$ approaches $G_0$ by showing $\zeta_r(s)\rightarrow 1$ and the large sample consistency.
Lemma~\ref{lemma:cnvrg_discovery} establishes the convergence using the above strategies. 
Here \textit{var.}, \textit{est.}, \textit{norm.},
\textit{ind.}, \textit{app.}, and \textit{consis.}
are abbreviations for ``variance'', ``estimation'', ``normalization'', ``independently'', ``approximation'' and ``consistency'' respectively.}
\label{fig:big_pic_EB}
\end{figure}
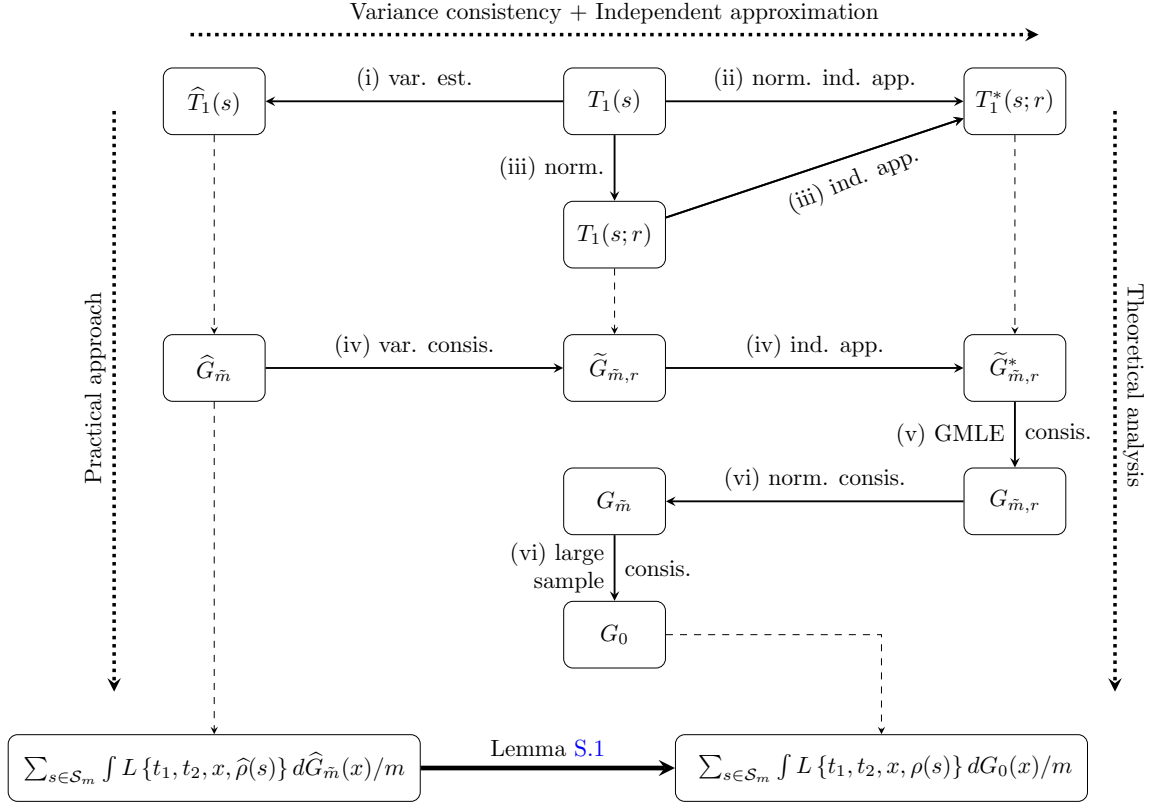

Now recall from \eqref{eqn:tStarDef} of the main paper that
\begin{equation*}
    T_1^*(s;r)=\Eb\left[T_1(s)\mid \mathfrak{F}\left\{\cup_{v\in \Ncal(s)}B(v;r)\right\}\right]/\zeta_r(s),
\end{equation*}
where $\zeta_r^2(s)=\var\left(\Eb\left[T_1(s)\mid \mathfrak{F}\left\{\cup_{v\in \Ncal(s)}B(v;r)\right\}\right]\right)$.
We call $\zeta_r(s)$ the normalization term since it ensures $\var\{T_1^*(s;r)\}=1$.
It can be directly shown that under Assumption~\ref{ass:cond_norm} of the main paper, $\{T^*_1(s;r):s\in\tScal_m\}$ are independent and normally distributed random variables with unit variance, provided that \eqref{equ:req_ind} of the main paper is satisfied for the subset $\tScal_m$. 
This result fulfills the commonly-used independence assumption in the theory of nonparametric empirical Bayes (NPEB); see \cite{Zhang2009,Jiang2009} for details.
We will prove that $T^*_1(s;r)$ and $\whT_{1}(s)$ used in our implementation are close by using the near epoch dependency (NED, Assumption \ref{ass:NED} of the main paper), the consistency of variance (i.e., showing $\whtau(s)\rightarrow \tau(s)$), and the convergence of the normalization term (i.e., showing $\zeta_r(s)\rightarrow 1$). 
More precisely, we will show that $\whT_{1}(s)$ and $T^*_1(s;r)$ are close to each other through the following approximations:
$$
\whT_{1}(s)\approx T_1(s) \approx T_1(s;r) \approx T_1^*(s;r),
$$
where 
$T_1(s)$ is defined in \eqref{eq:defT1} of the main paper (the auxiliary statistics with true variance); 
and $T_1(s;r)=T_1(s)/\zeta_r(s)$ is an intermediate variant involving the normalization term $\zeta_r(s)$ defined above.
The difference between $T_1(s)$ and $\whT_1(s)$ lies oin whether the normalization term is the true standard deviation of $\sum_{v\in\mathcal{N}(s)}X(v)$ or its estimate. As for $T_1(s)$ and $T_1(s;r)$, they again differ by the normalization terms ($\tau(s)$ versus $\zeta_r(s)$). The difference between $T_1(s;r)$ and its conditional version $T_1^*(s;r)$ is controlled by NED. 
The following intermediate quantities are the corresponding GMLEs based on $\whT_1(s)$, $T_1(s;r)$, and $T_1^*(s;r)$:
\begin{itemize}
    \item The GMLE based on $\{\whT_1(s),s\in\tScal_m\}$ is denoted as $$\whG_{\tm}(u) = \sum_{i=1}^{\whl} \whpi_i \boldsymbol{1}\left\{\whv_i\leq u\right\},$$ 
    where $|\whv_i|\leq \sup_{s\in\tScal_m}|\whT_1(s)|$, $\whl\leq\tm+1$, $\sum_{i=1}^{\whl} \whpi_i=1$, and $\whpi_i > 0$;
    \item The GMLE based on $\{T_1(s;r):s\in\tScal_m\}$ is denoted as $$\wtG_{\tm,r}(u) = \sum_{i=1}^{\wtl} \wtpi_i \boldsymbol{1}\left\{\wtv_i\leq u\right\},$$ 
    where $|\wtv_i|\leq \sup_{s\in\tScal_m}|T_1(s;r)|$, $\wtl\leq\tm+1$, $\sum_{i=1}^{\wtl} \wtpi_i=1$, and $\wtpi_i > 0$;
    \item The GMLE based on $\{T^*_1(s;r):s\in\tScal_m\}$ is denoted as $$\wtG_{\tm,r}^*(u) = \sum_{i=1}^{\wtl^*} \wtpi^*_i \boldsymbol{1}\left\{\wtv^*_i\leq u\right\},$$ 
    where $|\wtv^*_i|\leq \sup_{s\in\tScal_m}|T^*_1(s;r)|$, $\wtl^*\leq\tm+1$, $\sum_{i=1}^{\wtl^*} \wtpi^*_i=1$, and $\wtpi^*_i > 0$.
\end{itemize}
Here we suppress the dependence of $ \wtpi_i$, $\wtv_i$, $\wtpi^*_i$, and $\wtv^*_i$ on $r$ to simplify the presentation. 
We now describe the key idea and steps for proving Lemma~\ref{lemma:cnvrg_discovery} 
as visualized in Figure~\ref{fig:big_pic_EB}. The key idea is to prove that $\whG_{\tm}$ converges to $G_0$ in terms of the Hellinger distance when the subset $\tScal_m$ satisfies \eqref{equ:req_ind} of the main paper.
The details will be shown in Lemma~\ref{lemma:cnvrg_emp_bayes_cp}. To arrive at this result, we need the following steps.  
On the one hand, (iv) and (v) in Figure~\ref{fig:big_pic_EB} state that $\whG_{\tm}$ approximates
\begin{equation}\label{eqn:gTmRDef}
    G_{\tm,r}(u)=\frac{1}{\tm}\sumstSm\boldsymbol{1}\{\xi(s)/\zeta_r(s)\leq u\},
\end{equation}
with high accuracy (see Lemmas~\ref{lemma:cnvrg_GMLE} and \ref{lemma:cnvrg_emp_bayes}).
On the other hand, 
(\rNum{6}) in Figure~\ref{fig:big_pic_EB} states that $G_{\tm,r}(u)$ approaches
$$G_{\tm}(u)=\frac{1}{\tm}\sumstSm\boldsymbol{1}\{\xi(s)\leq u\}$$
by showing that $\zeta_r(s)$ tends to $1$ as $r\rightarrow\infty$ (see Lemma \ref{lemma:cnvrg_tau}). 
Together with Assumption~\ref{ass:emp_bayes} of the main paper, $G_{\tm,r}(u)$ converges to $G_{0}(u)$ as $\tm\rightarrow\infty$, 
which will be shown in Lemma \ref{lemma:f_wtG_cnvrg}. 
The detailed statements of Lemmas~\ref{lemma:cnvrg_tau}--\ref{lemma:cnvrg_emp_bayes_cp} are presented in the next subsection.

\subsection{Some Preliminary Lemmas}\label{sec:intro_lemmas}

Following the proof sketch in Section \ref{sec:conn_NPEB}, we now introduce some preliminary lemmas.
These lemmas are in accordance with the strategies depicted in Figure \ref{fig:big_pic_EB}.
In particular, Lemmas~\ref{lemma:cnvrg_tau} to \ref{lemma:cnvrg_GMLE} focus on a series of properties derived from NED and the consistency of variance. Lemma~\ref{lemma:cnvrg_emp_bayes} provides a large deviation inequality that gives the convergence rate of $\whG_{\tm}$ to $G_{\tm,r}$ as defined in \eqref{eqn:gTmRDef}. 
Lemmas~\ref{lemma:f_wtG_cnvrg} and \ref{lemma:cnvrg_emp_bayes_cp} show that $d_H(f_{\whG_{\tm}},f_{G_0})=\op(1)$, which is an important step in the proof of Lemma~\ref{lemma:cnvrg_discovery}. {Lemma~\ref{lemma:cnvrg_true} establishes the law of large numbers for the number of false/true discoveries.} The relations among Theorem~\ref{thm:cnvrg_fdr} of the main paper and Lemmas~\ref{lemma:cnvrg_discovery}--\ref{lemma:cnvrg_true} are depicted in Figure~\ref{fig:lemma_rela}. 

\begin{figure}[H]
\begin{tikzpicture}[node distance=3cm]
\tikzstyle{every node}=[font=\footnotesize]
\node (l_3) [Rl_item] {Lemma~\ref{lemma:cnvrg_tau}};
\node (l_4) [Rl_item, right of =l_3, xshift=1.5cm] {Lemma~\ref{lemma:cnvrg_max_sum}};
\node (l_5) [Rl_item, below of=l_4,yshift=1.5cm] {Lemma~\ref{lemma:cnvrg_GMLE}};
\node (l_8) [Rl_item, below of=l_5,yshift=1.5cm] {Lemma~\ref{lemma:cnvrg_emp_bayes_cp}};
\node (l_6) [Rl_item, left of=l_8, xshift=-1.5cm] {Lemma~\ref{lemma:f_wtG_cnvrg}};
\node (l_1) [Rl_item, right of=l_8, xshift=1.5cm] {Lemma~\ref{lemma:cnvrg_discovery}};
\node (l_7) [Rl_item, below of=l_8,yshift=1.5cm] {Lemma~\ref{lemma:cnvrg_emp_bayes}};
\node (l_9) [Rl_item, above of=l_4,yshift=-1.5cm] {Lemma~\ref{lemma:cnvrg_true}};
\node (l_2) [Rl_item, right of=l_4,xshift=1.5cm] {Lemma~\ref{lemma:cnvrg_proportion}};
\node (t_1) [Rl_item, right of=l_5, xshift=4cm] {Theorem~\ref{thm:cnvrg_fdr}};
\draw [Pf_arrow] (l_3) -- (l_9);
\draw [Pf_arrow] (l_9) -- (l_2);
\draw [Pf_arrow] (l_3) -- (l_4);
\draw [Pf_arrow] (l_3) -- (l_5);
\draw [Pf_arrow] (l_3) -- (l_6);
\draw [Pf_arrow] (l_4) -- (l_5);
\draw [Pf_arrow] (l_5) -- (l_8);
\draw [Pf_arrow] (l_6) -- (l_8);
\draw [Pf_arrow] (l_7) -- (l_8);
\draw [Pf_arrow] (l_8) -- (l_1);
\draw [Pf_arrow] (l_1) -- (t_1);
\draw [Pf_arrow] (l_2) -- (t_1);
\end{tikzpicture}
\caption{The relations among Lemmas~\ref{lemma:cnvrg_discovery}--\ref{lemma:cnvrg_emp_bayes_cp} and Theorem~\ref{thm:cnvrg_fdr}.}
\label{fig:lemma_rela}
\end{figure}

Now, we start to present Lemmas ~\ref{lemma:cnvrg_tau}--\ref{lemma:cnvrg_true}.
\begin{lemma}[Convergence of $\whtau^2(s)$]\label{lemma:cnvrg_tau}
Under Assumptions~\ref{ass:incr_dom}--\ref{ass:cond_norm} of the main paper, we have\\
\sublemma\label{lemma:cnvrg_tau:a} $\left|\zeta_r^2(s)-1\right|\leq C\psi^2(r)$ and $\left|1/\zeta_r^2(s)-1\right|\leq C\psi^2(r)$ uniformly for large enough $r$.\\
\sublemma\label{lemma:cnvrg_tau:b} $\supstSm\left|\whtau^2(s)/\tausq_r(s)-1\right|\leq C\psi^2(r) + C\tm^{-q}$ uniformly for large enough $r$ with probability tending to one as $\tm \to \infty$.\\
\sublemma\label{lemma:cnvrg_tau:c} $\sup_m\sup_{s\in\Scal_m} \eta(s;r)$ is uniformly bounded for large enough $r$.
\end{lemma}

\begin{remark}
\rm
According to the definition of NED (see Definition \ref{def:near_epoch} of the main paper), Lemma \ref{lemma:cnvrg_tau:a} implies that $\zeta_r(s) \to 1$ as $r \to \infty$.
It further implies
\begin{equation}\label{equ:T1nu1}
    \sup_{s\in\Scal_m}\Eb\left\{T^*_{1}(s;r)\right\}^2
    =1+ \sup_{s\in\Scal_m}\big[\Eb\{ T^*_{1}(s;r)\}\big]^2
\leq 1+\sup_{s\in\Scal_m}\nu_0^2/\zeta^2_r(s)\leq 1+2\nu_0^2
\defequ \nu_1^2
\end{equation}
due to Assumption~\ref{ass:emp_bayes} of the main paper and the unit variance of $ T^*_{1}(s;r)$ as defined in \eqref{eqn:tStarDef} of the main paper.

As for Lemma~\ref{lemma:cnvrg_tau:b}, we note that
\begin{equation}\label{equ:fractau}
\begin{aligned}
    \frac{\supstSm\left|\whtau^2(s)/\tausq_r(s)-1\right|}{1-\supstSm\left|\whtau^2(s)/\tausq_r(s)-1\right|}
    &\leq \{C\psi^2(r)+C\tm^{-q}\}/[1-\{C\psi^2(r)+C\tm^{-q}\}] \\
    &\leq C\psi^2(r)+C\tm^{-q}
\end{aligned}
\end{equation}
with high probability when $r$ and $\tm$ are large. 
Both \eqref{equ:fractau} and \eqref{equ:T1nu1} will be useful in our theoretical analysis. 
\end{remark}

\begin{remark}\rm\label{rmk:T2}
A similar conclusion holds for primary statistics $T_2(s)$. Specifically, define
$$
    T_2^*(s;r)=\Eb\left[T_2(s)\mid \mathfrak{F}\left\{\cup_{v\in \Ncal(s)}B(v;r)\right\}\right]/\breve{\zeta}_r(s),
$$
where $\breve{\zeta}_r^2(s)=\var\left(\Eb\left[T_2(s)\mid \mathfrak{F}\left\{\cup_{v\in \Ncal(s)}B(v;r)\right\}\right]\right)$. We have $\left|\breve{\zeta}_r^2(s)-1\right|\leq C\psi^2(r)$ and $\left|1/\breve{\zeta}_r^2(s)-1\right|\leq C\psi^2(r)$ uniformly and $\sup_m\sup_{s\in\Scal_m} \breve{\eta}(s;r)$ is uniformly bounded for large enough $r$.
\end{remark}

\begin{lemma}\label{lemma:cnvrg_max_sum}
Under Assumptions~\ref{ass:incr_dom}--\ref{ass:cond_norm} and \ref{ass:emp_bayes} of the main paper, for $\tm$ large enough and any $\delta_1,\delta_2>0$, with probability at least
\begin{equation}\label{equ:prob_T}
    1- \tm\exp(-\nu_1^2\delta_1^2/2) - C\tm\psi^p(r)/\nu_1^p\delta_1^p - C\psi^p(r)/\delta_2^p,
\end{equation}
the following events occur simultaneously. 
\\ \sublemma 
$$
    \supstSm\left|T^*_{1}(s;r)\right| 
    \leq \nu_1(1+\delta_1)
	\quad \mbox{and} \quad \supstSm\left|T_{1}(s;r)\right| 
	\leq \nu_1(1+2\delta_1).
$$
\sublemma 
$$
    \tm^{-1}\sumstSm\left|T_{1}(s;r)-T_1^*(s;r)\right| 
    \leq \delta_2
    \quad \mbox{and} \quad \tm^{-1}\left|\sumstSm T_{1}^2(s;r)-\left\{T_1^*(s;r)\right\}^2\right| 
    \leq \nu_1\delta_2(2+3\delta_1).
$$
\sublemma
$$
    \tm^{-1}\sumstSm \left|T_{1}(s;r) \right|
    \leq \nu_1\left(1+\delta_1\right)+\delta_2,
$$
and
$$
    \tm^{-1}\sumstSm T_{1}^2(s;r) 
    \leq \nu_1^2 + \nu_1\delta_2(2+3\delta_1) + \nu_1^2\delta_1\left(2+\delta_1\right).
$$
\end{lemma}
\begin{remark}
    The three tail probabilities in \eqref{equ:prob_T} respectively correspond to the events of 
    $\supstSm |T^*_1(s;r)-\Eb \{T^*_1(s;r)\}|\leq \nu_1\delta_1$, $\supstSm |T_1^*(s;r)-T_1(s;r)|\leq \nu_1\delta_1$, and \\ $\tm^{-1}\sumstSm \vert T_1(s;r)-T_1^*(s;r)\vert \leq \delta_2$. 
    In the proof, we will show that these events imply the upper bounds in (a)--(c) of Lemma \ref{lemma:cnvrg_max_sum}.
\end{remark}

\begin{lemma}\label{lemma:cnvrg_GMLE}
Under Assumptions~\ref{ass:incr_dom}--\ref{ass:cond_norm} and \ref{ass:emp_bayes} of the main paper, with probability at least \eqref{equ:prob_T}, the difference of generalized log-likelihood
\begin{equation}\label{equ:summation}
	\frac{1}{\tm}\left|\sumstSm\log\left[
	\frac{f_{\wtG^*_{\tm,r}}\{T_{1}(s;r)\}}{f_{\wtG^*_{\tm,r}}\{T^*_{1}(s;r)\}}
	\frac{f_{\wtG_{\tm,r}}\{\whT_{1}(s)\}}{f_{\wtG_{\tm,r}}\{T_{1}(s;r)\}}
	\frac{f_{\whG_{\tm}}\{T^*_{1}(s;r)\}}{f_{\whG_{\tm}}\{T_{1}(s;r)\}}
	\frac{f_{\whG_{\tm}}\{T_{1}(s;r)\}}{f_{\whG_{\tm}}\{\whT_{1}(s)\}}
	\right] \right|
\end{equation}
is upper bounded by 
$$
    C\nu^2_1\left\{\psi^2(r) + \tm^{-q}\right\}
	(1+\delta^2_1)+
	C\nu_1\delta_2\left(1+\delta_1\right).
$$
\end{lemma}

\begin{lemma}\label{lemma:cnvrg_emp_bayes}
	Under Assumptions~\ref{ass:neigh}, \ref{ass:cond_norm}, and \ref{ass:emp_bayes} of the main paper, conditioning on $\Eb|T_1^*(s;r)|\leq\nu_1$, if $\whG_{\tm}$ satisfies
	\begin{equation}\label{equ:cnvrg_whG_GtS}
	\prod_{s\in\tScal_m}\left[\frac{f_{\whG_{\tm}}\left\{T_{1}^*(s;r)\right\}}{f_{G_{\tm,r}}\left\{T^*_{1}(s;r)\right\}}\right] \geq e^{-2 t^{2} \tm c_{\tm}^{2} / 15},
	\end{equation}
	where
	\begin{equation}\label{equ:c_tm}
	c_{\tm}=\sqrt{\frac{\log (\tm)}{\tm}}\left\{\tm^{1 / b} \sqrt{\log \tm}\left(1\vee \nu_1\right)\right\}^{b /(2+2 b)}
	\end{equation}
	for some $b>0$, then there exists a universal constant $t^{*}$ such that for all $t \geq t^{*}$ and $\log \tm \geq 2 / b$,
	$$
	\Pb\left\{d_{H}\left(f_{\whG_{\tm}}, f_{G_{\tm,r}}\right) \geq t c_{\tm} \right\}
	\leq \exp \left\{-\frac{t^{2} \tm c_{\tm}^{2}}{2 \log \tm}\right\} 
	\leq e^{-t^{2} \log \tm}.
	$$
\end{lemma}

Lemma~\ref{lemma:cnvrg_emp_bayes} is a direct consequence of Theorem~1 in \cite{Zhang2009}. It states that $d_H(f_{\whG_{\tm}}, f_{G_{\tm,r}})$ would be small enough, as long as the generalized likelihood is nearly maximized in the sense of \eqref{equ:cnvrg_whG_GtS}. This result together with Lemma \ref{lemma:cnvrg_GMLE} allows us to replace $T_1^*(s;r)$ with $\whT_1(s)$ in estimating $G_{\tm,r}$. 

\begin{lemma}\label{lemma:f_wtG_cnvrg}
Under Assumptions~\ref{ass:incr_dom}--\ref{ass:cond_norm}, and \ref{ass:emp_bayes} of the main paper, for $\tm\rightarrow\infty$ and arbitrary $\delta_3>0$, 
we have
$$
    d_H(f_{G_{\tm,r}}, f_{G_0})\leq C\nu_1^{1/2}\psi(r) + \delta_3,
$$
with probability tending to one.
\end{lemma}

\begin{lemma}\label{lemma:cnvrg_emp_bayes_cp}
Under Assumptions~\ref{ass:incr_dom}--\ref{ass:emp_bayes} of the main paper, the GMLE based on $\{\whT_1(s),s\in\tScal_m\}$ satisfies $d_{H}\left(f_{\whG_{\tm}}, f_{G_{0}}\right)=\op(1)$ as $m\rightarrow\infty$. 
\end{lemma}

Lemma~\ref{lemma:cnvrg_emp_bayes_cp} is a consequence of Lemmas \ref{lemma:cnvrg_GMLE}--\ref{lemma:f_wtG_cnvrg}. Its proof uses the convergence result of $d_H(f_{G_{\tm,r}},f_{G_0})$ in Lemma~\ref{lemma:f_wtG_cnvrg}, and shows the convergence rate in Lemma~\ref{lemma:cnvrg_GMLE} is fast enough to ensure that $\whG_{\tm}$ is an approximate GMLE of $G_{\tm,r}$ and the condition in Lemma~\ref{lemma:cnvrg_emp_bayes} is fulfilled.

\begin{lemma}\label{lemma:cnvrg_true}
Under Assumptions~\ref{ass:incr_dom}--\ref{ass:cond_norm}, and \ref{ass:cnvrg_true} of the main paper, we have 
$$\begin{aligned}
	&\frac{\sum_{s\in\Scal_{0,m}} \1f\left\{T_{1}(s)\geq t_{1},T_{2}(s) \geq t_{2}\right\}}{m_0}:=\frac{V_{m}(t_1,t_2)}{m_0} \stackrel{p}{\longrightarrow} K_{0}\left(t_{1}, t_{2}\right), \\
	&\frac{\sum_{s\in\Scal_{1,m}} \1f\left\{T_{1}(s) \geq t_{1},T_{2}(s) \geq t_{2}\right\}}{m_1}:=\frac{S_{m}(t_1,t_2)}{m_1}  \stackrel{p}{\longrightarrow} K_{1}\left(t_{1}, t_{2}\right),
	\end{aligned}
	$$
as $m_0$ and $m_1$ goes to infinity.
\end{lemma}
Lemma~\ref{lemma:cnvrg_true} addresses the challenge posed by the non-Lipschitz nature of $\1f\{T_1(s)\geq t_1,T_2(s)\geq t_2\}$, which complicates the application of the NED-based law of large numbers by a new approach.

\subsection{Detailed Proofs of Lemmas~\ref{lemma:cnvrg_discovery} and \ref{lemma:cnvrg_proportion}}\label{sec:prf_1_2}
In this subsection, we complete the proofs of Lemmas~\ref{lemma:cnvrg_discovery} and \ref{lemma:cnvrg_proportion} using Lemmas~\ref{lemma:cnvrg_tau}--\ref{lemma:cnvrg_true}.

\begin{proof}[Proof of Lemma \ref{lemma:cnvrg_discovery}]
Due to the triangular inequality, the LHS of \eqref{eqn:k0FirstStep} can be written as
$$\begin{aligned}
	&\left|\frac{1}{m}\sumsSm \left\{\int L\left\{ t_{1}, t_{2},x, \whrho(s)\right\} d \whG_{\tm}(x)
	-
	\int L\left\{ t_{1}, t_{2},x,\rho(s)\right\} d G_{0}(x)\right\}\right| \\
	& \qquad \leq
	\left|\frac{1}{m}\sum_{s\in\Scal_m} \int \left[ L\left\{ t_{1}, t_{2},x,\whrho(s)\right\} - L\left\{ t_{1}, t_{2},x,\rho(s)\right\}\right] d \whG_{\tm}(x)\right|\\
	& \qquad \qquad \qquad + 
	\left|\frac{1}{m} \sumsSm \int L\left\{ t_{1}, t_{2},x,\rho(s)\right\} \left\{d \whG_{\tm}(x) - d G_{0}(x)\right\} \right|.
\end{aligned}$$
Thus, we only need to prove
\begin{equation}\label{equ:L_cov}
		\sup _{|t_{1}| \leq t_{1}^{\prime}, |t_{2}| \leq t_{2}^{\prime}}
		\left|\frac{1}{m}\sumsSm \int \left[L\left\{ t_{1}, t_{2},x,\whrho(s)\right\} - L\left\{ t_{1}, t_{2},x,\rho(s)\right\}\right] d \whG_{\tm}(x)\right|=\op(1)
\end{equation}
and 
\begin{equation}\label{equ:L_G}
		\sup _{|t_{1}| \leq t_{1}^{\prime}, |t_{2}| \leq t_{2}^{\prime}}\left|\frac{1}{m} \sumsSm \int L\left\{ t_{1}, t_{2},x,\rho(s)\right\} \left\{d \whG_{\tm}(x) - d G_{0}(x)\right\} \right|=\op(1).
\end{equation}
To simplify our presentation, we will prove \eqref{equ:L_cov} and \eqref{equ:L_G} under the condition that $L\left( t_{1}, t_{2},x, c\right)$ is uniformly continuous over $\Rb^3\times[-1,1]$. The justification of the uniform continuity of $L\left( t_{1}, t_{2},x, c\right)$ is given later.

(i) Now, we prove \eqref{equ:L_cov} with the uniform continuity of $L\left( t_{1}, t_{2},x, c\right)$. 
For an arbitrary $\epsilon>0$, there exists $0<\delta<2$ such that 
\begin{equation}\label{equ:L_unif_cnvrg}
    \sup _{(t_1,t_2,x)\in\Rb^3}\left|L\left( t_{1}, t_{2}, x, c_1\right)- L\left( t_{1}, t_{2}, x,c_2\right)\right|<\epsilon,
\end{equation}
for any $c_1,c_2\in[-1,1]$ satisfying $\left|c_1-c_2\right|<\delta$. Thus, we have
$$
	\sup _{(t_1,t_2,x)\in\Rb^3}\int\left|L\left(t_{1}, t_{2}, x, c_1\right)- L\left( t_{1}, t_{2}, x,c_2\right)\right|d\whG_{\tm}(x)<\epsilon.
$$
It further implies that
\begin{align*}
	&\Pb\left(
		\left|\frac{1}{m}\sumsSm \int \left[L\left\{t_{1}, t_{2},x,\whrho(s)\right\} - L\left\{ t_{1}, t_{2},x,\rho(s)\right\}\right] d \whG_{\tm}(x)\right|>\epsilon\right)\\
		& \; \leq
		\Pb\left\{
		\frac{1}{m}\sumsSm \int \left|L\left\{ t_{1}, t_{2}, x,\whrho(s)\right\} - L\left\{t_{1}, t_{2},x,\rho(s)\right\} \right| d \whG_{\tm}(x)>\epsilon;
		\sup_{s\in\Scal_m}\left|\rho(s)-\whrho(s)\right|<\delta\right\}\\
		& \qquad \qquad \qquad \qquad +
		\Pb\left\{
		\sup_{s\in\Scal_m}\left|\rho(s)-\whrho(s)\right|>\delta\right\}\\
		& \; =
		\Pb\left\{
		\sup_{s\in\Scal_m}\left|\rho(s)-\whrho(s)\right|>\delta\right\}
		\cnvrg 0,
\end{align*}
where the convergence in the last step is due to Assumptions~\ref{ass:neigh} and \ref{ass:cnvrg_cov:c} of the main paper.

(ii) To prove \eqref{equ:L_G}, it is sufficient to show that 
\begin{equation}\label{eqn:k0Part2}
	\sup _{|t_{1}| \leq t_{1}^{\prime}, |t_{2}| \leq t_{2}^{\prime}}\sup_{c\in[-1,1]}\left|\int L\left( t_{1}, t_{2},x,c\right) \left\{d \whG_{\tm}(x) - d G_{0}(x)\right\}\right|=\op(1).
\end{equation}
To show \eqref{eqn:k0Part2}, we note that the following pointwise convergence,
\begin{equation}\label{equ:pwise_L_G}
	\int L\left( t_{1}, t_{2},x, c\right) \left\{d \whG_{\tm}(x) - d G_{0}(x)\right\}
	= \op(1),
\end{equation}
can directly b obtained according to the proof of (44) in \cite{Zhang2021} when $(t_1,t_2,c)$ is fixed.
For any $\epsilon>0$, we can split $[-t_1^\prime,t_1^\prime]\times[-t_2^\prime,t_2^\prime]\times[-1,1]$ into $B$ disjoint finite sets $\cup_{1\leq k\leq B} \Ccal_k$ such that
$$\begin{aligned}
	&\sup_{(t_1,t_2,c),(\tilde{t}_1,\tilde{t}_2,\tilde{c})\in\Ccal_k}
	\left|\int \left\{L\left( t_{1}, t_{2},x,c\right)
	-
	L\left( \tilde{t}_{1}, \tilde{t}_{2},x,\tilde{c}\right)
	\right\} d G_{0}(x)\right|\leq \epsilon/2, \\
	&\mbox{and } ~~\sup_{(t_1,t_2,c),(\tilde{t}_1,\tilde{t}_2,\tilde{c})\in\Ccal_k}
	\left|\int \left\{L\left( t_{1}, t_{2},x,c\right)
	-
	L\left( \tilde{t}_{1}, \tilde{t}_{2},x,\tilde{c}\right)
	\right\} d \whG_{\tm}(x)\right|\leq \epsilon/2,
\end{aligned}$$
according to the uniform continuity of $L(t_1,t_2,x,c)$.
Fixing $(t_1^k,t_2^k,c^k)\in \Ccal_k,1\leq k\leq B$, we have
\begin{align*}
	&\sup _{|t_{1}| \leq t_{1}^{\prime}, |t_{2}| \leq t_{2}^{\prime}}
	\sup_{c\in[-1,1]}
	\left|\int L\left( t_{1}, t_{2},,x,c\right) \left\{d \whG_{\tm}(x) - d G_{0}(x)\right\}\right|\\
	& \qquad  = \sup _{(t_1,t_2,c)\in\cup_{1\leq k\leq B}\Ccal_k}
	\left|\int L\left( t_{1}, t_{2},x,c\right) \left\{d \whG_{\tm}(x) - d G_{0}(x)\right\}\right|\\
	& \qquad \leq
	\max_{1\leq k \leq B}
	\sup_{(t_1,t_2,c)\in\Ccal_k}
	\left|\int L\left( t_{1}, t_{2},x,c\right) -
	L\left( t^k_{1}, t^k_{2},x,c^k\right)
	d G_{0}(x)\right|\\
	& \qquad \qquad \qquad +
	\max_{1\leq k \leq B}
	\sup_{(t_1,t_2,c)\in\Ccal_k}
	\left|\int L\left( t_{1}, t_{2},x,c\right) - L\left( t^k_{1}, t^k_{2},x,c^k\right)
	d \whG_{\tm}(x)\right|\\
	& \qquad \qquad \qquad +
	\max_{1\leq k\leq B}
	\left|\int L\left( t^k_{1}, t^k_{2},x,c^k\right) \left\{d \whG_{\tm}(x) - d G_{0}(x)\right\}\right|
	\\
	& \qquad \leq 
	\epsilon+
	\max_{1\leq k\leq B}
	\left|\int L\left( t^k_{1}, t^k_{2},x,c^k\right) \left\{d \whG_{\tm}(x) - d G_{0}(x)\right\}\right|.
\end{align*}
Due to the pointwise convergence of \eqref{equ:pwise_L_G} and the arbitrariness of $\epsilon>0$, the uniform convergence of \eqref{eqn:k0Part2} stands from the eabove displayed inequality. 
Finally, the definition of $K_0(t_1,t_2)$ in Assumption~\ref{ass:cnvrg_true} together with \eqref{eqn:k0FirstStep} implies the conclusion of Lemma~\ref{lemma:cnvrg_discovery}.\\

We now justify the uniform continuity to complete the proof of Lemma~\ref{lemma:cnvrg_discovery}. 
We first prove $L\left( t_{1}, t_{2},x, c\right)$ is uniformly continuous over $\Rb^3\times[-1+\delta_\rho,1-\delta_\rho]$ for any $0<\delta_\rho<1$. Denote by $f(u,v;c)$ the bivariate normal density with mean zero, variance one and correlation $c\in(-1,1)$. Then we have 
$$\begin{aligned}
		L(t_1,t_2,x,c) 
		=&
		\int_{t_2-x}^\infty 
		\int_{t_1}^\infty f(u,v;c)dvd u\\
		=&
		\int_{t_2-x}^\infty d u
		\int_{t_1}^\infty \frac{1}{2\pi\sqrt{(1-c^2)}}
		\exp\left\{-\frac{u^2-2cuv+v^2}{2\left(1-c^2\right)}\right\}dv.
\end{aligned}$$
	
Next, we prove that given $\delta_\rho, M>0$, $c_1$ and $c_2$ such that $\left|c_1-c_2\right|<\delta$ and $0<\delta<2-2\delta_\rho$, the following inequalities hold
\begin{align*}
		&\sup _{(t_1,t_2,x)\in\Rb^3;-1+\delta_\rho\leq c_1,c_2\leq1-\delta_\rho}\left|L(t_1,t_2,x,c_1)-L(t_1,t_2,x,c_2)\right|\\
		& \qquad \leq
		\sup _{-1+\delta_\rho\leq c_1,c_2\leq1-\delta_\rho}
		\int_{-\infty}^\infty
		\int_{-\infty}^\infty
		\left| f(u,v;c_1)-f(u,v;c_2)\right|dudv\\
		& \qquad \leq
		8\int_{M}^\infty \phi(u) du 
		+
		\sup _{-1+\delta_\rho\leq c_1,c_2\leq1-\delta_\rho}
		\int_{-M}^M
		\int_{-M}^M
		\left| \frac{\partial f(u,v;c)}{\partial c}|_{c=\tilde{c}(c_1,c_2)}\right|\left|c_1-c_2\right|dudv\\
		& \qquad \leq
		8\int_{M}^\infty \phi(u) du 
		+
		\delta\int_{-M}^M
		\int_{-M}^M
		\sup_{c\in[-1+\delta_\rho,1-\delta_\rho]}\left|\frac{\partial f(u,v;c)}{\partial c} \right|dudv\\
		& \qquad \leq
		8\int_{M}^\infty \phi(u) du 
		+
		\delta C(\delta_\rho,M),
\end{align*}	
	where the second inequality is achieved by covering $(-\infty,\infty)^2$ with five regions, namely $[-M,M]^2$, $(-\infty,-M)\times(-\infty,\infty)$,$(M,\infty)\times(-\infty,\infty)$,$(-\infty,\infty)\times(M,\infty)$ and $(-\infty,\infty)\times(-\infty,-M)$ and then applying the mean value theorem to the first region and using the fact $\int_{-\infty}^\infty f(u,v;c) dv = \phi(u)$ (and $\int_{-\infty}^\infty f(u,v;c) du = \phi(v)$) for the remaining four regions, and the last inequality stands because $\partial f(u,v;c)/\partial c$ is continuous and 
	$$C(M,\delta_\rho)=4M^2
		\sup_{(u,v)\in[-M,M]^2,c\in[-1+\delta_\rho,1-\delta_\rho]}\left|\frac{\partial f(u,v;c)}{\partial c} \right|$$
	is a positive constant depending only on $M$ and $\delta_\rho$. For any $\epsilon>0$, we can choose $M>0$ large enough such that $8\int_{M}^\infty \phi(u) du<\epsilon/2$. We then take $\delta(M,\delta_\rho)=\min\{2-2\delta_\rho, \epsilon/2C(\delta_\rho,M)\}>0$ to fulfill $\left|L(t_1,t_2,x,c_1)-L(t_1,t_2,x,c_2)\right|\leq\epsilon$ when $\left|c_1-c_2\right|<\delta(M,\delta_\rho)$. 
	We extend the uniform continuity for the correlation parameter $c$ in $[-1+\delta_\rho,1-\delta_\rho]$ to $[-1,1]$. To this end, it is sufficient to show that for any $\epsilon>0$, there exists $\delta_\rho >0$ such that 
	$$\begin{aligned}
        &\sup _{(t_1,t_2,x)\in\Rb^3}\left|L(t_1,t_2,x,1-\delta)-L(t_1,t_2,x,1)\right|\leq \epsilon,\\
	   	&\mbox{and } ~~ \sup _{(t_1,t_2,x)\in\Rb^3}\left|L(t_1,t_2,x,-1+\delta)-L(t_1,t_2,x,-1)\right|\leq \epsilon,
	\end{aligned}$$
	for any $\delta$ satisfying $0< \delta\leq \delta_\rho$. 
	We present the proof of the first inequality above.
	The proof for the second equality is similar and thus omitted. Letting $t_1-x\leq t_2$, we have
	$$
	\begin{aligned}
	\left|L(t_1,t_2,x,1-\delta)-L(t_1,t_2,x,1)\right|
	&=
	\left|L(t_1,t_2,x,1-\delta)-\Pb(V_2(s)\geq t_2)\right|\\
	&=
	\left|\Pb \left\{V_{1}(s)+x\geq t_{1},V_{2}(s) \geq t_{2} \right\}-\Pb\{V_2(s)\geq t_2\}\right|\\
	&=
	\Pb \left\{V_{1}(s)< t_{1}-x,V_{2}(s) \geq t_{2} \right\}\\
	&<
	\Pb \left\{V_{1}(s)<V_{2}(s)\right\}
	,\\
	\end{aligned}
	$$
	where $(V_1(s),V_2(s))$ follows a bivariate normal distribution with variance one and correlation $1-\delta$. Similarly, when $t_1-x>t_2$, we have
	$$
	\left|L(t_1,t_2,x,1-\delta)-L(t_1,t_2,x,1)\right|
	<
	\Pb \left\{V_{1}(s)>V_{2}(s)\right\},
	$$ 
	which implies 
	$$
	\left|L(t_1,t_2,x,1-\delta)-L(t_1,t_2,x,1)\right|
	<
	\Pb \left\{V_{1}(s)\not =V_{2}(s)\right\}.
	$$ As $\Pb \left\{V_{1}(s)\not=V_{2}(s)\right\}$ as a function of the correlation $\delta$ is right continuous at $\delta=0$, a proper $\delta_\rho>0$ can always be selected.
	To sum up, we have verified the uniform continuity and the proof is thus completed.
\end{proof}

\begin{proof}[Proof of Lemma~\ref{lemma:cnvrg_proportion}] 
	To show the uniform convergence of $m_0^{-1} \whV_{m}(t_{1}, t_{2})$ in \eqref{eqn:lemS2Eq1}, we begin with the pointwise convergence. 
	In other words, we first show $|m_0^{-1}  \whV_{m}(t_{1}, t_{2}) - K_{0}(t_{1}, t_{2})|=\op(1)$ for any fixed $(t_1,t_2)$. It suffices to show that
	\begin{equation}\label{eqn:lemm2Upp}
        \Pb\left\{m_0^{-1} \whV_{m}\left(t_{1}, t_{2}\right) \leq K_{0}\left(t_{1}, t_{2}\right)+\delta_0\right\}\rightarrow 1
    \end{equation}
    and 
    \begin{equation}\label{eqn:lemm2Low}
        \Pb\left\{m_0^{-1} \whV_{m}\left(t_{1}, t_{2}\right) \geq K_{0}\left(t_{1}, t_{2}\right)-\delta_0\right\}\rightarrow 1
    \end{equation}
    as $m\rightarrow\infty$.

    We now focus on \eqref{eqn:lemm2Upp}. Observe that 
	\begin{align*}
	    \whV_{m}\left(t_{1}, t_{2}\right)
	    & = \sum_{s\in\Scal_{0,m}} \1f\left\{\whT_{1}(s) \geq t_{1},\whT_{2}(s) \geq t_{2}\right\}\\
	    &=
	    \sum_{s\in\Scal_{0,m}} \1f\left\{T_{1}(s) \geq \whr^{(1)}(s) t_{1},T_{2}(s) \geq \whr^{(2)}(s)t_{2}\right\}\\
	    &:=	
	    V_{m}\left(\widehat{\rf}^{(1)}t_{1},\widehat{\rf}^{(2)} t_{2}\right),
	\end{align*}
	where $\whr^{(1)}(s)=\whsig(s)/\sigma(s)$, $\whr^{(2)}(s)=\whtau(s)/\tau(s)$ for $s\in\Scal_{0,m}$, and $\widehat{
	\rf}^{(k)}=(\whr^{(k)}(s))_{s\in\Scal_{0,m}}$ for $k=1,2$. 
	For an arbitrary $\delta>0$, we define three events $\Acal_{k,\delta}=	\{\sup_{s\in\Scal_{0,m}}\left|\whr^{(k)}(s)-1\right|\leq\delta\}$ for $k=1,2$, and 
	$$\Acal_{3,\delta}(t_1,t_2)=\left\{\left|m_0^{-1}V_{m}(t_1,t_2)-K_0(t_1,t_2)\right|\leq\delta\right\}.$$
It can be seen that $\Pb(\Acal_{k,\delta})\rightarrow 1$ for $k=1,2$ due to Assumptions~\ref{ass:neigh} and \ref{ass:cnvrg_cov}. 
Similarly, $\Pb\{\Acal_{3,\delta}(t_1,t_2)\}\rightarrow 1$ for any fixed $(t_1,t_2)\in\Rb^2$ as $m\rightarrow\infty$, because of Assumptions~\ref{ass:neigh}, \ref{ass:pi0}, and Lemma~\ref{lemma:cnvrg_true}.
To verify \eqref{eqn:lemm2Upp}, we notice that
	\begin{align*}
	    &\left\{\frac{1}{m_0}  \whV_{m}\left(t_{1}, t_{2}\right) \leq K_{0}\left(t_{1}, t_{2}\right)+\delta_0\right\}\\
	    & \qquad \supseteq\left\{\frac{1}{m_0} V_{m}\left(\widehat{\rf}^{(1)}t_1,\widehat{\rf}^{(2)}t_2\right) \leq K_{0}\left(t_{1}, t_{2}\right)+\delta_0 \right\}\cap
	    \left(\Acal_{1,\delta}\cap\Acal_{2,\delta}\right)\\
	    & \qquad \supseteq
	    \left\{\frac{1}{m_0}  V_{m}\left(\widehat{\rf}^{(1)}t_1,\widehat{\rf}^{(2)}t_2\right) \leq\frac{1}{m_0}  V_{m}\left((1-\delta)t_1,(1-\delta)t_2\right) \right\}
	    \cap
	    \left(\Acal_{1,\delta}\cap\Acal_{2,\delta}\right)\\
	    &\qquad \qquad \qquad\qquad  \cap\left\{\left|\frac{1}{m_0}  V_{m}\left((1-\delta)t_1,(1-\delta)t_2\right)-K_0\left((1-\delta)t_1,(1-\delta)t_2\right) \right|\leq \delta_0/2 \right\}\\
	    &\qquad \qquad \qquad\qquad \cap\left\{
	    \left| K_0\left(\left(1-\delta\right)t_1,\left(1-\delta\right)t_2\right)-K_0(t_1,t_2) \right|\leq \delta_0/2 \right\}\\
	    & \qquad \supseteq 
	    \left(\Acal_{1,\delta}\cap\Acal_{2,\delta}\right)\cap  \Acal_{3,\delta_0/2}((1-\delta)t_1,(1-\delta)t_2)\\
	    &\qquad \qquad \qquad\qquad
	    \cap\left\{\left| K_0((1-\delta)t_1,(1-\delta)t_2)-K_0(t_1,t_2) \right|\leq \delta_0/2 \right\},
	\end{align*}
	where the first inclusion is because of 
	$\whV_{m}\left(t_{1}, t_{2}\right)=	V_{m}\left(\widehat{\rf}^{(1)}t_{1},\widehat{\rf}^{(2)} t_{2}\right)$; the second inclusion is due to the triangle inequality; 
	as for the third inclusion, we used the fact that $V_{m}$ is monotonically decreasing with respect $(t_1,t_2)$ and 
	$$\left\{\frac{1}{m_0}  V_{m}\left(\widehat{\rf}^{(1)}t_1,\widehat{\rf}^{(2)}t_2\right) \leq\frac{1}{m_0}  V_{m}\left((1-\delta)t_1,(1-\delta)t_2\right) \right\}$$
	holds true under $\Acal_{1,\delta}\cap\Acal_{2,\delta}$,	and the definition of $\Acal_{3,\delta}$. 
	Thus for an arbitrary $\delta_0>0$, we can choose $\delta>0$ to guarantee 
	$$\left| K_0((1-\delta)t_1,(1-\delta)t_2)-K_0(t_1,t_2) \right|\leq \delta_0/2,$$
	because $K_0(t_1,t_2)$ is continuous, and \eqref{eqn:lemm2Upp} holds due to $$\Pb\left(\Acal_{1,\delta}\cap\Acal_{2,\delta}\cap  \Acal_{3,\delta_0/2}\left((1-\delta)t_1,(1-\delta)t_2\right)\right)\rightarrow 1.$$
	For \eqref{eqn:lemm2Low}, it can be shown similarly and thus the pointwise convergence of $m_0^{-1} \whV_{m}(t_{1}, t_{2})$ holds. 
	The uniform convergence in \eqref{eqn:lemS2Eq1} can be derived similarly as in the proof of Lemma~\ref{lemma:cnvrg_discovery} after getting the pointwise convergence of $m_0^{-1} \whV_{m}(t_{1}, t_{2})$.
	As for \eqref{eqn:lemS2Eq2} and \eqref{eqn:lemS2Eq3}, these two results can be proved analogously as \eqref{eqn:lemS2Eq1} and thus their proofs are omitted.
\end{proof}

\section{Proofs of Lemmas S.3--S.9}\label{sec:prf_lemmas}
This section is organized as follows. Section~\ref{sec:relation_tm_r} discusses how to choose a set $\tScal_m$ used for NPEB
such that Assumption~\ref{ass:num_tSm} of the main paper is satisfied. 
Section~\ref{sec:prf_3_8} presents the detailed proofs of Lemmas~\ref{lemma:cnvrg_tau}--\ref{lemma:cnvrg_true}.

\subsection{Justification of Assumption \ref{ass:num_tSm} of the main paper}\label{sec:relation_tm_r}

Observe that $\{T^*_1(s;r):s\in\tScal_m\}$ defined in \eqref{eqn:tStarDef} of the main paper are independent once \eqref{equ:req_ind} of the main paper is fulfilled. Due to Assumptions~\ref{ass:incr_dom} and \ref{ass:neigh} of the main paper, \eqref{equ:req_ind} of the main paper holds as long as
\begin{equation}\label{equ:safe_dis}
    \dist(s,s^\prime)\geq2 N_{nei}\Delta_u+2r,\quad s,s^\prime\in\tScal_m.
\end{equation} 
It is because for any $w\in\cup_{v\in \Ncal(s)}B(v;r)$, $\dist(s,w)\leq\dist(s,v)+\dist(v,w)\leq N_{nei}\Delta_u+r$, where we have used Assumption~\ref{ass:neigh} of the main paper that $\mathcal{N}(s)$ is the set of nearest neighbors with uniformly bounded cardinality for each location $s$. With \eqref{equ:safe_dis}, $r$ can be taken as large as $\widetilde{\Delta}_{l,m}/2-N_{nei}\Delta_u$ to ensure the independence of $\{T^*_1(s;r):s\in\tScal_m\}$, where $\widetilde{\Delta}_{l,m}$ is defined as in \eqref{eqn:npebDist} of the main paper.
In other words, $r$ and $\widetilde{\Delta}_{l,m}$ are of the same order.
We show that $\tScal_m$ can be chosen such that
\begin{equation}\label{equ:tm_m}
    \tm =|\tScal_m| \asymp  m/r^K.
\end{equation}
To this end, we pick all possible $s\in\Scal_m$ into $\tScal_m$ which satisfies \eqref{equ:safe_dis}
for any $s^\prime\in\tScal_m$ until no more locations satisfy the condition. A typical choice is
\begin{equation}\label{equ:max_tm}
    \tScal_m=\argmax_{\tScal_m\in\widetilde{\mathscr{S}}_m}|\tScal_m|
\quad\text{with}\quad \widetilde{\mathscr{S}}_m=\left\{\tScal_m\subset\Scal_m:\tScal_m\text{ satisfies \eqref{equ:safe_dis}}\right\},
\end{equation}
which can be viewed as the largest $2N_{nei}\Delta_u+2r$-packing, motivated by the definition of packing number \citep[see e.g., Definition 5.4 of][]{WainWright2019}. Borrowing the idea of volume comparison lemma, we construct $m$ $K$-dimensional cubes centered at the locations in $\Scal_m$ with the length $\Delta_l/2$, which are non-overlapping due to Assumption \ref{ass:incr_dom}, and $\tm$ $K$-dimensional cubes centered at the locations in $\tScal_m$ with the length $2(2N_{nei}\Delta_u+\Delta_u+2r)$. It is straightforward to verify the cubes centered at $\tScal_m$ cover all cubes centered at $\Scal_m$; otherwise, \eqref{equ:max_tm} will be violated.
Thus $m$, $\tm$, and $r$ satisfy
$$
m \Delta_l^K/2^K\leq \tm 2^K(2N_{nei}\Delta_u+\Delta_u+2r)^K,
$$
where the LHS is the total volume of cubes centered at $\Scal_m$ and the RHS is the total volume of cubes centered at $\tScal_m$. 
Hence, we obtain
$$
\tm \geq m \frac{\Delta_l^K} {4^K(2N_{nei}\Delta_u+\Delta_u+2r)^K} \propto m/r^K.
$$
To satisfy the other side of \eqref{equ:tm_m}, we just need to pick fewer locations into the $\tScal_m$. This completes the proof of \eqref{equ:tm_m}. 

Next, we show that $\tScal_m$ can be constructed such that $\tm^{1/(\lambda p)}\{\log(\tm)\}^{-1/(2\lambda )}=o(\widetilde{\Delta}_{l,m})$ to fulfill \eqref{equ:phi_r_restri} and $\tm \rightarrow \infty$ as $m\rightarrow\infty$, where $\lambda, p >0$ are associated with the $L_p$-NED in Assumption \ref{ass:NED}.
The above requirement on $\tScal_m$ can be guaranteed when we take, for simplicity, $\widetilde{\Delta}_{l,m}\asymp\tm^{1/(\lambda p)}$, which implies $r\asymp\tm^{1/(\lambda p)}$. 
Combining with \eqref{equ:tm_m}, we get
$$\tm\asymp m^{\frac{\lambda p}{K+\lambda p}},$$
which tends to infinity and $\tm^{1/(\lambda p)}\{\log(\tm)\}^{-1/(2\lambda )}=o(\widetilde{\Delta}_{l,m})$ as $m\rightarrow\infty$. 

\subsection{Detailed Proofs of Lemmas~\ref{lemma:cnvrg_tau}--\ref{lemma:cnvrg_true}}\label{sec:prf_3_8}

In this section, we provide the detailed proofs of Lemmas~\ref{lemma:cnvrg_tau}--\ref{lemma:cnvrg_emp_bayes_cp} stated in Section \ref{sec:intro_lemmas}.
We first present some results about the difference between $T_1(s;r)$ and $T_1^*(s;r)$. Suppose $X$ is $L_p(\mathbf{d})$-NED on the random field $Y=\{Y(s),s\in \Vcal_m\}$. 
Then the $L_p$ norm of the difference between $T_1(s)$ and $\Eb\left[T_1(s)\mid \mathfrak{F}\left\{\cup_{v\in \Ncal(s)}B(v;r)\right\}\right]$ is controlled by
\begin{align}\label{equ:NED_T1_prop}
    &\left\|T_1(s)-\Eb\left[T_1(s)\mid \mathfrak{F}\left\{\cup_{v\in \Ncal(s)}B(v;r)\right\}\right]\right\|_p \nonumber \\
    &\qquad \qquad \leq \sum_{v\in\Ncal(s)}\frac{\left\|X(v)-\Eb\left[X(v)\mid \mathfrak{F}\left\{\cup_{v\in \Ncal(s)}B(v;r)\right\}\right]\right\|_p}{\tau(s)} \nonumber \\
    &\qquad \qquad \leq \sum_{v\in\Ncal(s)}\frac{d_m(v)}{\tau(s)} \psi(r)
\end{align}
due to the triangular inequality and the generalized non-decreasing property.
Accordingly, the difference between the normalized variants is controlled by
\begin{equation}\label{equ:NED_T1}
    \left\|T_1(s;r)-T_1^*(s;r)\right\|_p
    \leq \sum_{v\in\Ncal(s)}\frac{d_m(v)}{\tau(s)\zeta_r(s)} \psi(r): = \eta(s;r)\psi(r),
\end{equation}
where $\eta(s;r)=\sum_{v\in\Ncal(s)}d_m(v)/\{\tau(s)\zeta_r(s)\}$. 
Thus, it can be seen that
$$\begin{aligned}
    \Pb\left\{
	\left|T_{1}(s;r)-T_1^*(s;r)\right|>\delta\right\}
    \leq 
    \frac{1}{\delta^{p}}\left\|T_{1}(s;r)-T_1^*(s;r)\right\|_p^p
    \leq 
    \frac{\eta^p(s;r)\psi^p(r)}{\delta^p}
\end{aligned}$$
for any $\delta>0$ due to the Markov inequality. Subsequently, we can establish the convergence of
$$
	\supstSm
	\left|T_{1}(s;r)-T^*_1(s;r)\right|
	\quad
	\text{and}
	\quad
	\frac{1}{\tm}\sumstSm
	\left|T_{1}(s;r)-T^*_1(s;r)\right|.
$$

In particular, for a fixed $\delta>0$, the maximum difference between $T_{1}(s;r)$ and $T^*_1(s;r)$
is controlled by
\begin{equation}\label{equ:maxTT*}
\begin{aligned}
\Pb\left\{\supstSm
	\left|T_{1}(s;r)-T_1^*(s;r)\right|\leq\delta\right\}
	&=
	\Pb\left\{\cap_{s\in\tScal_m}
	\left|T_{1}(s;r)-T_1^*(s;r)\right|\leq\delta\right\}\\
    &\geq
    1-\frac{1}{\delta^p}\sumstSm\eta^p(s;r)\psi^p(r).
\end{aligned}
\end{equation}
The mean difference is controlled by
\begin{align}
    \Pb\left\{\frac{1}{\tm}\sumstSm
	\left|T_{1}(s;r)-T_1^*(s;r)\right|>\delta\right\}
    & \leq 
    \frac{1}{\tm^p\delta^{p}}\left\|\sumstSm
	\left|T_{1}(s;r)-T_1^*(s;r)\right|\right\|_p^p  \nonumber \\
    & \leq 
    \frac{1}{\tm^p\delta^{p}}\left\{\sumstSm
	\left\|T_{1}(s;r)-T_1^*(s;r)\right\|_p\right\}^p \label{equ:meanTT*} \\
    & \leq 
    \frac{1}{\tm^p\delta^{p}} \left\{\sumstSm\eta(s;r)\psi(r)\right\}^p , \nonumber
\end{align}
where the first, second, and last inequalities are due to the Markov inequality, the triangular inequality, and \eqref{equ:NED_T1}, respectively. 

\begin{proof}[Proof of Lemma~\ref{lemma:cnvrg_tau}]
First note that $X$ is uniformly $L_p$-NED on $Y$ for $ p\geq 2$ implies that $X$ is uniformly $L_2$-NED on $Y$.

(a) 
Recall $\zeta_r^2(s)=\var\left(\Eb\left[T_1(s)\mid \mathfrak{F}\left\{\cup_{v\in \Ncal(s)}B(v;r)\right\}\right]\right)$ and $\var\left\{T_1(s)\right\}=1$. Then we have
\begin{equation}\label{equ:zeta_1side}
    \zeta^2_r(s)=\var\left(\Eb\left[T_1(s)\mid \mathfrak{F}\left\{\cup_{v\in \Ncal(s)}B(v;r)\right\}\right]\right)\leq \var\left\{T_1(s)\right\}=1
\end{equation}
according to the law of total variance. Further, setting $p=2$ in \eqref{equ:NED_T1_prop}, we obtain
$$
    \left\|T_1(s)-\Eb\left[T_1(s)\mid \mathfrak{F}\left\{\cup_{v\in \Ncal(s)}B(v;r)\right\}\right]\right\|_2
    \leq 
    \sum_{v\in\Ncal(s)}\frac{d_m(v)}{\tau(s)}\psi(r)\leq C\psi(r)
$$
where the last inequality holds by Assumptions~\ref{ass:neigh}, \ref{ass:cnvrg_cov:a} and \ref{ass:NED} of the main paper. Therefore, we get
$$\begin{aligned}
    \left|\zeta_r^2(s)-1\right|
    &=\left|\var\left(\Eb\left[T_1(s)\mid \mathfrak{F}\left\{\cup_{v\in \Ncal(s)}B(v;r)\right\}\right]\right)-\var\left\{T_1(s)\right\}\right|\\
    &=\left|\Eb\left(\Eb\left[T_1(s)\mid \mathfrak{F}\left\{\cup_{v\in \Ncal(s)}B(v;r)\right\}\right]-T_1(s)\right)^2\right|\\
    &=\left\|T_1(s)-\Eb\left[T_1(s)\mid \mathfrak{F}\left\{\cup_{v\in \Ncal(s)}B(v;r)\right\}\right]\right\|_2^2\\
    &\leq C\psi^2(r),
\end{aligned}$$
where the second equality stands because $\Eb\left(\Eb\left[T_1(s)\mid \mathfrak{F}\left\{\cup_{v\in \Ncal(s)}B(v;r)\right\}\right]\right)=\Eb\left\{ T_1(s)\right\}$. 
For the third equality, we have
\begin{align*}
    &\Eb\left(T_1(s)\Eb\left[T_1(s)\mid \mathfrak{F}\left\{\cup_{v\in \Ncal(s)}B(v;r)\right\}\right]\right)\\
    & \qquad =
    \Eb\left\{\Eb\left(T_1(s)\Eb\left[T_1(s)\mid \mathfrak{F}\left\{\cup_{v\in \Ncal(s)}B(v;r)\right\}\right]\mid \mathfrak{F}\left\{\cup_{v\in \Ncal(s)}B(v;r)\right\}\right)\right\}\\
    & \qquad =
    \Eb\left(\Eb\left[T_1(s)\mid \mathfrak{F}\left\{\cup_{v\in \Ncal(s)}B(v;r)\right\}\right]\right)^2,
\end{align*}
where $T_1(s)\Eb\left[T_1(s)\mid \mathfrak{F}\left\{\cup_{v\in \Ncal(s)}B(v;r)\right\}\right] \in L_1$ is due to the Gaussianity from Assumption~\ref{ass:cond_norm} of the main paper. 

Together with \eqref{equ:zeta_1side}, for large enough $r$, we have
$$
    \left|1/\zeta_r^2(s)-1\right|
    \leq
    \left\{1-C\psi^2(r)\right\}^{-1}C\psi^2(r)\leq C\psi^2(r)
$$
since $\left|1/x^2-1\right|$ decreases as $x^2\in(0,1]$ increases. This completes Part (a) of this lemma.

(b) Assumptions~\ref{ass:neigh} and \ref{ass:cnvrg_cov} of the main paper imply $|\whtau(s)/\tau(s)-1|=\op(\tm^{-q})$ and $|\whtau^2(s)/\tau^2(s)-1|=\op(\tm^{-q})$. 
For any $0<\delta=C\tm^{-q}<1$, combining it with the proved conclusion in (a), we have 
$$\begin{aligned}
    \left|\whtau^2(s)/\tausq_r(s)-1\right|
    &=
    \left|\whtau^2(s)/\tau^2(s)\zeta_r^2(s)-1\right|\\
    &\leq
    \left|\whtau^2(s)/\tau^2(s)\zeta_r^2(s)-\whtau^2(s)/\tau^2(s)\right|
    +
    \left|\whtau^2(s)/\tau^2(s)-1\right|
    \\
    &\leq
    \whtau^2(s)/\tau^2(s)\left|1/\zeta_r^2(s)-1\right|
    +
    \left|\whtau^2(s)/\tau^2(s)-1\right|
    \\
    &\leq
  \left(1+C\tm^{-q}\right)C\psi^2(r)
    +
    C\tm^{-q}\\
    &\leq
    C\psi^2(r)
    +
    C\tm^{-q}
    \end{aligned}
$$
for large enough $r$ with probability tending to one as $\tm\rightarrow\infty$.
    
For (c), we notice that
$$
    \sup_m\sup_{s\in\Scal_m} \eta(s;r)=\sup_m\sup_{s\in\Scal_m} \sum_{v\in\Ncal(s)}\frac{d_m(v)}{\tau(s)\zeta_r(s)}\leq C \sup_m\sup_{s\in\Scal_m} \frac{1}{\zeta_r(s)},
$$
where the last inequality stands due to Assumptions~\ref{ass:neigh}, \ref{ass:cnvrg_cov:a}, and \ref{ass:NED} of the main paper. Finally, $\sup_m\supstSm \eta(s;r)$ is bounded for large enough $r$ according to Part (a) of this lemma. 
\end{proof}

\begin{proof}[Proof of Lemma \ref{lemma:cnvrg_max_sum}]
Note that
\begin{equation}\label{equ:prob_joint}
\begin{aligned}
    &\Pb\left(
    \supstSm \left|T^*_1(s;r)-\Eb\left\{T^*_1(s;r)\right\}\right|>\nu_1\delta_1\ \mathrm{\cup}\ \supstSm \left|T_1^*(s;r)-T_1(s;r)\right|>\nu_1\delta_1
    \ \right.\\
    &\left.\qquad\qquad\qquad\qquad
    \mathrm{\cup} \  \tm^{-1}\sumstSm\left|T_1(s;r)-T_1^*(s;r)\right|> \delta_2
    \right)\\
    & \qquad \leq
    \Pb\left[
    \supstSm \left|T^*_1(s;r)-\Eb\left\{T^*_1(s;r)\right\}\right|>\nu_1\delta_1\right] \\
    &\qquad \qquad +\Pb\left\{ \supstSm \left|T_1^*(s;r)-T_1(s;r)\right|>\nu_1\delta_1\right\} \\
    &\qquad \qquad + \Pb\left\{ \tm^{-1}\sumstSm\left|T_1(s;r)-T_1^*(s;r)\right|>\delta_2\right\} \\
    & \qquad  \leq
    \tm\exp(-\nu_1^2\delta_1^2/2)
    +C\tm\psi^p(r)/\nu_1^p\delta_1^p
    +C\psi^p(r)/\delta_2^p
\end{aligned}
\end{equation}
due to \eqref{equ:maxTT*}, \eqref{equ:meanTT*}, Lemma~\ref{lemma:cnvrg_tau:c}, and the observation that
$$\begin{aligned}
    \Pb\left(\supstSm\left|T^*_{1}(s;r)-\Eb\left\{T^*_{1}(s;r)\right\}\right| >\nu_1\delta_1\right)
	& = 
	\Pb\left(\cup_{s\in\tScal_m}\left[T^*_{1}(s;r)-\Eb\left\{T^*_{1}(s;r)\right\} >\nu_1\delta_1\right]\right)\\
	& \leq 
	\tm\exp(-\nu_1^2\delta_1^2/2)
\end{aligned}$$ 
implied directly by Assumption~\ref{ass:cond_norm}. 
Thus, to obtain Parts (a)--(c) of Lemma~\ref{lemma:cnvrg_max_sum} with probability at least \eqref{equ:prob_T}, it suffices to show the desired upper bounds in Lemma~\ref{lemma:cnvrg_max_sum} can be derived from the complementary event of \eqref{equ:prob_joint}, i.e., 
\begin{equation}\label{equ:cond_maxT*}
    \supstSm \left|T^*_1(s;r)-\Eb\left\{T^*_1(s;r)\right\}\right|\leq\nu_1\delta_1, 
\end{equation}
\begin{equation}\label{equ:cond_maxT*T}
    \supstSm \left|T_1^*(s;r)-T_1(s;r)\right|\leq\nu_1\delta_1,
\end{equation}
and
\begin{equation}\label{equ:cond_meanTT*}
    \tm^{-1}\sumstSm\left|T_1(s;r)-T_1^*(s;r)\right|\leq\delta_2.
\end{equation}
Now let us turn to the cumbersome details.

(a) First, $\supstSm T_1^*(s;r)$ is upper bounded by $\nu_1(1+\delta_1)$ through
\begin{equation}\label{equ:maxT*}
	\supstSm\left|T^*_{1}(s;r)\right|
	 	  \leq
	 \supstSm\left|T^*_{1}(s;r)-\Eb\left\{T^*_{1}(s;r)\right\}\right|+\supstSm\left|\Eb\left\{T^*_{1}(s;r)\right\}\right| 
	 \leq \nu_1(1+\delta_1),
\end{equation}
where the second inequality is due to \eqref{equ:T1nu1} and \eqref{equ:cond_maxT*}. 
For $T_1(s;r)$, we have
\begin{equation}\label{equ:maxT}
	\supstSm\left|T_{1}(s;r)\right|
    \leq
   \supstSm\left|T^*_{1}(s;r)\right|+ \supstSm\left|T^*_{1}(s;r)-T_{1}(s;r)\right| 	
   \leq 
   \nu_1(1+2\delta_1),
  \end{equation}
where the second inequality is because of \eqref{equ:cond_maxT*T} and \eqref{equ:maxT*}.
    
(b) To measure the difference between $T_{1}^2(s;r)$ and $\left\{T_1^*(s;r)\right\}^2$, we have
\begin{equation}\label{equ:meanTT*2}
\begin{aligned}
    &\tm^{-1}\left|\sumstSm T_{1}^2(s;r)-\{T_1^*(s;r)\}^2\right|\\
    & \qquad = 
    \tm^{-1}\left|\sumstSm\left\{T_{1}(s;r)-T_1^*(s;r)\right\}\left\{T_{1}(s;r)+T_1^*(s;r)\right\}\right| \\
    &\qquad \leq
    \left\{\tm^{-1}\left|\sumstSm T_{1}(s;r)-T_1^*(s;r)\right|\right\}
    \left\{\supstSm\left|T_{1}(s;r)+T_1^*(s;r)\right|\right\}\\
    & \qquad \leq
    \left\{\tm^{-1}\sumstSm\left| T_{1}(s;r)-T_1^*(s;r)\right|\right\}
    \left\{\supstSm\left|T_1(s;r)\right|+
    \supstSm\left|T^*_{1}(s;r)\right|\right\}\\
    & \qquad \leq
    \nu_1\delta_2(2+3\delta_1),
\end{aligned}
\end{equation}
where the first inequality in the above holds by the H\"older's inequality and the last inequality is 
due to \eqref{equ:cond_meanTT*}--\eqref{equ:maxT}.
    
(c) The upper bound for $\tm^{-1}\sumstSm\left| T_{1}(s;r)\right|$ is obtained by
\begin{align}
	& \tm^{-1}\sumstSm \left|T_{1}(s;r)\right| \nonumber \\
	& \quad =
   	\tm^{-1}\sumstSm\left|T_{1}(s;r)-T^*_{1}(s;r)
	+
	T^*_{1}(s;r)-\Eb \{T^*_{1}(s;r)\}
	 +
	 \Eb \{T^*_{1}(s;r)\}\right| \nonumber \\
	& \quad \leq
   	\tm^{-1}\sumstSm\left|T_{1}(s;r)-T^*_{1}(s;r)\right|
	+
	 \supstSm\left|T^*_{1}(s;r)-\Eb\{T^*_{1}(s;r)\} \right| 
	 +\supstSm\left|\Eb \{T^*_{1}(s;r)\}\right| \nonumber \\
	 & \quad \leq 
	 \nu_1\left(1+\delta_1\right)+\delta_2 \label{equ:meanT}
\end{align}
where the last step in \eqref{equ:meanT} is due to \eqref{equ:T1nu1}, \eqref{equ:cond_maxT*}, and \eqref{equ:cond_meanTT*}. 
As for $\tm^{-1}\sumstSm T^2_{1}(s;r)$, the H\"older inequality implies
\begin{align}
    & \tm^{-1}\sumstSm \left[\left\{T_{1}^*(s;r)\right\}^2-\Eb\left\{T_{1}^*(s;r)\right\}^2\right] \nonumber \\
    & \qquad \leq 
    \left|\tm^{-1}\sumstSm \left[T_{1}^*(s;r)-\Eb\left\{T_{1}^*(s;r)\right\}\right]\right|    \times 
    \supstSm\left|T_{1}^*(s;r) + \Eb\left\{T_{1}^*(s;r)\right\}\right|  \nonumber \\
    & \qquad \leq
    \supstSm \left|T_{1}^*(s;r)-\Eb\left\{T_{1}^*(s;r)\right\}\right| \times 
    \left\{\supstSm\left|T_{1}^*(s;r)\right| + \supstSm\Eb\left|T_{1}^*(s;r)\right|\right\} \nonumber \\
    & \qquad \leq
    \nu_1^2\delta_1(2+\delta_1), \label{eqn:lem4Holder2}
\end{align}
where the last line is due to \eqref{equ:T1nu1}, \eqref{equ:cond_maxT*}, and \eqref{equ:maxT*}. 
We then have
\begin{align}
    &\tm^{-1}\sumstSm T_{1}^2(s;r) \nonumber \\
    & \quad =
	\tm^{-1}\sumstSm \left[T_{1}^2(s;r)-\left\{T_{1}^*(s;r)\right\}^2\right]
   +
	\tm^{-1}\sumstSm \left[\left\{T_{1}^*(s;r)\right\}^2-\Eb\left\{T_{1}^*(s;r)\right\}^2\right] \nonumber \\
   & \qquad \qquad \qquad \qquad +
	\tm^{-1}\sumstSm \Eb\left\{T_{1}^*(s;r)\right\}^2
    \nonumber \\
    & \quad \leq
	\tm^{-1} \left| \sumstSm T_{1}^2(s;r)-\left\{T_{1}^*(s;r)\right\}^2\right| 
    +
	\tm^{-1}\sumstSm \left[\left\{T_{1}^*(s;r)\right\}^2-\Eb\left\{T_{1}^*(s;r)\right\}^2\right] \nonumber \\
    & \qquad \qquad \qquad \qquad +
	\supstSm \Eb\left\{T_{1}^*(s;r)\right\}^2 \nonumber \\
	& \quad \leq
	 \nu_1^2 + \nu_1\delta_2(2+3\delta_1) + \nu_1^2\delta_1\left(2+\delta_1\right), \label{equ:meanT2}
\end{align}
where we have used \eqref{equ:T1nu1}, \eqref{equ:meanTT*2}, and \eqref{eqn:lem4Holder2} to obtain the last inequality in \eqref{equ:meanT2}. The proof is thus completed.
\end{proof}

\begin{proof}[Proof of Lemma \ref{lemma:cnvrg_GMLE}]
The proof of Lemma \ref{lemma:cnvrg_max_sum} has shown that with probability at least \eqref{equ:prob_T}, the events of \eqref{equ:cond_meanTT*}--\eqref{equ:meanT2} hold. It is thus sufficient to show that \eqref{equ:summation} is upper bounded by $C\nu^2_1\left\{\psi^2(r) + \tm^{-q}\right\} (1+\delta^2_1)+ C\nu_1\delta_2\left(1+\delta_1\right)$ under the events of \eqref{equ:cond_meanTT*}--\eqref{equ:meanT2}.
To show this, 
we divide \eqref{equ:summation} into four components and examine their upper bounds under the events of \eqref{equ:cond_meanTT*}--\eqref{equ:meanT2} separately. 

To begin with, it can be shown
	\begin{align*}
	    &\left|	\frac{1}{\tm}\sumstSm\log f_{\wtG_{\tm,r}}\{T_{1}(s;r)\}
	    -
	    \frac{1}{\tm}\sumstSm\log f_{\wtG_{\tm,r}}\{\whT_{1}(s)\} \right|  \\
	    & \quad = 
	   \left|\frac{1}{\tm}\sumstSm\log\left[ \frac{\sumitl \wtpi_i\phi\{T_{1}(s;r)-\wtv_i\}}{\sumitl \wtpi_i\phi\{\whT_{1}(s)-\wtv_i\}}\right]\right| \\
	    & \quad = 
	    \Bigg|\frac{1}{\tm}\sumstSm 
	    \log\Bigg( 
	    e^{-\left\{T^2_{1}(s;r)-\whT^2_{1}(s)\right\} \big/2} \\ 
	    & \qquad \qquad \qquad \qquad \qquad \times \frac{\sumitl \wtpi_i
	    \exp\left\{\whT_{1}(s)\wtv_i-\whv_{i}^2/2\right\}
		\exp\left[ \left\{T_{1}(s;r)-\whT_{1}(s)\right\}\wtv_i\right]}
	    {\sumitl \wtpi_i\exp\left\{\whT_{1}(s)\wtv_i -\wtv_i^2/2\right\}}
	    \Bigg)\Bigg| \\
	    & \quad \leq
	    \left|\frac{1}{\tm}\sumstSm 
	    \left\{\frac{\left|\whT^2_{1}(s)-T_{1}^2(s;r)\right|}{2}
	    +
	    \left|\whT_{1}(s)-T_{1}(s;r)\right|\max_{i=1,\cdots,\wtl}\left|\wtv_i\right|
	    \right\}
	    \right| \\
	    & \quad \leq 
	    \left|\frac{1}{\tm}\sumstSm 
	    \left\{\frac{\left|\whT_{1}^2(s)-T_{1}^2(s;r)\right|}{2}
	    +
	    \left|\whT_{1}(s)-T_{1}(s;r)\right|\supstSm\left|T_{1}(s;r)\right|
	    \right\}
	    \right| \\
	    & \quad \leq 
	    \frac{\supstSm\left|\whtau^2(s)/\tausq_r(s)-1\right|}{2\left\{1-\supstSm\left|\whtau^2(s)/\tausq_r(s)-1\right|\right\}}
	    \left\{
		\frac{1}{\tm}\sumstSm T_{1}^2(s;r)
		+
		\supstSm\left|T_{1}(s;r)\right|
		\left|\frac{1}{\tm}\sumstSm T_{1}(s;r)\right|
	    \right\},
	\end{align*}
	where the second inequality follows from the fact that the support of $\widetilde{G} _{\tm}$ is within the range of $\big\{T_1(s;r):s\in \tScal_m \big\}$. 
	Plugging in \eqref{equ:maxT}, \eqref{equ:meanT}, and \eqref{equ:meanT2}, 
 we observe that
	$$\begin{aligned}
	    &\left|\frac{1}{\tm}\sumstSm T_{1}^2(s;r)\right|
		+
		\supstSm\left|T_{1}(s;r)\right|
		\left|\frac{1}{\tm}\sumstSm T_{1}(s;r)\right|\\
		& \qquad \leq
		\nu_1^2 + \nu_1\delta_2(2+3\delta_1) + \nu_1^2\delta_1\left(2+\delta_1\right)
		+
		\nu_1\left(1+2\delta_1\right)
		\left\{\nu_1\left(1+\delta_1\right)+\delta_2\right\}\\
		& \qquad =
		2\nu_1^2+\nu_1\delta_2(3+5\delta_1)+\nu_1^2\delta_1(5+3\delta_1).
	\end{aligned}$$
	Combining the above with \eqref{equ:fractau}, we immediately obtain that the first component of the target difference \eqref{equ:summation} is bound by
	$$
	    C\nu_1\left\{\psi^2(r)+\tm^{-q}\right\}
        \left\{ 
         2\nu_1+\delta_2(3+5\delta_1)+\nu_1\delta_1(5+3\delta_1)
        \right\}.
	$$
	Secondly, we have
	\begin{align*}
	&\left|\frac{1}{\tm}\sumstSm\log f_{\wtG^*_{\tm,r}}\{T_{1}(s;r)\}
	-
	\frac{1}{\tm}\sumstSm\log f_{\wtG^*_{\tm,r}}\{T^*_{1}(s;r)\}
	\right| \\
	& \quad = 
	\left|\frac{1}{\tm}\sumstSm\log\left[\frac{\sum_{i=1}^{\wtl^*} \wtpi^*_i\phi\{T_{1}(s;r)-\wtv^*_i\}}{\sum_{i=1}^{\wtl^*}  \wtpi^*_i\phi\{T^*_{1}(s;r)-\wtv^*_i\}}\right]\right| \\
	& \quad =
	\Bigg|\frac{1}{\tm}\sumstSm 
	\log\Bigg(
	e^{-\left[T_{1}^2(s;r)-\left\{T_{1}^*(s;r)\right\}^2\right] 
	\big/2}  \\
	& \quad \qquad \qquad \qquad \qquad \times  \frac{\sum_{i=1}^{\wtl^*}  \wtpi^*_i
		\exp\left\{T^*_{1}(s;r)\wtv^*_i-\left(\wtv_{i}^*\right)^2/2\right\}
		\exp\left[ \left\{T_{1}(s;r)-T^*_{1}(s;r)\right\}\wtv^*_i\right]}
	{\sum_{i=1}^{\wtl^*}  \wtpi^*_i\exp\left\{T^*_{1}(s;r)\wtv^*_i -\left(\wtv_i^*\right)^2/2\right\}}
	\Bigg)\Bigg| \\
	& \quad \leq
	\left|\frac{1}{\tm}\sumstSm 
	\frac{\left\{T^*_{1}(s;r)\right\}^2-T_{1}^2(s;r)}{2}\right|
	+
	\frac{1}{\tm}\sumstSm
	\left|T_{1}(s;r)-T^*_{1}(s;r)\right|\max_{i=1,\cdots,\wtl^*}\left|\wtv^*_i\right|\\
	& \quad \leq
	\left|\frac{1}{\tm}\sumstSm 
	\frac{\left\{T^*_{1}(s;r)\right\}^2-T^2_{1}(s;r)}{2}\right|
	+
	\frac{1}{\tm}\sumstSm
	\left|T_{1}(s;r)-T^*_{1}(s;r)\right|\supstSm\left|T^*_{1}(s;r)\right|,
	\end{align*}
	where the second inequality follows the fact that the support of $\widetilde{G}^*_{\tm,r}$ is within the range of $\big\{T^*_1(s;r):s\in \tScal_m\big\}$. Using \eqref{equ:cond_meanTT*}, \eqref{equ:maxT*}, and \eqref{equ:meanTT*2}, 
	we further obtain that $\nu_1\delta_2\left(3+4\delta_1\right)$ is an upper bound for the above difference. 
	As for the third component 
	of the target \eqref{equ:summation}, we have
	\begin{align*}
    	&\left|\frac{1}{\tm}\sumstSm\log f_{\whG_{\tm}}\{\whT_{1}(s)\}
    	-
    	\frac{1}{\tm}\sumstSm\log f_{\whG_{\tm}}\{T_{1}(s;r)\}
	    \right|\\
	    & \quad \leq
	    \frac{\supstSm\left|\whtau^2(s)/\tausq_r(s)-1\right|}{2\left\{1-\supstSm\left|\whtau^2(s)/\tausq_r(s)-1\right|\right\}}
	    \left\{
	    \left|\frac{1}{\tm}\sumstSm T_{1}^2(s;r)\right|
	    +
	    \supstSm\left|
	    \whT_{1}(s)\right|
	    \left|\frac{1}{\tm}\sumstSm T_{1}(s;r)\right|\right\} \\
	    & \quad \leq
	    \frac{\supstSm\left|\whtau^2(s)/\tausq_r(s)-1\right|}{2\left\{1-\supstSm\left|\whtau^2(s)/\tausq_r(s)-1\right|\right\}} \\
        & \qquad \qquad \qquad \qquad \times \left\{
	    \left|\frac{1}{\tm}\sumstSm T_{1}^2(s;r)\right|
	    +
	    \frac{\supstSm\left|
	    T_{1}(s;r)\right|}{1-\supstSm\left|\whtau^2(s)/\tausq_r(s)-1\right|}
	    \left|\frac{1}{\tm}\sumstSm T_{1}(s;r)\right|\right\} \\
    	& \quad \leq 
	    \frac{\supstSm\left|\whtau^2(s)/\tausq_r(s)-1\right|}{2\left\{1-\supstSm\left|\whtau^2(s)/\tausq_r(s)-1\right|\right\}} \\
         & \qquad \qquad \qquad \qquad \times \left\{
	    \left|\frac{1}{\tm}\sumstSm T_{1}^2(s;r)\right|
	    +C\supstSm\left|
	    T_{1}(s;r)\right|
	    \left|\frac{1}{\tm}\sumstSm T_{1}(s;r)\right|\right\},
	\end{align*}
	where the last inequality holds for the same reason as for \eqref{equ:fractau}.
	Combining \eqref{equ:fractau}, \eqref{equ:maxT}, \eqref{equ:meanT}, and \eqref{equ:meanT2}, we get the upper bound
	$$\begin{aligned}
	&C
	\left\{\psi^2(r)+\tm^{-q}\right\}
    \left\{
	\nu_1^2 + \nu_1\delta_2(2+3\delta_1) + \nu_1^2\delta_1\left(2+\delta_1\right)
	+C\nu_1(1+2\delta_1) \left\{\nu_1(1+\delta_1)+\delta_2\right\}
    \right\}\\
    & \; =
	C\left\{\psi^2(r)+\tm^{-q}\right\}
	\left[
	\nu_1^2\left(1+C\right)
	+
	\nu_1\delta_2\left\{2+C+\delta_1\left(3+2C\right)\right\}
	+
	\nu_1^2\delta_1\left\{2+3C+\delta_1\left(1+2C\right)\right\}
    \right]\\
    & \; \leq
    C\nu_1
	\left\{\psi^2(r)+ \tm^{-q}\right\}
	\left\{
	\nu_1
	+
	\delta_2\left(1+\delta_1\right)
	+
	\nu_1\delta_1\left(1+\delta_1\right)
    \right\}.
	\end{aligned}$$
	For the fourth component, we obtain
	\begin{align*}
	    &\left|\frac{1}{\tm}\sumstSm\log f_{\whG_{\tm}}\{T^*_{1}(s;r)\}
        -
    	\frac{1}{\tm}\sumstSm\log f_{\whG_{\tm}}\{T_{1}(s;r)\} \right|\\
	    &\quad \leq
	    \left|\frac{1}{\tm}\sumstSm 
	    \frac{\left\{T_{1}^*(s;r)\right\}^2-T_{1}^2(s;r)}{2}\right|
	    + 
	    \frac{1}{\tm}\sumstSm
	    \left|T_{1}(s;r)-T^*_{1}(s;r)\right|\supstSm\left|\whT_{1}(s)\right|\\
	    & \quad \leq 
	    \left|\frac{1}{\tm}\sumstSm
	    \frac{\left\{T_{1}^*(s;r)\right\}^2-T_{1}^2(s;r)}{2}\right|
	    +
	    \frac{\supstSm\left|
    	T_{1}(s;r)\right|}{1-\supstSm\left|\whtau^2(s)/\tausq_r(s)-1\right|} \\
    	& \qquad \qquad \qquad \qquad \qquad \qquad \qquad \qquad \qquad \qquad \times \left\{
    	\frac{1}{\tm}\sumstSm
	    \left|T_{1}(s;r)-T^*_{1}(s;r)\right|\right\}\\
	    & \quad \leq 
	    \left|\frac{1}{\tm}\sumstSm
	    \frac{\left\{T_{1}^*(s;r)\right\}^2-T_{1}^2(s;r)}{2}\right|
	    +
	    C\supstSm\left|
	    T_{1}(s;r)\right|\left\{
	    \frac{1}{\tm}\sumstSm
	    \left|T_{1}(s;r)-T^*_{1}(s;r)\right|\right\}.
	\end{align*}
	Similarly, due to \eqref{equ:cond_meanTT*}, \eqref{equ:maxT}, and \eqref{equ:meanTT*2}, the upper bound of the above is
	$$\begin{aligned}
	\nu_1\delta_2\left(2+3\delta_1\right)+C \nu_1\delta_2 \left(1+2\delta_1\right)
	\leq
	C\nu_1\delta_2\left(1+\delta_1\right).
	\end{aligned}$$
Finally, combining the four components with the triangular inequality, we get the upper bound
$$\begin{aligned}
    &	\frac{1}{\tm}\left|\sumstSm\log\left[
	\frac{f_{\wtG^*_{\tm,r}}\{T_{1}(s;r)\}}{f_{\wtG^*_{\tm,r}}\{T^*_{1}(s;r)\}}
	\frac{f_{\wtG_{\tm,r}}\{\whT_{1}(s)\}}{f_{\wtG_{\tm,r}}\{T_{1}(s;r)\}}
	\frac{f_{\whG_{\tm}}\{T^*_{1}(s;r)\}}{f_{\whG_{\tm}}\{T_{1}(s;r)\}}
	\frac{f_{\whG_{\tm}}\{T_{1}(s;r)\}}{f_{\whG_{\tm}}\{\whT_{1}(s)\}}
	\right] \right| \\
	& \qquad \leq
    C\nu_1^2\left\{\psi^2(r)+ \tm^{-q}\right\}
	\left\{
	1
	+
	\delta_1\left(1+\delta_1\right)
    \right\}+
	C\nu_1\delta_2\left(1+\delta_1\right)\\
	& \qquad \leq
    C\nu_1^2
	\left\{\psi^2(r)+\tm^{-q}\right\}
	\left(1+\delta^2_1\right)+
	C\nu_1\delta_2\left(1+\delta_1\right),
\end{aligned}$$
which finishes the proof.
\end{proof}



Lemma~\ref{lemma:cnvrg_emp_bayes} is a direct consequence of Theorem~1 in \cite{Zhang2009}. Interested readers are referred to \cite{Zhang2009} and references therein for more details.

\begin{proof}[Proof of Lemma \ref{lemma:f_wtG_cnvrg}]
First, notice that $d_{H}^{2}(f_{G_{\tm,r}}, f_{G_{\tm}}) \leq d_{T V}(f_{G_{\tm,r}}, f_{G_{\tm}})$, where $d_{TV}$ denotes the total variation distance
$$
    d_{TV}\left(f_{G_{\tm,r}}, f_{G_{\tm}}\right) =
    \frac{1}{2}\int \left|\int \phi(v-u) \{d G_{\tm,r}(u) -d G_{\tm}(u)\}\right|dv.
$$
Plugging the definitions of $G_{\tm,r}(u)$ and $G_{\tm}(u)$ into the above equation, $d_{TV}\left(f_{G_{\tm,r}}, f_{G_{\tm}}\right)$ can be written as
\begin{align*}
    &\frac{1}{2\tm}\int \left|\sumstSm\int \phi(v-u) [d \boldsymbol{1}\{\xi(s)/\zeta_r(s)\leq u\} -d \boldsymbol{1}\{\xi(s)\leq u\}]\right|dv\\
    & \qquad =
    \frac{1}{2\tm}\int \left|\sumstSm \phi\{v-\xi(s)/\zeta_r(s)\} -\phi\{v-\xi(s)\}\right|dv\\
    & \qquad \leq
    \frac{1}{2\tm}\sumstSm \int \left|\phi\{v-\xi(s)/\zeta_r(s)\} -\phi\{v-\xi(s)\}\right|dv.
\end{align*}
Note that $\phi\{v-\xi(s)/\zeta_r(s)\}$ and $\phi\{v-\xi(s)\}$ are symmetric about $x=\xi(s)\{\zeta_r(s)+1\}/2\zeta_r(s)$. The above integral is equivalent to
\begin{align*}
    &\frac{1}{\tm}\sumstSm  \left|\int^\infty_{\xi(s)\{1+\zeta_r(s)\}/2\zeta_r(s)} [\phi\{v-\xi(s)/\zeta_r(s)\} -\phi\{v-\xi(s)\}]dv\right|\\
    & \qquad =
    \frac{1}{\tm}\sumstSm  \big|\Phi[\xi(s)\{1-\zeta_r(s)\}/2\zeta_r(s)] -\Phi[\xi(s)\{\zeta_r(s)-1\}/2\zeta_r(s)] \big|\\
    & \qquad =\frac{1}{\tm}\sumstSm  \left|\phi(u(s))\right|\left|\xi(s)\{1-\zeta_r(s)\}/\zeta_r(s)\right|,
\end{align*}
where $u(s)$ is some value between $\xi(s)\{1-\zeta_r(s)\}/2\zeta_r(s)$ and $\xi(s)\{\zeta_r(s)-1\}/2\zeta_r(s)$ due to the mean value theorem. Finally, by Lemma~\ref{lemma:cnvrg_tau:a}, \eqref{equ:T1nu1}, and the uniform boundedness of $\phi(x)$, $d_{TV}\left(f_{G_{\tm,r}}, f_{G_{\tm}}\right)$ is upper bounded by
$C\nu_1\psi^2(r)$. Taking the square root, we get
$$d_H(f_{G_{\tm,r}}, f_{G_{\tm}})\leq d^{1/2}_{TV}\left(f_{G_{\tm,r}}, f_{G_{\tm}}\right) \leq  C\nu_1^{1/2}\psi(r).$$

Assumption~\ref{ass:emp_bayes} states that $d_{H}(f_{G_{\tm}},f_{G_0})\leq\delta_3$ for any $\delta_3>0$ with probability tending to one. 
The desired result follows from the triangular inequality.
\end{proof}


\begin{proof}[Proof of Lemma \ref{lemma:cnvrg_emp_bayes_cp}]
Due to the triangular inequality, the Hellinger distance $d_H:=d_{H}\left(f_{\whG_{\tm}}, f_{G_{0}}\right)$ is bounded by the summation of the following two components,
$$d_{H,1}:=d_{H}\left(f_{\whG_{\tm}}, f_{G_{\tm,r}}\right)\quad \text{and}\quad d_{H,2}:=d_{H}\left(f_{G_{\tm,r}}, f_{G_{0}}\right)$$ 
for any $r>0$. Thus, for an arbitrary $\epsilon>0$, it shows that
\begin{align}\label{eqn:lem8Decom}
	\Pb\left(d_{H} \geq \varepsilon\right) 
	&\leq \Pb\left(d_{H,1} \geq \epsilon/2\right)+ \Pb\left(d_{H,2} \geq\epsilon/2\right)  \nonumber \\
	&\leq \Pb\left(d_{H,1} \geq t c_{\tm}\right)+ \Pb\left\{d_{H,2} \geq C\nu_1\phi^2(r)+\epsilon/4\right\},
\end{align}
where the first inequality is because of the triangular inequality and the second inequality is achieved once (\rNum{1}) choosing $r$ such that $C\nu_1\phi^2(r)\leq \epsilon/4$; (\rNum{2}) $\tm$ is sufficiently large that induces arbitrarily small $c_{\tm}$ with fixed $b>0$ defined at \eqref{equ:c_tm} in Lemma~\ref{lemma:cnvrg_emp_bayes}. The second term of the RHS in \eqref{eqn:lem8Decom} tends to zero by setting $\delta_3=\epsilon/4$ in Lemma~\ref{lemma:f_wtG_cnvrg}.
As for the first term of the RHS in \eqref{eqn:lem8Decom}, we have
\begin{align}\label{equ:cnvrg_dH}
	\Pb\left(d_{H,1} \geq t c_{\tm}\right)
	&=
	\Pb\left(d_{H,1} \geq t c_{\tm},
	\frac{1}{\tm}\sumstSm\log\left[\frac{f_{\whG_{\tm}}\left\{T^*_{1}(s;r)\right\}}{f_{G_{\tm,r}}\left\{T^*_{1}(s;r)\right\} }\right] \geq -2 t^{2} c_{\tm}^{2} / 15 
	\right) \nonumber \\
	& \qquad \qquad +
	\Pb\left(
	\frac{1}{\tm}\sumstSm\log\left[\frac{f_{\whG_{\tm}}\left\{T^*_{1}(s;r)\right\}}{f_{G_{\tm,r}}\left\{T^*_{1}(s;r)\right\}}\right] <-2 t^{2} c_{\tm}^{2} / 15 
	\right) \nonumber\\
	& \leq
	C\tm^{-2} 
	+
	\Pb\left(
	\frac{1}{\tm}\sumstSm\log\left[\frac{f_{\whG_{\tm}}\left\{T^*_{1}(s;r)\right\}}{f_{G_{\tm,r}}\left\{T^*_{1}(s;r)\right\}}\right] <-2 t^{2} c_{\tm}^{2} / 15 
	\right)
\end{align}
when $t$ is large enough due to Lemma \ref{lemma:cnvrg_emp_bayes}. As for the second term of the RHS in \eqref{equ:cnvrg_dH},
notice that
\begin{align}
	&\frac{1}{\tm}\sumstSm\log\left[\frac{f_{\whG_{\tm}}\{T^*_{1}(s;r)\}}{f_{G_{\tm,r}}\{T_{1}^*(s;r)\}}\right] \nonumber\\
	& \quad = \frac{1}{\tm}\sumstSm\log\Bigg[
	\frac{f_{\wtG^*_{\tm,r}}\{T^*_{1}(s;r)\}}{f_{G_{\tm,r}}\{T^*_{1}(s;r)\}}
	\frac{f_{\wtG^*_{\tm,r}}\{T_{1}(s;r)\}}{f_{\wtG^*_{\tm,r}}\{T^*_{1}(s;r)\}}
	\frac{f_{\wtG_{\tm,r}}\{T_{1}(s;r)\}}{f_{\wtG^*_{\tm,r}}\{T_{1}(s;r)\}}  \nonumber\\
	& \qquad \qquad \qquad \qquad \qquad \times \frac{f_{\wtG_{\tm,r}}\{\whT_{1}(s)\}}{f_{\wtG_{\tm,r}}\{T_{1}(s;r)\}}
	\frac{f_{\whG_{\tm}}\{\whT_{1}(s)\}}{f_{\wtG_{\tm,r}}\{\whT_{1}(s)\}}
	\frac{f_{\whG_{\tm}}\{T^*_{1}(s;r)\}}{f_{\whG_{\tm}}\{\whT_{1}(s)\}}
	\Bigg] \nonumber\\
	& \quad \geq 
	\frac{1}{\tm}\sumstSm\log \left[
	\frac{f_{\wtG^*_{\tm,r}}\{T_{1}(s;r)\}}{f_{\wtG^*_{\tm,r}}\{T^*_{1}(s;r)\}}
	\frac{f_{\wtG_{\tm,r}}\{\whT_{1}(s)\}}{f_{\wtG_{\tm,r}}\{T_{1}(s;r)\}}
	\frac{f_{\whG_{\tm}}\{T^*_{1}(s;r)\}}{f_{\whG_{\tm}}\{T_{1}(s;r)\}}
	\frac{f_{\whG_{\tm}}\{T_{1}(s;r)\}}{f_{\whG_{\tm}}\{\whT_{1}(s)\}}
	\right], \label{eqn:lem8Prod}
\end{align}
where the last inequality is due to the definitions of $\wtG^*_{\tm,r}$, $\wtG_{\tm,r}$, and $\whG_{\tm}$.
In addition, \eqref{equ:c_tm} implies
\begin{equation}\label{eqn:lem8Upper}
    2t^2c_{\tm}^2/15=2t^2\tm^{-b/(1+b)} \left(\log \tm\right)^{(2+3b)/(2+2b)}\left(1\vee \nu_1\right)^{b/(1+b)}/15.
\end{equation}
Plugging \eqref{eqn:lem8Prod} and \eqref{eqn:lem8Upper} into the second probability of the RHS of \eqref{equ:cnvrg_dH} yields
$$
	\Pb\left(
	\left|
	\frac{1}{\tm}\sumstSm\log \left[
	\frac{f_{\wtG^*_{\tm,r}}\{T_{1}(s;r)\}}{f_{\wtG^*_{\tm,r}}\{T^*_{1}(s;r)\}}
	\frac{f_{\wtG_{\tm,r}}\{\whT_{1}(s)\}}{f_{\wtG_{\tm,r}}\{T_{1}(s;r)\}}
	\frac{f_{\whG_{\tm}}\{T^*_{1}(s;r)\}}{f_{\whG_{\tm}}\{T_{1}(s;r)\}}
	\frac{f_{\whG_{\tm}}\{T_{1}(s;r)\}}{f_{\whG_{\tm}}\{\whT_{1}(s)\}}
	\right]  \right|
	> C\tm^{-a}
	\right),
$$
for some $a$ such that $b/(1+b)<a<\min(q,1/p)$ when $\tm$ is large enough.

Next, we shall show that the above probability goes to zero as $m\rightarrow\infty$ when the subset $\tScal_m$ used for NPEB fulfills Assumption~\ref{ass:num_tSm} of the main paper. 
Using Lemma~\ref{lemma:cnvrg_GMLE}, we only need to show for $\delta>0$,
$$\tm\exp(-\nu_1^2\delta_1^2/2)\leq\delta, \quad C\tm\psi^p(r)/\nu_1^p\delta_1^p\leq\delta,\quad \text{and}\quad C\psi^p(r)/\delta_2^p \leq \delta,$$
as well as 
$$C\nu_1^2\left\{\psi^2(r) + \tm^{-q}\right\}
	\left(1+\delta^2_1\right)=O(\tm^{-a}) \quad \text{and}\quad C\nu_1\delta_2\left(1+\delta_1\right)=O(\tm^{-a})$$
for some $\delta_1,\delta_2>0$. 
Note that $\tm\exp(-\nu_1^2\delta_1^2/2)\leq \delta$ is equivalent to $\delta_1\geq \sqrt{2\ln(\tm/\delta)}/\nu_1$. We set
$$\delta_1=\frac{1}{\nu_1}\sqrt{2\ln(\tm/\delta)}$$
in the following calculations.
For $\delta_2$, $C\nu_1\delta_2\left(1+\delta_1\right)\leq\tm^{-a}$ implies $\delta_2\leq C\tm^{-a}/\nu_1(1+\delta_1)$. 
With the choice of $\delta_1$, we set
$$\delta_2=\frac{C\tm^{-a}}{\nu_1+\sqrt{2\ln(\tm/\delta)}}.$$
Similarly, with the above choice of $\delta_1$, the constraint $C\nu_1^2 \tm^{-q}\left(1+\delta^2_1\right)\leq\tm^{-a}$ becomes
$$\tm^{-q}\leq \frac{C\tm^{-a}}{\nu_1^2+2\ln(\tm/\delta)},$$
which can be achieved according to the upper and lower bounds of $a$. 
As
$C\tm\psi^p(r)/\nu_1^p\delta_1^p\leq \delta$, $C\psi^p(r)/\delta_2^p \leq\delta$, and $C\nu_1^2\psi^2(r) \left(1+\delta^2_1\right)\leq\tm^{-a}$, we require
\begin{equation}\label{equ:phi_r_restri}
\begin{aligned}
    r^{-\lambda}
     \asymp \psi(r) \leq \min 
    \bigg[C\delta\tm^{-1/p}\sqrt{\ln(\tm/\delta)} , ~
    &C\delta^{1/p}\tm^{-a} \big/ \Big\{\nu_1+\sqrt{2\ln(\tm/\delta)}\Big\}, \\
    &  \tm^{-a/2} \big/ \sqrt{\nu_1^2+2\ln(\tm/\delta)}\bigg]
\end{aligned}
\end{equation}
after plugging in $\delta_1$ and $\delta_2$ with Assumption~\ref{ass:NED}. 
Again, according to the upper bounds of $a$, the RHS of \eqref{equ:phi_r_restri} is dominated by its first term when $\tm$ is large enough. 
Thus \eqref{equ:phi_r_restri} is a consequence of Assumption~\ref{ass:num_tSm} when we take $r=\widetilde{\Delta}_{l,m}/2$, which completes the proof. 
\end{proof}

In the proof of Lemma~\ref{lemma:cnvrg_emp_bayes_cp}, \eqref{equ:phi_r_restri} requires $r$ increases as $\tm$ increases for non-degenerating $\psi(r)$. A special case is m-dependent, that is $\psi(r)=0$, whenever $r>R$ for some $R>0$. According to \eqref{equ:phi_r_restri}, the choice of $r$ is free of $\tm$ and only depends on $R$ in this special case.

{
\begin{proof}[Proof of Lemma~\ref{lemma:cnvrg_true}]
    We prove the first convergence and the second can be proved similarly. For arbitrary $\epsilon>0$, we can choose $m_0$ large enough such that 
    \begin{align}
&\Pb\left(\left| \frac{V_m\left(t_1, t_2\right)}{m_0} -K_0(t_1,t_2)\right|>\epsilon\right)\nonumber\\
\leq &
\Pb\left(\left|\frac{V_m\left(t_1, t_2\right)}{m_0}- \Eb\left\{\frac{V_m\left(t_1, t_2\right)}{m_0}\right\} \right|>\epsilon/2\right) + \Pb\left(\left|\Eb\left\{\frac{V_m\left(t_1, t_2\right)}{m_0}\right\} -K_0(t_1,t_2)\right|>\epsilon/2\right)\nonumber\\
\leq &
\frac{4\var\left\{V_m\left(t_1, t_2\right)\right\}}{\epsilon^2 m_0^2},
\label{equ:probVm}    \end{align}
where the first inequality is due to the triangular inequality, the first probability in line 2 is bound by Chebyshev inequality and the second probability in line 2 holds by taking large enough $m_0$ and applying Assumption~\ref{ass:cnvrg_true} of the main paper. We just need to show that line 3 of the above formula can be arbitrarily small by taking sufficient large $m_0$. Note that the numerator can be expressed as follows 
\begin{equation}\label{equ:cov_val}
\begin{aligned}
&\sum_{s\in\Scal_{0,m}}\left[\sum_{\stackrel{s^\prime\in\Scal_{0,m}}{\dist(s,s^\prime)\leq \widetilde{r}}}\cov\left\{1\left\{T_1(s) \geq t_1, T_2(s) \geq t_2\right\},1\left\{T_1(s^\prime) \geq t_1, T_2(s^\prime) \geq t_2\right\}\right\}\right.\\
&\qquad\qquad+\left.\sum_{\stackrel{s^\prime\in\Scal_{0,m}}{\dist(s,s^\prime)> \widetilde{r}}}\cov\left\{1\left\{T_1(s) \geq t_1, T_2(s) \geq t_2\right\},1\left\{T_1(s^\prime) \geq t_1, T_2(s^\prime) \geq t_2\right\}\right\}\right].
\end{aligned}
\end{equation}

The idea is to choose an appropriate $\widetilde{r}$ so that the first summation does not have many terms and the covariances in the second summation are small enough. 
For the first part, we adopt the same idea in Section~\ref{sec:relation_tm_r}. Fixing $s\in\Scal_{0,m}$, we construct $K$-dimensional non-overlapping cubes centered at the locations $\{s^\prime\in\Scal_{0,m}:\dist(s,s^\prime)\leq \widetilde{r}\}$ with the length $\Delta_l/2$, and one big $K$-dimensional cube centered at location $s$ with the length $2\widetilde{r}+\Delta_l/2$. Since the small cubes are covered by the big cube, we have 
\begin{equation}\label{equ:comp_ms}
    m(s)\leq (4\widetilde{r}/\Delta_l+1)^K,\quad \text{because} \quad m(s)(\Delta_l/2)^K\leq (2\widetilde{r}+\Delta_l/2)^K,
\end{equation}
where $m(s)=\left|\{s^\prime\in\Scal_{0,m}:\dist(s,s^\prime)\leq \widetilde{r}\}\right|$. 

For the second part, we are going to replace $T_1(s)$ and $T_2(s)$ with their normalized conditional statistics $T_1^*(s;r)$ and $T_2^*(s;r)$. 
Similar to the discussion of $T^*_1(s;r)$, $ \big(T_1^*(s;r),T_2^*(s;r)\big)$ and $\big(T_1^*(v;r),T_1^*(v;r)\big)$ are independent if $s,s^\prime\in\Scal$ satisfy \eqref{equ:req_ind} of the. W.l.o.g, we can define 
\begin{equation}
\label{equ:r_tr}    r=\widetilde{r}/2- N_{nei}\Delta_u,
\end{equation}
 where $\widetilde{r}$ large enough so that $r>0$. Similar to the derivation of \eqref{equ:NED_T1_prop} and \eqref{equ:NED_T1}, for any $\delta>0$, we have 
 \begin{equation}\label{equ:NED_T2}
  \Pb\left\{\left|T_2(s) -{T}_2^*(s;r) \right|>\delta\right\}
 \leq
 \frac{1}{\delta^p} \left\|T_2(s)-{T}_2^*(s;r) \right\|_p^p \leq \frac{\breve{\eta}^p(s;r)\psi^p(r)}{\delta^p},
 \end{equation}
 where $\breve{\eta}(s;r)= {d_m(s)}/\{\tau(s)\breve{\zeta}_r(s)\}$. Then, we can obtain that
$$
\begin{aligned}
    &\Pb\left\{T_1(s) \geq t_1, T_2(s) \geq t_2,T_1(s^\prime) \geq t_1, T_2(s^\prime) \geq t_2\right\}\\
\leq&
\Pb\left\{T_1(s) \geq t_1, T_2(s) \geq t_2,T_1(s^\prime) \geq t_1, T_2(s^\prime) \geq t_2, \right.\\
&\left.\left|T_1(s)- T_1^*(s;r)\right|\leq\delta
, \left|T_1(s^\prime)- T_1^*(s^ \prime;r)\right|\leq\delta
, \left|T_2(s)- T_2^*(s;r)\right|\leq\delta
, \left|T_2(s^\prime)- T_2^*(s^\prime;r)\right|\leq\delta\right\}\\
+&\Pb\left\{\left|T_1(s)- T_1^*(s;r)\right|>\delta\right\}
+\Pb\left\{\left|T_1(s^\prime)- T_1^*(s^\prime;r)\right|>\delta\right\}\\
+&\Pb\left\{\left|T_2(s)- T_2^*(s;r)\right|>\delta\right\}
+\Pb\left\{\left|T_2(s^\prime)- T_2^*(s^\prime;r)\right|>\delta\right\}\\
\leq&\Pb\left\{T_1^*(s;r) \geq t_1-\delta, T_2^*(s;r) \geq t_2-\delta\right\}\Pb\left\{T_1^*(s^\prime;r) \geq t_1-\delta, T_2^*(s^\prime;r)\geq t_2-\delta\right\}+
C{\psi^p(r)}/{\delta^p},
\end{aligned}
$$
where the last inequality holds by \eqref{equ:NED_T1}, \eqref{equ:NED_T2}, Lemma~\ref{lemma:cnvrg_tau:c}, and Remark~\ref{rmk:T2}. Similarly, we can derive that
$$
\begin{aligned}
 &\Pb\left\{T_1^*(s;r) \geq t_1+\delta, T_2^*(s;r) \geq t_2+\delta\right\}\\
\leq&\Pb\left\{T_1^*(s;r) \geq t_1+\delta, T_2^*(s;r) \geq t_2+\delta,\left|T_1(s)- T_1^*(s;r)\right|\leq\delta
,\left|T_2(s)- T_2^*(s;r)\right|\leq\delta
\right\}
\\
+&\Pb\left\{\left|T_1(s)- T_1^*(s;r)\right|>\delta
\right\}+\Pb\left\{\left|T_2(s)- T_2^*(s;r)\right|>\delta
\right\}\\
\leq&
\Pb\left\{T_1(s) \geq t_1, T_2(s) \geq t_2\right\}
+
C{\psi^p(r)}/{\delta^p}.
\end{aligned}
$$
Therefore, we have 
\begin{align*}
    &
\cov\left\{1\left\{T_1(s) \geq t_1, T_2(s) \geq t_2\right\},1\left\{T_1(s^\prime) \geq t_1, T_2(s^\prime) \geq t_2\right\}\right\}
\\
=&\Pb\left\{T_1(s) \geq t_1, T_2(s) \geq t_2,T_1(s^\prime) \geq t_1, T_2(s^\prime) \geq t_2\right\}\\
&\qquad\qquad\qquad\qquad\qquad-\Pb\left\{T_1(s) \geq t_1, T_2(s) \geq t_2\right\}\Pb\left\{T_1(s^\prime) \geq t_1, T_2(s^\prime) \geq t_2\right\}\\
\leq 
&\Pb\left\{T_1^*(s;r) \geq t_1-\delta, T_2^*(s;r) \geq t_2-\delta\right\}\Pb\left\{T_1^*(s^\prime;r) \geq t_1-\delta, T_2^*(s^\prime;r)\geq t_2-\delta\right\}\\
- &
\Pb\left\{T_1^*(s;r) \geq t_1+\delta, T_2^*(s;r) \geq t_2+\delta\right\}\Pb\left\{T_1^*(s^\prime;r) \geq t_1+\delta, T_2^*(s^\prime;r)\geq t_2+\delta\right\}+
C{\psi^p(r)}/{\delta^p}\\
:=&
\epsilon_\delta + 
C{\psi^p(r)}/{\delta^p}.
\end{align*}
Similarly, the covariance can also be lower bounded by $
-\epsilon_\delta - 
C{\psi^p(r)}/{\delta^p}$. 
Combine the two parts, and we can bound \eqref{equ:cov_val} by
$$
\sum_{s\in\Scal_{0,m}}
\left[\sum_{\stackrel{s^\prime\in\Scal_{0,m}}{\dist(s,s^\prime)\leq \widetilde{r}}} 1 + 
\sum_{\stackrel{s^\prime\in\Scal_{0,m}}{\dist(s,s^\prime)>\widetilde{r}}}\left\{
\epsilon_\delta + C\psi^p(r)/\delta^p\right\}
\right]
\leq 
\sum_{s\in\Scal_{0,m}}
m(s) + m_0^2 \epsilon_\delta + Cm_0^2\psi^p(r)/\delta^p.
$$ 
Choose a moderate large $\widetilde{r}=(\Delta_l(m_0)^{a/K}-1)/4$ for some $0<a<1$. Then, the probability \eqref{equ:probVm} is bounded by 
$$
4\left[m_0^{-(1-a)} + \epsilon_\delta + C \left\{\Delta_l(m_0^{a/K}-1)/8 - N_{nei}\Delta_u\right\}^{-\lambda p}/\delta^p\right] /\epsilon^2,
$$ 
which can be arbitrarily small by choosing a $\delta>0$ such that $\epsilon_\delta$ be small enough and let $m_0$ tend to infinity. The proof is completed. 
\end{proof}
}
{
\section{Power Analysis}\label{sec:pow}
In this section, we establish the power enhancement of the 2d-SMT procedure as outlined in Theorem~\ref{rmk:power} of the main paper and explore the factors contributing to power improvement.

\subsection{Proof of Theorem~\ref{rmk:power} of the Main Paper}\label{sec:pow_proof}

\begin{proof}[Proof of Theorem~\ref{rmk:power} of the main paper]
(i) Recall that
$$
(t^{2d}_1,t^{2d}_2)=\argmax_{(t_1,t_2)\in\Fcal^{2d}_{q,\infty}} K(t_1,t_2)
\quad\text{and}\quad
t^{1d}_2=\argmax_{t_2\in\Fcal^{1d}_{q,\infty}} K(-\infty,t_2),
$$
where $\Fcal^{2d}_{q,\infty}=
\{(t_1,t_2):\FDP_{\lambda}^\infty(t_1,t_2)\leq q\}$ and $\Fcal^{1d}_{q,\infty}=
\{t_2:\FDP_{\lambda}^\infty(-\infty,t_2)\leq q\}.$ The corresponding percentages of true discoveries in the asymptotic sense are $\PTD^{1d}=K_1(-\infty,t^{1d}_2)$ and $\PTD^{2d}=K_1(t^{2d}_1,t^{2d}_2)$, respectively. 

According to the definition of $\Fcal^{1d}_{q,\infty}$ and $\Fcal^{2d}_{q,\infty}$, we have $K(t^{2d}_1,t^{2d}_2)\geq K(-\infty,t^{1d}_2)$. 
Note that $K(t_1,t_2)=\pi_0 K_0(t_1,t_2) + (1-\pi_0)K_1(t_1,t_2)$. If $K_0(t^{2d}_1,t^{2d}_2)< K_0(-\infty,t^{1d}_2)$, then we can directly conclude $\PTD^{2d}> \PTD^{1d}$. Otherwise, in the case that $K_0(t^{2d}_1,t^{2d}_2)\geq K_0(-\infty,t^{1d}_2)$, we begin by dividing the numerator and denominator of 
$\FDP_{\lambda}^\infty\left(t_{1}, t_{2}\right)$ by $K_0(t_1,t_2)$ and expressing $K(t_1,t_2)$ in terms of $K_0(t_1,t_2)$ and $K_1(t_1,t_2)$ as follows
$$
\FDP_{\lambda}^\infty\left(t_{1}, t_{2}\right)=
\frac{F(\lambda)}{\Phi(\lambda)}\frac{\left(\lim_{m\rightarrow\infty}\int \sumsSm L(t_1,t_2,x,\rho(s))d G_0(x)/m\right)/K_0(t_1,t_2)}{\pi_0 +(1-\pi_0)K_1(t_1,t_2)/K_0(t_1,t_2)}.
$$
According to Assumption~\ref{ass:cnvrg_true}, we have 
\begin{equation}\label{equ:alt_null}
\frac{\lim_{m\rightarrow\infty}\int \sumsSm L(t_1,t_2,x,\rho(s))d G_0(x)/m}{K_0(t_1,t_2)}\geq 1.
\end{equation}
Setting the first threshold as $-\infty$, we find that
$$
\lim_{m\rightarrow\infty}\int \sumsSm L(-\infty,t_2,x,\rho(s))d G_0(x)/m=K_0(-\infty,t_2)
$$
for any $t_2\in\Rb$. 
Therefore, we have 
\begin{equation}\label{equ:qFDPlim}
q=\FDP_{\lambda}^\infty\left(t^{2d}_{1}, t^{2d}_{2}\right)\geq \frac{F(\lambda)}{\Phi(\lambda)}\frac{1}{\pi_0 +(1-\pi_0)K_1(t^{2d}_{1}, t^{2d}_{2})/K_0(t^{2d}_{1}, t^{2d}_{2})}
\end{equation}
and 
$$
q=\FDP_{\lambda}^\infty\left(-\infty, t^{1d}_{2}\right)
=
\frac{F(\lambda)}{\Phi(\lambda)}\frac{1}{\pi_0 +(1-\pi_0)K_1(-\infty, t^{1d}_{2})/K_0(-\infty, t^{1d}_{2})}
$$
because $\FDP_{\lambda}^\infty(t_1,t_2)$ is continuous. Since $K_0(t^{2d}_1,t^{2d}_2)\geq K_0(-\infty,t^{1d}_2)$, we must have $\PTD^{2d}=K_1(t^{2d}_1,t^{2d}_2)\geq K_1(-\infty,t^{1d}_2)=\PTD^{1d}$. 

(ii) 
The proof is similar to that of part (i), except that \eqref{equ:qFDPlim} becomes strict inequality. Therefore, whether $K_0(t^{2d}_1,t^{2d}_2)< K_0(-\infty,t^{1d}_2)$ or $K_0(t^{2d}_1,t^{2d}_2)\geq K_0(-\infty,t^{1d}_2)$, we have $\PTD^{2d}>\PTD^{1d}$. 
\end{proof}

Theorem~\ref{rmk:power} part (ii) of the main paper suggests that if the expectations of auxiliary statistics under the alternative are generally larger than those under the null, the 2d-SMT procedure is more powerful than the 1d-SMT procedure even when they have the same number of discoveries. This also means that the power improvement of the 2d-SMT procedure is not just because of making more rejections by enlarging the searching ranges of the cutoffs but also due to the auxiliary statistics' ability to differentiate between the null and alternative hypotheses.

\subsection{An Example}\label{sec:pow_ex}
In this section, we explicitly analyze the power improvement of the 2d-SMT in a specific example. 

Denote $m=|\Scal_m|$, $m_0=|\Scal_{0,m}|$, $m_1=|\Scal_{1,m}|$, $\Scal_{0,m}=\{1,2,\cdots,m_0\}$ and $\Scal_{1,m}=\{m_0+11,m_0+12,\cdots,m_0+m_1+10\}$. 
Suppose that $\mu(s)=0$ for $s\in\Scal_{0,m}$ and $\mu(s)=\mu_1>0$ for $s\in\Scal_{1,m}$. Since $\Ncal(s)$ is the $\kappa$-nearest
neighbors, we notice $\Ncal(s)\subseteq\Scal_{0,m}$ for all $s\in\Scal_{0,m}$ and $\Ncal(s)\subseteq\Scal_{1,m}$ for all $s\in\Scal_{1,m}$ for $0<\kappa<10$. Given some $\rho_X\in(0,1)$, we let $\sigma(s)\equiv 1$, $\corr(X(s),X(v))= \rho_X$ if $|s-v|\leq\kappa$, and $\corr(X(s),X(v))= 0$ if $|s-v|>\kappa$. In Section~\ref{sec:mov}, we have shown that $T_{1}(s)=\xi(s)+V_1(s)$ and $ T_{2}(s)=\sigma^{-1}(s)\mu(s)+V_{2}(s)$, 
where $\xi(s)=\tau^{-1}(s)\sum_{v\in \mathcal{N}(s)}\mu(v)$ and
$$\left(\begin{array}{c}
V_{1}(s) \\
V_{2}(s)
\end{array}\right)\sim
\mathcal{N}\left(
\left(\begin{array}{c}
0 \\
0
\end{array}\right),
\left(\begin{array}{cc}
1 & \rho(s) \\
\rho(s)  & 1
\end{array}\right)\right)$$
\noindent with $\rho(s)= \{\sigma(s)\tau(s)\}^{-1} \sum_{v\in \mathcal{N}(s)} \corr\{\epsilon(s),\epsilon(v)\}$. We have $\xi(s)=\sigma^{-1}(s)\mu(s)=0$ for $s\in\Scal_{0,m}$, and   $\xi(s)=\mu_1/\sqrt{\rho_X+(1-\rho_X)/\kappa}\defequ\xi_1$ as well as $\sigma^{-1}(s)\mu(s)=\mu_1$ for $s\in\Scal_1$ by calculation. Furthermore, locations $s$ not on the boundary satisfy $\tau^2(s)=\kappa+(\kappa^2-\kappa)\rho_X\defequ\tau$ and $\rho(s)=\rho_X/\sqrt{\rho_X+(1-\rho_X)/\kappa}\defequ\rho$. These conditions imply the expected proportions of true and false rejections in Assumption~\ref{ass:cnvrg_true} of the main paper are $$K_0(t_1,t_2) \approx \Pb\left\{V_1(s)\geq t_1,V_2(s)\geq t_1\right\}
\ \text{and}\ K_1(t_1,t_2) \approx \Pb\left\{V_1(s)+\xi_1\geq t_1,V_2(s)+\mu_1\geq t_1\right\}$$
because only a minority of $s$ deviate from $\tau$ and $\rho$. 

We can explicitly confirm that $ \widehat{\FDP}_{\lambda,\tScal}(t_{1}, t_{2})$ in \eqref{equ:def_tfdp} of the main paper, with known parameters $(\sigma(s),\tau(s),\rho(s))$, is a conservative estimator of $\FDP(t_1,t_2)$. 
As shown in \eqref{equ:cnvrg_FDP_K}, $\FDP(t_1,t_2)$ uniformly converges to $\pi_0K_0(t_1,t_2)/K(t_1,t_2)$. In 2d-SMT, the denominator $K(t_1,t_2)$ is estimated by the proportion of discoveries. The estimation of the numerator consists of two components. The first component $\pi_0$ can be conservatively estimated by $\whpi_0$ $$\frac{\sum_{s\in\Scal} \boldsymbol{1}\{T_{2}(s)<\lambda\}}{\left|\Scal\right|\Phi(\lambda)}\approx \pi_0 + (1-\pi_0)\frac{\Pb(V_2(s)+\mu_1<\lambda)}{\Phi(\lambda)}$$ 
according to \eqref{equ:est_pi0}. 
The second component $K_0(t_1,t_2)$ is approximated by 
$$
\begin{aligned}
&\lim_{m\rightarrow\infty}\int\frac{1}{m} \sumsSm L(t_1,t_2,x,\rho(s))d G_0(x)\\
&\qquad \qquad \approx
\pi_0K_0(t_1,t_2) + (1-\pi_0) \Pb\left\{V_1(s)+\xi_1\geq t_1,V_2(s)\geq t_1\right\},
\end{aligned}
$$
which serves as a conservative estimator of $K_0(t_1,t_2)$ because $\xi_1>0$. 

We then numerically calculated the theoretical power improvement of the 2d-SMT over the 1d-SMT, defined as $(\PTD^{2d}-\PTD^{1d})/\PTD^{1d}$, and examined how it changes with the magnitudes under the alternative $\mu_1$, the null proportion $\pi_0$, the number of neighbors $\kappa$, and the correlation $\rho_X$. 
The theoretical power improvement in Figures~\ref{fig:PowImpKappa}--\ref{fig:PowImpPi0} aligned with our previous numerical simulations. First, Figure~\ref{fig:PowImpKappa} demonstrates that the power increased alongside a growing number of neighbors. This tendency was also observed in Setups I and II from Section~\ref{sec:rob_nei_num}. Second, Figures~\ref{fig:PowImpRho} and \ref{fig:PowImpPi0} illustrate the significant power enhancement in scenarios with sparse signals and weak correlation. Third, the power improvement decreased as the $\PTD^{1d}$, which varied by altering $\mu_1$, increased across all configurations in this section. The second and third observations were consistent with the simulation of all setups in Section~\ref{sec:simu}. 

\begin{figure}
\centering
    \includegraphics[width=\textwidth]{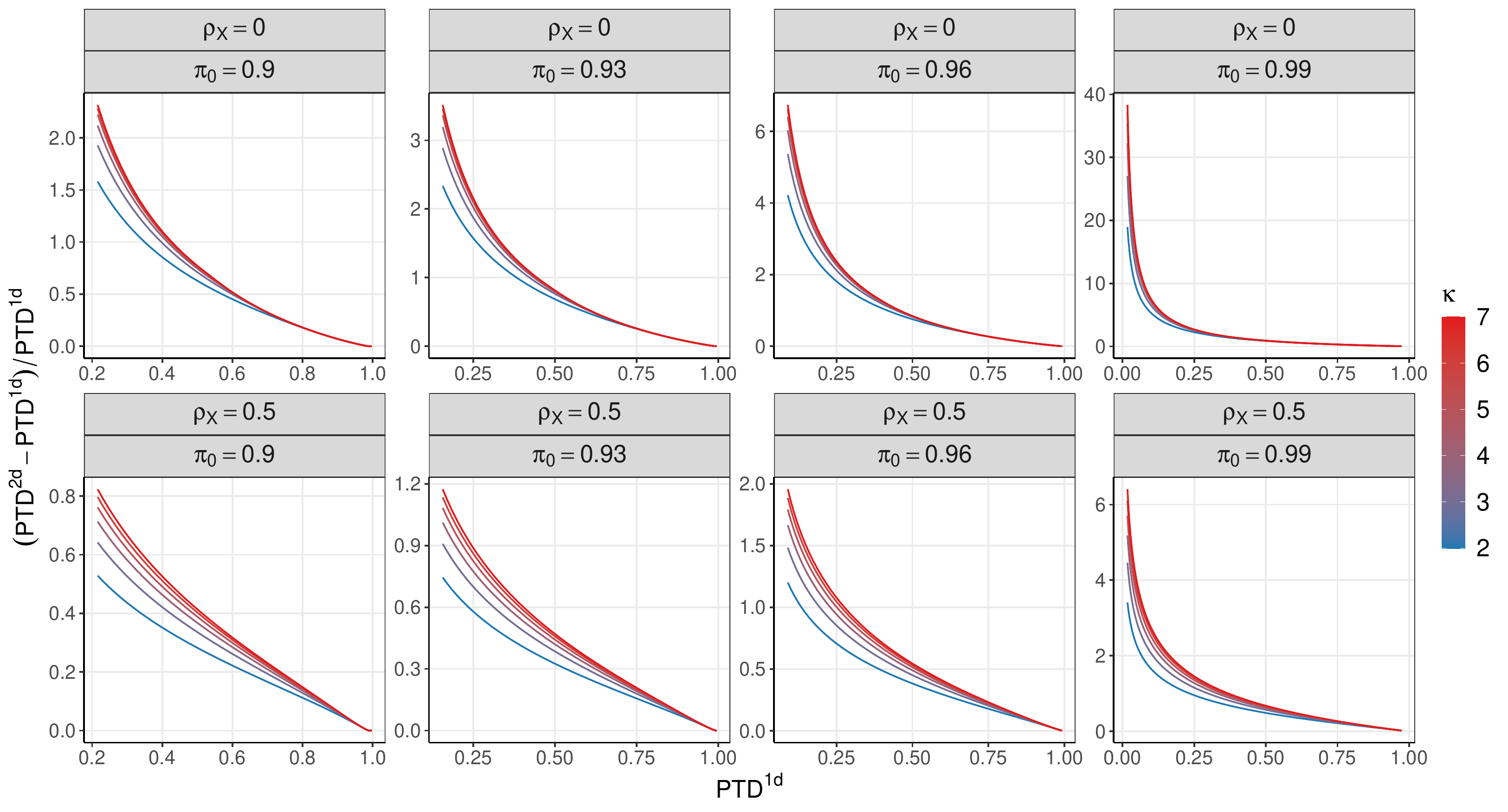}
    \caption{The theoretical power improvement of the 2d-SMT procedure with $\kappa\in\{2,3,\cdots,7\}$, $\mu_1\in[2,5]$, $\pi_0\in\{0.90,0.93,0.96,0.99\}$, $\rho\in\{0,0.5\}$.}
    \label{fig:PowImpKappa}
\end{figure}

\begin{figure}
    \centering
    \includegraphics[width=\textwidth]{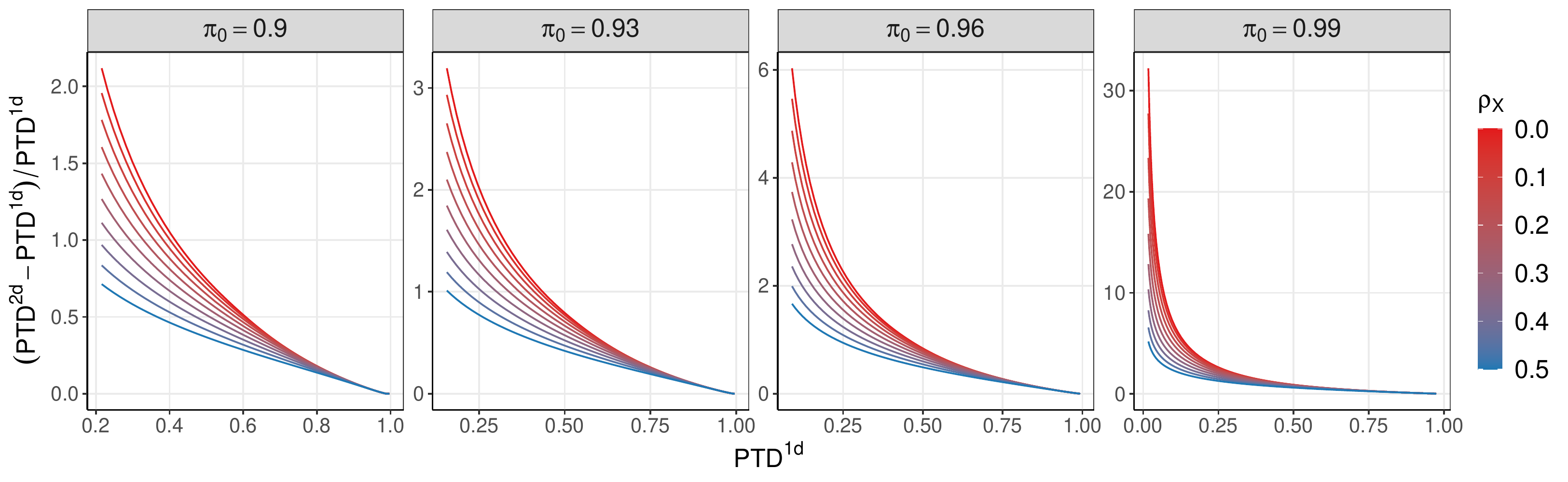}
    \caption{The theoretical power improvement of the 2d-SMT procedure with $\kappa=4$, $\mu_1\in[2,5]$, $\pi_0\in\{0.90,0.93,0.96,0.99\}$, $\rho\in[0,0.5]$.}
    \label{fig:PowImpRho}
\end{figure}

\begin{figure}
    \centering
    \includegraphics[width=0.8\textwidth]{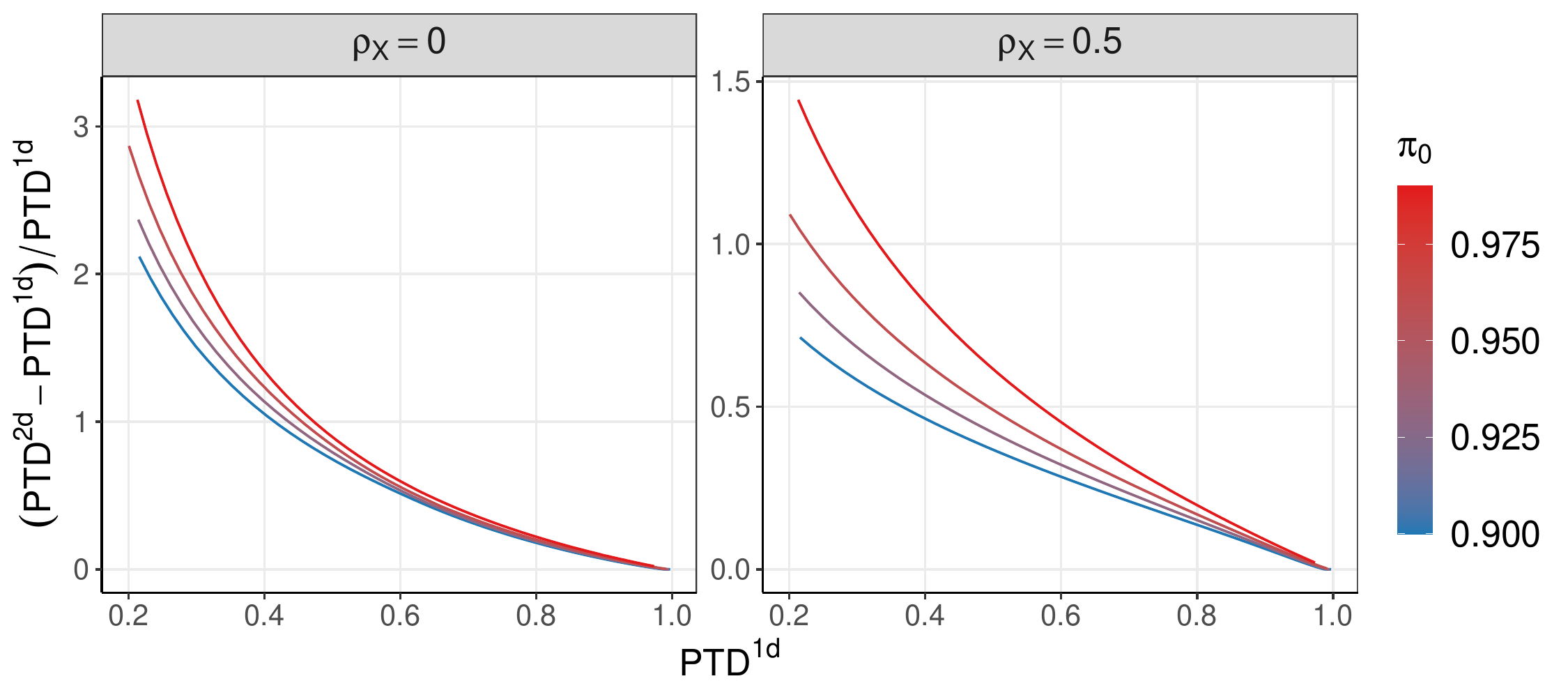}
    \caption{The theoretical power improvement of the 2d-SMT procedure with $\kappa=4$, $\mu_1\in[2,5]$, $\pi_0\in\{0.90,0.93,0.96,0.99\}$, $\rho\in\{0,0.5\}$.}
    \label{fig:PowImpPi0}
\end{figure}

}
\section{Covariance Estimation}\label{sec:cov_est}
We discuss the estimation of the covariance matrix. Suppose the error process has the decomposition
\begin{equation*}
	\epsilon(s)=\epsilon_1(s) + \epsilon_2(s),
\end{equation*}
where $\epsilon_1(s)$ is a spatial process modeling the spatial correlation and $\epsilon_2(s)$ is an independent error process, known as the nugget effect, modeling measurement error. It is common to assume that (i) $\epsilon_1(s)$ is a spatial Gaussian process with the covariance function $c(\cdot,\cdot)$; (ii) $\epsilon_2(s)$ independently follows the normal distribution with mean zero and variance $\gamma^2$ at every location. For example, a popular choice of the spatial correlation function is the Mat\'{e}rn family of stationary correlation functions, i.e.,
\begin{align*}
	\rho(s,s';\nu,\phi)=\frac{1}{2^{\nu-1}\Gamma(\nu)}\left\{\frac{2\nu^{1/2}\dist(s,s')}{\phi}\right\}^\nu J_{\nu}\left(\frac{2\nu^{1/2}\dist(s,s')}{\phi}\right), \quad \nu>0,\phi>0,    
\end{align*}
where $\Gamma$ is the Gamma function, $J_\nu (\cdot)$ is the modified Bessel function of the second kind with order $\nu$, and $\dist(\cdot,\cdot)$ denotes the Euclidean distance. Here, $\nu$ controls the degree of smoothness and $\phi$ is the range parameter.

Generally, we can specify $c(\cdot,\cdot)=\sigma^2\rho(\cdot,\cdot;\theta)$ for a correlation function $\rho$ parameterized by $\theta$, e.g., $\theta=(\nu,\phi)$ for the Mat\'{e}rn family. Recall that $\mathcal{S}_m=\{s_1,\dots,s_m\}$ and let $C_m(\theta,\sigma^2)=(\sigma^2 \rho(s_i,s_j;\theta))^{m}_{i,j=1}$. The log-likelihood function of $\mathbf{X}=(X(s_1),\dots,X(s_m))^\top$ is 
\begin{align*}
	l_m(\boldsymbol{\mu},\theta,\sigma^2,\gamma^2)&=-\frac{m}{2}\log(2\pi)-\frac{1}{2}\log \text{det}\{C_m(\theta,\sigma^2)+\gamma^2 I_m\}\\
	&\qquad -\frac{1}{2}(\mathbf{X}-\boldsymbol{\mu})^\top\{C_m(\theta,\sigma^2)+\gamma^2 I_m\}^{-1}(\mathbf{X}-\boldsymbol{\mu}), 
\end{align*}
where $\boldsymbol{\mu}=(\mu_1,\dots,\mu_m)^\top$.  
Jointly estimating $(\bmu, \theta,\sigma^2,\gamma^2)$ without special structure for $\bmu$ is challenging. 
Some constraints are typically needed to regularize the form of $\bmu$. For example, when external covariates $\Zf=(\Zf(s_1),\ldots,\Zf(s_m))^\top$ with $\Zf(s_j)=(Z_1(s_j),\ldots, Z_p(s_j))^\top$ is available, we can model the signal by $\bmu=\Zf\bbeta$ for $\bbeta\in\Rb^p$. Then, the parameters are estimated by
\begin{equation*}
	\argmin_{\bbeta,\theta,\sigma^2,\gamma^2} \;
	\log \text{det}\{C_m(\theta,\sigma^2)+\gamma^2 I_m\}
	+(\mathbf{X}-\Zf\bbeta)^\top\{C_m(\theta,\sigma^2)+\gamma^2 I_m\}^{-1}(\mathbf{X}-\Zf\bbeta).
\end{equation*}
Profiling out $\bbeta$, we can obtain the estimates for the target parameters  $(\theta,\sigma^2,\gamma^2)$ by solving the following problem
\begin{equation*}
\begin{aligned}
	\argmin_{\theta,\sigma^2,\gamma^2} \;
	\log \text{det} &\{C_m(\theta,\sigma^2)+\gamma^2 I_m\}
	+\{\mathbf{X}-\Zf\widehat{\bbeta}(\theta,\sigma^2,\gamma^2)\}^\top \\
	& \qquad \qquad \qquad \qquad \qquad \qquad \times \{C_m(\theta,\sigma^2)+\gamma^2 I_m\}^{-1}\{\mathbf{X}-\Zf\widehat{\bbeta}(\theta,\sigma^2,\gamma^2)\},
\end{aligned}
\end{equation*}
where $\widehat{\bbeta}(\theta,\sigma^2,\gamma^2)=\argmin_{\bbeta}
	(\mathbf{X}-\Zf\bbeta)^\top\{C_m(\theta,\sigma^2)+\gamma^2 I_m\}^{-1}(\mathbf{X}-\Zf\bbeta)$.

\begin{remark}{\label{rmk:cov_multi_obs}}
{\rm 
In some applications, we have multiple observations from the spatial random field
\begin{align*}
    X_i(s)=\mu(s)+\epsilon_i(s),\quad i=1,2,\dots,n.    
\end{align*}
The joint log-likelihood function is given by
\begin{align*}
    l_{n,m}(\boldsymbol{\mu},\theta,\sigma^2,\gamma^2)& =-\frac{mn}{2}\log(2\pi)-\frac{n}{2}\log \text{det}\{C_m(\theta,\sigma^2)+\gamma^2 I_m\} \\
    & \qquad -\frac{1}{2}\sum^{n}_{i=1}(\mathbf{X}_i-\boldsymbol{\mu})^\top\{C_m(\theta,\sigma^2)+\gamma^2 I_m\}^{-1}(\mathbf{X}_i-\boldsymbol{\mu}),    
\end{align*}
with $\mathbf{X}_i=(X_i(s_1),\dots,X_i(s_m))^\top$. In this case, we can estimate the covariance parameters by solving the problem
\begin{align*}
    \argmin_{\theta,\sigma^2,\gamma^2} \; n\log \text{det}\{C_m(\theta,\sigma^2)+\gamma^2 I_m\}+\sum^{n}_{i=1}(\mathbf{X}_i-\bar{\mathbf{X}})^\top\{C_m(\theta,\sigma^2)+\gamma^2 I_m\}^{-1}(\mathbf{X}_i-\bar{\mathbf{X}})
\end{align*}
with $\bar{\mathbf{X}}=n^{-1}\sum^{n}_{i=1}\mathbf{X}_i$. 
}
\end{remark}

\section{Searching Algorithm}\label{ref:searching_alg}

This section provides more details about Algorithm \ref{alg:fast_alg} in Section~\ref{sec:imp_detail} of the main paper, and examines its computational complexity through numerical studies. 
Finding the optimal thresholds requires solving the constrained optimization problem \eqref{eq-opt} of the main paper. Due to the discrete nature of the problem, the solution can be obtained if we replace $\mathcal{F}_q$ by
\begin{align*}
	\{(t_1,t_2)\in\mathcal{T}:\widehat{\FDP}_{\lambda,\tScal}(t_1,t_2)\leq q \},
\end{align*}
where $\mathcal{T}=\{(\whT_1(s),\whT_2(s')): s,s'\in \mathcal{S}\}$ is the set of all candidate cutoff values. A naive grid search algorithm would require evaluating $\widehat{\FDP}_{\lambda,\tScal}$ at $|\mathcal{S}|^2$ different values, which is computationally prohibitive for a large number of spatial locations. Interestingly, we show that there exists a faster algorithm that 
retains an exact maximization of \eqref{eq-opt} shown in the main paper. We derive our algorithm in three steps utilizing the specific structure of the optimization problem. For ease of presentation, we assume that there is no tie among $\{\widehat{T}_j(s):s\in\mathcal{S}\}$ for $j=1,2.$

\textbf{Step 1.} We partition the candidate set $\mathcal{T}$ into $I$ subsets (say $\{\mathscr{S}_i\}^{I}_{i=1}$) such that the rejection set remains unchanged using the cutoffs within the same subset. For example, let $\Dcal(t_1,t_2)=\{s\in\Scal:\whT_1(s)\geq t_1,\whT_2(s)\geq t_2\}$ be the set of locations rejected using the cutoff $(t_1,t_2)$. Then we have $\Dcal(t_1,t_2)=\Dcal(t_1',t_2')$ for $(t_1,t_2),(t_1',t_2')\in \mathscr{S}_i$. As $ L\left(t_{1}, t_{2}, x, \whrho(s)\right)$ is a non-increasing function of $(t_1,t_2)$, we know that $\widehat{\FDP}_{\lambda,\tScal}(t_1,t_2)$ is a non-increasing function within each $\mathscr{S}_i$ (note that $\widehat{R}(t_1,t_2)$ is a constant over $\mathscr{S}_i$).
For $(t_1,t_2)\in\mathscr{S}_i$, $\widehat{\FDP}_{\lambda,\tScal}(t_1,t_2)$ achieves its minimum value at $(\max_{(t_1,t_2)\in\mathscr{S}_{i}}t_1,	\max_{(t_1,t_2)\in\mathscr{S}_{i}}t_2)$. Thus we can reduce the candidate set from $\mathcal{T}$ to 
\begin{equation}\label{equ:intui_Tcal}
\Tcal'=\left\{\left(	\max_{(t_1,t_2)\in\mathscr{S}_{i}}t_1,	\max_{(t_1,t_2)\in\mathscr{S}_{i}}t_2\right):1\leq i\leq I\right\},
\end{equation}
where the maximum is defined to be infinity when $\mathscr{S}_i=\emptyset$. For each $\mathscr{S}_i$, we let $\mathcal{D}_i\subseteq \mathcal{S}$ be the set of locations associated with the hypotheses being rejected.
It is not hard to verify that 
\begin{equation}\label{equ:intui_Tcal-2}
\left(	\max_{(t_1,t_2)\in\mathscr{S}_{i}}t_1,	\max_{(t_1,t_2)\in\mathscr{S}_{i}}t_2\right)=\left(	\min_{s\in\mathcal{D}_{i}}\whT_{1}(s),	\min_{s\in\mathcal{D}_{i}}\whT_{2}(s)\right).
\end{equation}

Below we derive an alternative expression for $\mathcal{T}'$ which facilitates the implementation of our fast algorithm. Without loss of generality, let us assume that $\Scal = \{s_1,s_2,\cdots,s_m\}$ and 
\begin{align}\label{equation-order}
\whT_{2}(s_1)>\whT_{2}(s_2)>\cdots>\whT_2(s_m).    
\end{align}
We claim that
\begin{equation}\label{equ:cal_Tcal}
 \Tcal'=
 \left\{ (\whT_{1}(s_l),\whT_{2}(s_k))
    :\whT_{1}(s_l)\leq \whT_{1}(s_k) \mbox{ and } l\leq k, \, k=1,2,\ldots,m\right\} \cup\left\{(\infty,\infty)\right\}.
\end{equation}
To prove the above result, let us consider any cutoff $(\whT_{1}(s_l),\whT_{2}(s_k))$ from the set defined in the RHS of 
\eqref{equ:cal_Tcal}. There exists a $1\leq i\leq I$ such that the corresponding set of rejected locations is given by $\mathcal{D}_{i}$. Then we have
\begin{equation}\label{eqn:convenSupp}
	\min_{s\in\mathcal{D}_{i}}\whT_{1}(s)\geq \whT_{1}(s_l) \quad \mbox{and} \quad \min_{s\in\mathcal{D}_{i}}\whT_{2}(s)\geq \whT_{2}(s_k).
\end{equation}
Also note that $\mathcal{H}_{s_l}$ and $\mathcal{H}_{s_k}$ are both rejected as $\whT_{2}(s_l)\geq \whT_{2}(s_k)$ for $l\leq k$ and $\whT_{1}(s_l)\leq \whT_{1}(s_k)$ by the requirement in (\ref{equ:cal_Tcal}). Hence, both inequalities in \eqref{eqn:convenSupp} become equalities, i.e., $\min_{s\in\mathcal{D}_{i}}\whT_{1}(s)=\whT_{1}(s_l)$
and $\min_{s\in\mathcal{D}_{i}}\whT_{2}(s)= \whT_{2}(s_k)$, which shows the RHS of (\ref{equ:cal_Tcal}) belongs to $\mathcal{T}'$. To show the other direction, 
we note that for any $1\leq i\leq I$, there exist $1\leq k,l\leq m$ such that $s_{k},s_{l}\in\mathcal{D}_i$, $\whT_1(s_{l})=\min_{s\in\mathcal{D}_i} \whT_1(s)$ and $\whT_2(s_{k})=\min_{s\in\mathcal{D}_i} \whT_2(s)$. Therefore, we get $\whT_{1}(s_{l})\leq \whT_{1}(s_{k})$ and $\whT_{2}(s_{k})\leq \whT_{2}(s_{l})$. In view of (\ref{equation-order}), we must have $l\leq k$ and thus $(\whT_1(s_{l}),\whT_2(s_{k}))$ is an element of the set in the RHS of \eqref{equ:cal_Tcal}.

\begin{figure}
    \centering
    \includegraphics[width=0.95\textwidth]{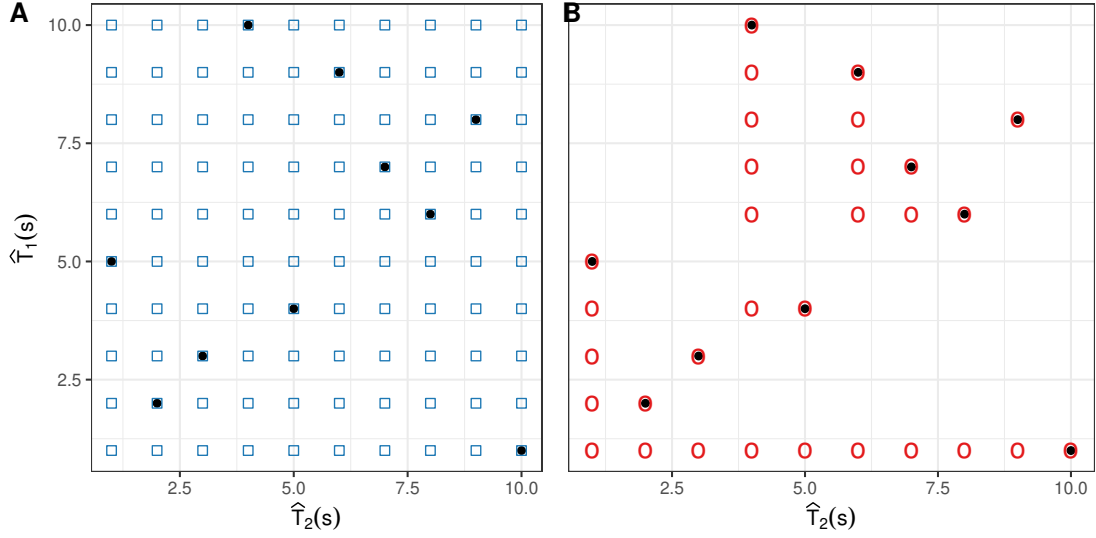}
    \caption{The black points denote the values of the test statistics. The blue squares in Panel A denote all the candidate cutoffs in $\Tcal$. The red circles in Panel B correspond to the cutoffs in $\mathcal{T}'\setminus\{(\infty,\infty)\}$.}
    \label{fig:all_final}
\end{figure}

To understand the complexity of $\mathcal{T}'$, we point out two extreme cases. In the first case, we assume that $\whT_{1}(s_1)>\whT_{1}(s_2)>\cdots>\whT_1(s_m)$. Then we have $\mathcal{T}'=\{ (\whT_{1}(s_k),\whT_{2}(s_k)):k=1,2,\dots,m\}\cup\{(\infty,\infty)\}$ and hence $|\mathcal{T}'|=m+1.$ In the second case, suppose $\whT_{1}(s_1)<\whT_{1}(s_2)<\cdots<\whT_1(s_m)$. Then $\Tcal'=\left\{ (\whT_{1}(s_l),\whT_{2}(s_k)): l\leq k, \, k=1,2,\ldots,m\right\} \cup\left\{(\infty,\infty)\right\}$ and $|\Tcal'|=m(m+1)/2+1$. In general, the cardinality of $\Tcal'$ is between these extreme cases. See
Figure~\ref{fig:all_final} for an illustration about $\mathcal{T}'$ in the intermediate case.

\textbf{Step 2.}
Suppose that we have found some cutoff $(t_1,t_2)\in\Rb^2$ such that $\widehat{\FDP}_{\lambda,\tScal}(t_1,t_2)\leq q$. Then we can reduce the candidate set by removing those cutoffs whose rejection numbers are less than $\widehat{R}(t_1,t_2)$. For example, let
$$
\widetilde{t}_2^{\#}=\min\argmax_{t_2:\widehat{\FDP}_{\lambda,\tScal}(-\infty,t_2)\leq q}\whR(-\infty,t_2),
$$
which can be efficiently computed using the BH procedure. Clearly, $\widehat{R}(t_1,t_2)\leq  \widehat{R}(-\infty,\widetilde{t}_2^{\#})$ for $t_2>\widetilde{t}_2^{\#}$. Thus we can further reduce $\mathcal{T}'$ to 
$
\mathcal{T}''=\big\{(t_1,t_2)\in\Tcal':t_2\leq \widetilde{t}_{2}^{\#}\big\}
$.

\textbf{Step 3.} Searching over $\mathcal{T}''$ can still be computationally expensive. Here we show that pruning can be used to increase the computational efficiency whilst still ensuring that the method finds a global optimum of \eqref{eq-opt} of the main paper. The essence of pruning in this context is to remove those cutoffs that can never deliver the best number of rejections.
To introduce pruning, we denote the elements in $\mathcal{T}''$ by $(t_{1,i,j},t_{2,i})$ for $i=1,2\dots,\breve{m}$ and $1\leq j\leq m_i$, where $\breve{m}$ is the number of $\widehat{T}_2(s)$'s that are no larger than $\widetilde{t}_{2}^{\#}$. Suppose the points are sorted in the following way: (1) $t_{2,1}>t_{2,2}>\cdots>t_{2,\breve{m}}$; (2) 
$t_{1,i,1}>t_{1,i,2}>\cdots>t_{1,i,m_i}$ for all $1\leq i\leq \breve{m}$.
We then examine the candidate cutoffs in a sequential way. In the outer loop of Algorithm \ref{alg:fast_alg} of the main paper below, we consider $t_{2,i}$ for $i$ running from 1 to $\breve{m}$. In the inner loop, for a given $i$, we consider $t_{1,i,j}$ for $j$ running from 1 to $m_i$.
Let $(t_1^{C},t_2^{C})$ be the best cutoff value we have found so far that delivers the largest number of rejections while controlling $\widehat{\text{FDP}}_{\lambda,\tScal}$ at the desired level $q$.
We skip all the remaining cutoffs whose rejection numbers are less than $\whR(t_1^{C},t_2^{C})$. We replace $(t_1^{C},t_2^{C})$ by $(t_{1,i,j},t_{2,i})$ if one of the following two conditions is satisfied:
\begin{enumerate}
    \item $\whR(t_{1,i,j},t_{2,i})=\whR(t_1^{C},t_2^{C})$ and $\widehat{\FDP}_{\lambda,\tScal}(t_{1,i,j},t_{2,i})<\widehat{\FDP}_{\lambda,\tScal}(t_1^{C},t_2^{C})\leq q$; 
    \item $\whR(t_{1,i,j},t_{2,i})>\whR(t_1^{C},t_2^{C})$ and $\widehat{\FDP}_{\lambda,\tScal}(t_{1,i,j},t_{2,i})\leq q$.
\end{enumerate}
When the cutoff $(t_{1,i,j},t_{2,i})$ being examined satisfies neither condition, instead of moving directly to the next cutoff on the list, we can use the statistics we have computed in the current step to decide the minimum number of rejections required for the next cutoff. 
Specifically, note that
$$
	\frac{\int L\left(t_{1,i,j},t_{2,i}, x, \whrho(s)\right) d \whG_{\tScal}(x)}{\whR(t_{1,i,j'},t_{2,i})}
	\leq
	\frac{\int L\left(t_{1,i,j'},t_{2,i}, x, \whrho(s)\right) d \whG_{\tScal}(x)}{\whR(t_{1,i,j'},t_{2,i})} =
	\widehat{\FDP}_{\lambda,\tScal}(t_{1,i,j'},t_{2,i}).
$$
For a cutoff to be valid, we need $\widehat{\FDP}(t_{1,i,j'},t_{2,i})\leq q$, which implies that 
$$\begin{aligned}
    \whR(t_{1,i,j'},t_{2,i}) &\geq \left\lceil\frac{1}{q}\int L\left(t_{1,i,j},t_{2,i}, x, \whrho(s)\right) d \whG_{\tScal}(x) \right\rceil \\
    &=\left\lceil\frac{1}{q}\widehat{\FDP}_{\lambda,\tScal}(t_{1,i,j},t_{2,i})\widehat{R}(t_{1,i,j},t_{2,i})\right\rceil \\
    &:=R_{\text{req}},
\end{aligned}$$
where $\lceil\cdot\rceil$ denotes the ceiling function. When there is no tie, increasing $j$ by $k$ brings $k$ more rejections. Thus we must have
\begin{equation}\label{equ:j_to_j'}
    j'-j\geq \max\left\{1,R_{\text{req}}-\whR(t_{1,i,j},t_{2,i})\right\}.
\end{equation}
This observation is used to prune the candidate set, which improves efficiency. Combining the insights from the above discussions, we propose Algorithm~\ref{alg:fast_alg} of the main paper, the fast searching algorithm.

{
\textbf{Step 4.(Optional)} We notice that iterating $i$ from 1 to $\breve{m}$ in the outer loop is overly time-consuming. To address this, we set $m_{\mathrm{stop}}\geq 0$ and terminate our searching procedure if we encounter an $i^{\star}$ such that $\widehat{\FDP}_{\lambda,\tilde{\Scal}}(t_{1,i^{\star},j},t_{2,i^{\star}})\leq q$ for some $j\in[m^\star_i]$ and $\widehat{\FDP}_{\lambda,\tilde{\Scal}}(t_{1,i^{\star}+i,j},t_{2,i^{\star}+i})> q$ for all $j\in[m_{i^\star+i}]$ and $i= 1,\cdots, m_{\mathrm{stop}}$. Although this approach limits iterations in the outer loop and lacks a theoretical guarantee of identifying the optimal cutoff, forthcoming numerical results illustrate that the cutoffs identified after Step 3 and Step 4 are always the same.
}

\begin{remark}
Algorithm~\ref{alg:fast_alg} of the main paper can be modified to find the optimal cutoffs for the 2d procedure coupled with various weighted BH procedures (wBH) as discussed in Section~\ref{sec:vary_null} of the main paper. 
Recall that the rejection rule for the hypothesis at location $s$ is given by $p_1(s)\leq \min\{\tau,w(s)t_1\}$ and $p_2(s)\leq \min\{\tau,w(s)t_2\}$ (or equivalently $p_1(s)\leq \tau$, $p_2(s)\leq \tau$, $p_1(s)/w(s)\leq t_1$, and $p_2(s)/w(s)\leq t_2$), where $p_j(s)=1-\Phi(\whT_j(s))$ for $j=1,2$. 
Thus the set of all candidate cutoff values is 
$$
    \Big\{\big(p_1(s)/w(s),p_2(s)/w(s)\big):p_1(s)\leq\tau, p_2(s)\leq\tau,s\in\Scal\Big\}.
    $$
Algorithm~\ref{alg:fast_alg} of the main paper can be modified to find the optimal thresholds $(t_1,t_2)$ for the weighted p-values.  
\end{remark}

\begin{remark}\label{rmk:with_ties}
{\rm
When there exist ties among $\{\whT_j(s):s\in\Scal\}$ for $j=1,2$, Algorithm \ref{alg:fast_alg} of the main paper still manages to find the target threshold. The key here is to argue that line 11 of Algorithm \ref{alg:fast_alg} of the main paper does not skip any cutoff whose rejection number is no less than $R_{\text{req}}$. 
To see this, suppose 
$$t_{1,i,1}=\cdots= t_{1,i,j_1}> t_{1,i,j_1+1} =\cdots= t_{1,i,j_2}> \cdots >t_{1,i,j_{m'_i-1}+1}=  \cdots= t_{1,i,j_{m'_i}},$$ 
where $j_0=0$ and $j_{m_i'}=m_i$. 
Obviously, the rejection numbers of the nearby cutoffs satisfy that $\whR( t_{1,i,j},t_{2,i})=\whR( t_{1,i,j_{k}},t_{2,i})$ for $j_{k-1}< j\leq j_{k}$ and $\whR( t_{1,i,j_{k}},t_{2,i})-\whR( t_{1,i,j_{k-1}},t_{2,i})=j_{k}-j_{k-1}$ for all $0< k\leq m'_i$. 
For non-nearby cutoffs, suppose they locate between $j_{k-1}<j=j_{k}-l\leq j_k$ and $j_{k'-1}<j'=j_{k'}-l'\leq j_{k'}$ for some $j<j'$ and $0<k,k'\leq m'_{i}$. 
Then, 
$$\begin{aligned}
    \whR( t_{1,i,j'},t_{2,i})-\whR( t_{1,i,j},t_{2,i})&=\whR( t_{1,i,j_{k'}},t_{2,i})-\whR( t_{1,i,j_k},t_{2,i})\\
    &=\sum_{l=k}^{k'-1}(j_{l+1}-j_{l})\\
    &=j_{k'}-j_{k}.
\end{aligned}$$
We now show that the rejection numbers of the skipped cutoffs are no more than $R_{\text{req}}$. 
Start with an arbitrary cutoff $(t_{1,i,j},t_{2,i})$ whose rejection number is less than $R_{\text{req}}$, i.e., $\whR(t_{1,i,j},t_{2,i})<R_{\text{req}}$.
Line 11 in Algorithm~\ref{alg:fast_alg} suggests $j'=j+R_{\text{req}}-\whR(t_{1,i,j},t_{2,i})$. 
As $j=j_{k}-l$, $j'=j_{k'}-l'$ and $\whR( t_{1,i,j'},t_{2,i})-\whR( t_{1,i,j},t_{2,i})=j_{k'}-j_{k}$, we have
$$\whR( t_{1,i,j'},t_{2,i})-\whR( t_{1,i,j},t_{2,i})=j_{k'}-j_k=l'-l+R_{\text{req}}-\whR(t_{1,i,j},t_{2,i}),$$
which implies that 
$$\whR( t_{1,i,j'},t_{2,i})= R_{\text{req}}+l'-l.$$
We thus only need to verify that the rejection number of $( t_{1,i,j_{k'-1}},t_{2,i})$ is less than $R_{\text{req}}$. 
Indeed, when $l'\leq l$, 
$$\whR( t_{1,i,j_{k'-1}},t_{2,i})<\whR(t_{1,i,j'},t_{2,i})\leq R_{\text{req}}.$$ 
When $l<l'$, 
$$\whR( t_{1,i,j_{k'-1}},t_{2,i})= R_{\text{req}}+l'-l-(j_{k'}-j_{k'-1})<R_{\text{req}}-l\leq R_{\text{req}}.$$
In both cases, the skipped cutoffs induce no more than $R_{\text{req}}$ rejections.
}
\end{remark}

We briefly analyze the computational complexity of our fast searching algorithm. The computational complexity of estimating the FDP in \eqref{equ:def_tfdp} of the main paper for a single cutoff is $O(m)$ because the number of supporting points of $\widehat{G}_{\tilde{\Scal}}(x)$ is fixed in practice. The total computational complexity primarily depends on the number of candidate cutoffs required for the FDP estimation. Considering the stochastic nature of primary and auxiliary statistics, we numerically investigated the relationship between the number of candidate cutoffs and the number of locations. This analysis was conducted within the framework of Setup I in Section~\ref{sec:simu} of the main paper, focusing on scenarios with medium signal strength, medium correlation, and $\gamma=2$. We varied the location size $m$ from $100$ to $2000$ and defined the associated domain as $\Scal=[0, m/30]$.

\begin{figure}[!h]
     \centering
    \includegraphics[width=\textwidth]{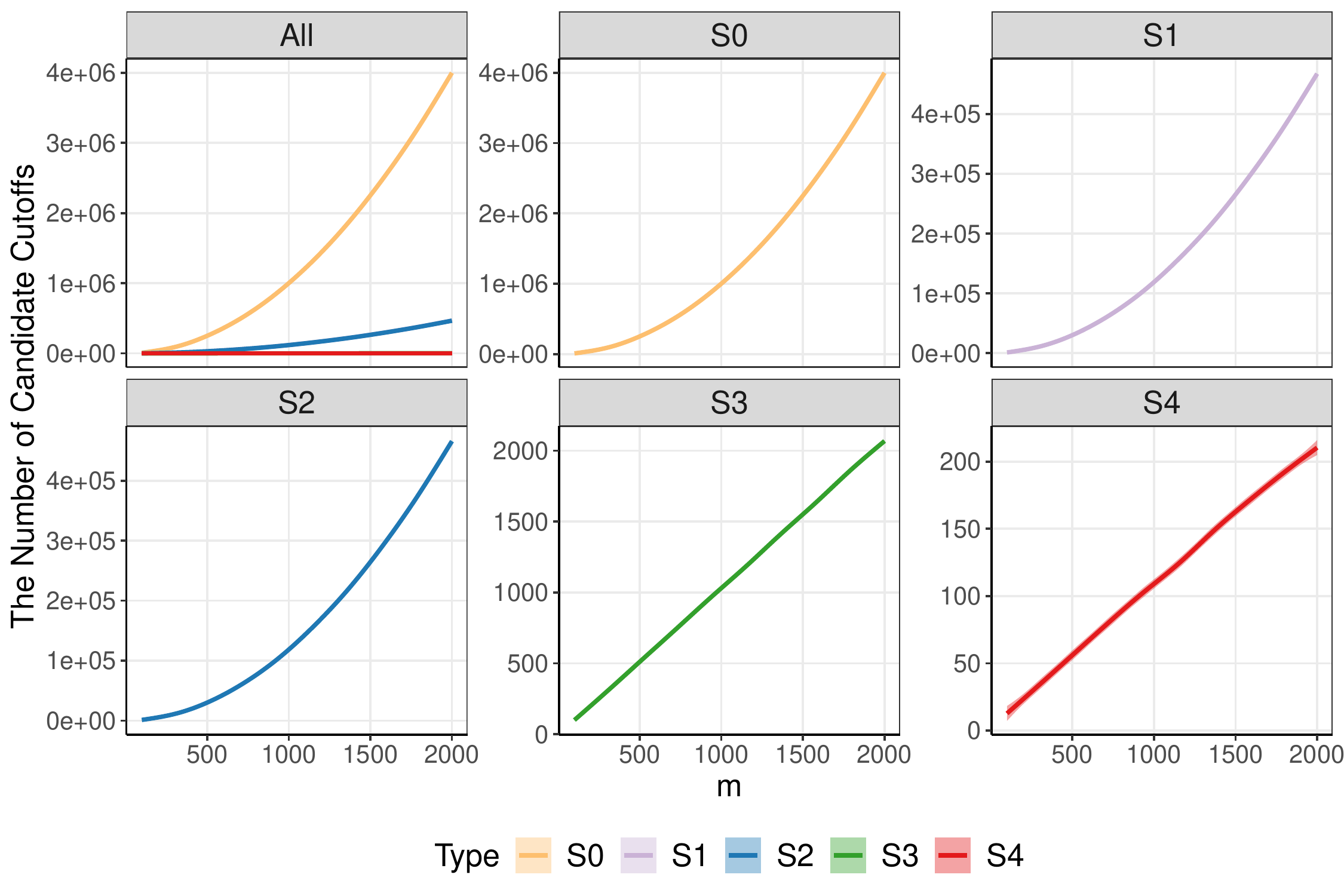}
        \caption{The number of candidate cutoffs for Setup I with the medium signal, medium dependency, $\gamma=2$, and $m$ locations. Different colors represent the number of candidate cutoffs for the naive grid search (S0, $m^2$), after Step 1 (S1, $|\Tcal^\prime|$); after Step 2 (S2, $|\Tcal^{\prime\prime}|$); after Step 3 (S3), and after Step 4 (S4). 
        }
        \label{fig:Time}
\end{figure}

Figure~\ref{fig:Time}~displays the number of candidate cutoffs for different searching strategies: the naive grid search (S0, $m^2$), after Step 1 (S1, $|\Tcal^\prime|$); after Step 2 (S2, $|\Tcal^{\prime\prime}|$); after Step 3 (S3), and after Step 4 (S4).  The number of candidate cutoffs for strategies S0, S1, and S2 appeared to scale quadratically with the number of locations. The difference between S0 and S1 highlighted the efficiency gains from implementing Step 1. In contrast, the nearly identical performances of S1 and S2 suggested that Step 2 did not significantly enhance computational speed. However, for strategies S3 and S4, the number of candidate cutoffs increased linearly with the number of locations, indicating substantial computational acceleration due to Step 3. Finally, the cutoffs determined after executing Steps 1–3 were identical to those obtained after completing Steps 1–4, which showcases that Step 4 could accelerate searching without sacrificing accuracy. To sum up, Step 3 in our fast searching algorithm can drastically reduce the computational complexity and be further improved by Step 4.

\bibhang=1.7pc
\bibsep=2pt
\fontsize{9}{14pt plus.8pt minus .6pt}\selectfont
\renewcommand\bibname{\large \bf References}
\expandafter\ifx\csname
natexlab\endcsname\relax\def\natexlab#1{#1}\fi
\expandafter\ifx\csname url\endcsname\relax
  \def\url#1{\texttt{#1}}\fi
\expandafter\ifx\csname urlprefix\endcsname\relax\def\urlprefix{URL}\fi

\bibhang=1.7pc
\bibsep=2pt
\fontsize{9}{14pt plus.8pt minus .6pt}\selectfont
\renewcommand\bibname{\large \bf References}
\expandafter\ifx\csname
natexlab\endcsname\relax\def\natexlab#1{#1}\fi
\expandafter\ifx\csname url\endcsname\relax
\def\url#1{\texttt{#1}}\fi
\expandafter\ifx\csname urlprefix\endcsname\relax\def\urlprefix{URL}\fi

\bibliographystyle{asa}
\bibliography{mybib}